 \documentclass[final,authoryear,5p]{elsarticle}

 \usepackage{graphicx}
\usepackage{amssymb}

\usepackage[a4paper=true,%
breaklinks=true,%
colorlinks=true,%
pdfauthor={Inigo Arregui},%
pdftitle={Bayesian Coronal Seismology}%
]{hyperref}

\journal{Advances in Space Research}

\newcommand{\thetabold}{\mbox{\boldmath$\theta$}}
\newcommand{\Thetabold}{\mbox{\boldmath$\Theta$}}

\newcommand{\Omegabold}{\mbox{\boldmath$\Omega$}}

\begin{document}

%%%%%%%%%%%%%%%%%%%%%%%%%%%%%%%%%%%%%%%%%%%%%%%%%%%%%%%%%%%%%%%%%%%%%%%%%%%%%
%% Frontmatter
\begin{frontmatter}


% Use the tnoteref command within \title and fnref within \author or \address for footnotes;
% use the corref command within \author for corresponding author footnotes;
% \author{Name\corref{cor1}\fnref{label2}}
% \address{Address\fnref{label3}}

\title{Bayesian Coronal Seismology}
\tnotetext[footnote1]{This template can be used for all publications in Advances in Space Research.}

% Use optional labels to link authors explicitly to addresses:
\author{I\~nigo Arregui}
\address{Instituto de Astrof\'{\i}sica de Canarias, E- 38205 La Laguna, Spain\\Departamento de Astrof\'{\i}sica, Universidad de La Laguna, E-38206 La Laguna, Spain}
%\cortext[cor]{Corresponding author}
%\fntext[footnote2]{Additional information regarding the corresponding author}
\ead{iarregui@iac.es}

% Url can be given like this:
% \ead[url]{http://www.elsevier.com/wps/find/authorsview.authors/latex}

\begin{abstract}
In contrast to the situation in a laboratory, the study of the solar atmosphere has to be pursued without direct access to the 
physical conditions of interest. Information is therefore incomplete and uncertain and inference methods need to be employed 
to diagnose the physical conditions and processes. One of such methods, solar atmospheric seismology, makes use of observed 
and theoretically predicted properties of waves to infer plasma and magnetic field properties. A recent development 
in solar atmospheric seismology consists in the use of inversion and model comparison methods based on Bayesian analysis. 
In this paper, the philosophy and methodology of Bayesian analysis are first explained. Then, we provide an account of what 
has been achieved so far from the application of these techniques to solar atmospheric seismology and a prospect of possible 
future extensions.
\end{abstract}


\end{frontmatter}

\parindent=0.5 cm

%%%%%%%%%%%%%%%%%%%%%%%%%%%%%%%%%%%%%%%%%%%%%%%%%%%%%%%%%%%%%%%%%%%%%
%% Main text

%%%%%%%%%%%%%%%%%%%%%%%%%%%%%%%%%%%%%%%%%%%%%%%%%%%%%%%%%%%%%%%%%%%%%
\section{Introduction}
%%%%%%%%%%%%%%%%%%%%%%%%%%%%%%%%%%%%%%%%%%%%%%%%%%%%%%%%%%%%%%%%%%%%%

% (1) Solar atmospheric seismology - useful and successful
Solar atmospheric seismology aims to obtain information about difficult to measure physical parameters in the solar atmosphere
by a combination of observed and theoretical properties of magnetohydrodynamic (MHD) waves \citep{uchida70,rosenberg70,roberts84,nakariakov05,demoortel05,banerjee07,demoortel12,arregui12b}. The technique has been 
successful in the determination of a number of parameters in coronal, prominence, and chromospheric plasmas such as the 
magnetic field strength \citep{nakariakov01},  the radial density contrast and the Alfv\'en travel time \citep{goossens08a}, the coronal density scale height \citep{andries05a}, or the Alfv\'en speed \citep{verwichte13a}.

% (2) Nature of the inversion problem - need probabilistic approach
Seismology diagnostic studies are based on the adoption of a theoretical model to explain the observed oscillations, the solution of 
the forward problem to obtain the theoretically predicted wave properties as a function of the model parameters, the comparison with 
the observed wave properties, and the solution to the inverse problem to extract the magnetic and plasma parameters of interest.  
The forward problem does not involve difficulties other than those arising from the analytical/numerical solution of the MHD wave 
equations. The inverse process of extracting information on the physical conditions from measured wave properties  is a more involved task. The solution to such inverse problem might not exist or be unique. Measurements of 
wave properties have always a certain degree of uncertainty associated to e.g., the existence of noise.  Even if there is a unique 
analytical relation between one observable and one model parameter, the presence of noise can make the inversion completely useless. 
In other cases, inferring parameters from observables may consist in the solution of a mathematically ill-posed  problem in which the 
amount of unknowns outnumbers that of observables. In summary, the solution to the inverse problem has to be pursued under 
conditions in which direct access to the physical conditions of interest is not possible and indirect observational information is always incomplete and uncertain. For these reasons,  extracting information on physical parameters by comparison of theoretical model 
predictions with observed data has to be carried out in a probabilistic framework. This means that our conclusions will at best be 
probabilities, where probability refers to the quality that enables us to quantify uncertainty in terms of degree of belief.

% (3) Parameter inference is model dependent - need model comparison
Once the inversion problem is correctly solved,  one must realise that the resulting inference depends on the 
underlying theoretical model that has been assumed. This makes model comparison the next necessary 
task to be performed in order to assess the plausibility of any obtained inference. This requires to devise 
methods to perform a comparison between alternative hypotheses that enable us to present different theoretical models to the 
same data to assess, in a quantitative manner,  which model is favoured by the data.

% (4) The Bayesian approach 
A recent development in solar atmospheric seismology has been the adoption of the Bayesian approach to the inference and 
model comparison problems. This comes from the realisation that the Bayesian approach is the only correct way we have 
to obtain information about physical parameters from observations (inference) and to compare the relative  performance of 
alternative models to explain observed data (model comparison) \citep{vontoussaint11}. The method has been successfully 
applied to areas such as cosmology  or extrasolar planet detection, which are now mostly based on the application of the 
Bayesian paradigm \citep[see e.g.,][]{loredo92,gregory05,trotta08}. This approach provides a natural and principled way of 
combining prior information, model predictions, and observations, providing inferences that are conditional on the data. The 
final product, in the form of posterior probability density functions, provides interpretable statements, such as the probability 
of a parameter value falling in a given credible interval. Probability in this context has an unambiguous meaning as the 
measure of the grade of belief on the parameter of interest taking on a given value conditional on the observed data. 
For model comparison, the Bayesian formalism provides well defined levels of evidence for the plausibility of a given 
model with respect to an alternative explanation.

% (5) The aim of the paper - explain Bayes and see how it has been applied to solar atmospheric seismology
The aim of this paper is to explain the philosophy and methodology of the Bayesian approach and to show its feasibility for 
solar atmospheric seismology by describing some initial applications in the area of seismology of coronal loops and 
prominence fine structures using small amplitude MHD waves. We are motivated by the hope that these early success 
cases will encourage future Bayesian developments in this and other areas.

% (6) Layout of the article
The layout of the article is as follows. Section~\ref{method} describes the fundamentals of Bayesian analysis and its 
three levels of inference: parameter inference, model comparison, and model averaging,  together with a description of hierarchical multi-level models and methods to compute the probability density functions. In Section~\ref{apps}, a number of examples are shown in which these three levels of inference are used in combination with 
observations of wave dynamics and theoretical models to obtain information about physical conditions and their structuring in coronal and prominence plasmas. We finish presenting a summary in Section~\ref{summary}.

%%%%%%%%%%%%%%%%%%%%%%%%%%%%%%%%%%%%%%%%%%%%%%%%%%%%%%%%%%%%%%%%%%%%%
\section{Bayesian methodology}\label{method}
%%%%%%%%%%%%%%%%%%%%%%%%%%%%%%%%%%%%%%%%%%%%%%%%%%%%%%%%%%%%%%%%%%%%%

In the Bayesian framework, the solution to an inverse problem involves the calculation of a conditional probability using 
Bayes' Theorem \citep{bayes63,laplace74}. Consider the problem of inferring a set of parameters {\boldmath$\theta$} of a theoretical model $M$, conditional to observed data $d$. The Theorem states that the probability of the parameters taking on given values is a combination of how well the data are reproduced by the model parameters and of the probability of the parameters independently of the observed data. These quantities are related as follows

\begin{equation}\label{bayes}
p(\mbox{\boldmath$\theta$} | d,M)=\frac{p(d | \mbox{\boldmath$\theta$},M)p(\mbox{\boldmath$\theta$}|M)}{\int p(d|\mbox{\boldmath$\theta$}, M)p(\mbox{\boldmath$\theta$}| M)d\mbox{\boldmath$\theta$}}.
\end{equation}

\noindent
Here, $p(\mbox{\boldmath$\theta$} | d, M)$ is the posterior probability density function of the parameters conditional on the observed data and the assumed theoretical model;  $p(d| \mbox{\boldmath$\theta$}, M)$ is the likelihood function, which represents the probability of observing the data that have been obtained as a function of the parameters {\boldmath$\theta$}; and  $p(\mbox{\boldmath$\theta$}| M)$ is the prior probability density function, which encodes our belief on the parameter values before considering the observed data.  The denominator is the so-called evidence or marginal likelihood,  an integral of the likelihood over the prior distribution that normalises the likelihood and turns it into a probability.  Note that all quantities in Equation~(\ref{bayes}) are conditional on the assumed model $M$.
In the inference process, both prior and likelihood are directly assigned, while the posterior is computed.

Bayesian inference to calculate probabilities conditional on data makes use of all the available information: prior understanding, observed data, and the assumed theoretical model. Equation~(\ref{bayes}) also constitutes a mathematical expression for the process of learning from experience and of the logic of science, whereby prior understanding is combined with new data to update our knowledge. 
Bayesian inference makes prior assumptions explicit, which has the advantage that one has full control over the 
information that one is introducing a priori and can explicitly know its effects, if any.  

Two limiting situations may then arise wherein the posterior will eventually be dominated by either the likelihood or by the prior. This will depend on the characteristics of the problem, the information available on the data, and the strongly/weakly informative character of the chosen prior information. As a general rule, the prior will have little effect on the inference unless the prior is as informative as the data. 
When results are prior dependent, what we know after consideration of the data strongly depends on what 
we knew without the data.  If the posterior depends sensitively on the prior, we still learn that the 
data provide little information. In this case, the Bayesian approach automatically warns us when the data are uninformative.

The adoption of Bayesian analysis to solve inverse problems in solar atmospheric seismology has made use of  a number of different levels of inference and methods to compute the relevant probability density functions. These are next described.

\subsection{Parameter inference}

The first level of Bayesian analysis is parameter inference. Once the full posterior probability density function has been computed, the
marginal posterior distribution of a given parameter $\theta_i$ compatible with the observed data $d$ can be obtained by integration of the full posterior with respect to the remaining parameters,

\begin{equation}\label{marginalposterior}
p(\theta_i|d, M) = \int p(\mbox{\boldmath$\theta$} | d, M) d\theta_1 \ldots  d\theta_{i-1} d\theta_{i+1} \ldots d\theta_{N}.
\end{equation}

\noindent
The marginal posterior distribution encodes all information for model parameter $\theta_i$ available in the priors and the data. Furthermore, it correctly propagates uncertainty in the rest of the parameters to the one of interest. Once the marginal posterior distribution is known, the probability density function may be summarised in different ways, by providing the mean; the mode; the median; etc. Another convenient way is to provide the maximum a posteriori estimate of the parameter of interest, $\theta^{\rm MAP}_i$, the value of $\theta_i$ that makes the posterior as large as possible.

\subsection{Model comparison}

Any inference result depends on the particular model that has been adopted.  The second level of Bayesian analysis is model comparison, which enables us to quantify the relative plausibility between competing theoretical models when explaining a given observation. 

A first data analysis tool to evaluate the plausibility of a model in explaining observed data is the marginal likelihood, which quantifies the probability of the observed data $d$, given that the assumed model $M$ is true. The marginal likelihood  tells us how well the observed data are predicted by the model and is computed using the expression

\begin{eqnarray}\label{marginallikelihood}
p(d|M)&=&\int p(d,\mbox{\boldmath$\theta$}|M)d\mbox{\boldmath$\theta$}\nonumber\\
&=&\int p(d|\mbox{\boldmath$\theta$},M)p(\mbox{\boldmath$\theta$}|M)d\mbox{\boldmath$\theta$}.
\end{eqnarray}
This is an integral of the product of the likelihood function and the prior probability density over the full parameter space.

When two alternative models are proposed to explain a given set of observations, Bayesian model comparison offers a way to quantify how much plausible one of them is in comparison to the other. This is done by computing posterior ratios. Applying Equation~(\ref{bayes}) to 
two models, $M_1$ and $M_2$, the posterior ratio is given by

\begin{equation}\label{bayesfactor}
\frac{p(M_1|d)}{p(M_2|d)}=\frac{p(d|M_1)}{p(d|M_2)}\frac{p(M_1)}{p(M_2)}=B_{12}\frac{p(M_1)}{p(M_2)},
\end{equation}
with $B_{12}$ the Bayes factor. If we consider that both models are equally probable a priori, $p(M_1)=p(M_2)$, the prior ratio is unity and the posterior ratio reduces to the Bayes factor, i.e., the ratio of marginal likelihoods for both models.

Once the Bayes factor is computed, the level of evidence for one model in front of the alternative is assessed by relating  the numerical value of the Bayes factor to different levels of evidence. To do so, empirical evidence classification tables such as the one by \cite{kass95} shown in Table~\ref{kasstable} are used.

\begin{table}
\caption{Model evidence classification by \cite{kass95}}
\begin{tabular}{ll}
\hline
$2\ log_{e}B_{12}$&Evidence for $M_1$\\
\hline
0-2 & Inconclusive Evidence (IE)\\
2-6 & Positive Evidence (PE)\\
6-10 & Strong Evidence (SE)\\
$>$10 & Very Strong Evidence (VSE)\\
\hline
\end{tabular}
\label{kasstable}
\end{table}

\subsection{Model averaging}
Sometimes, we may face a situation in which the evidence quantified using Bayes factors is not large enough for any of the proposed models. Still, the Bayesian approach provides us with a tool to perform the most general possible inference. This is accomplished by computing a model-averaged posterior distribution for a given parameter $\theta_i$, conditional on the observed data $d$ and weighted with the probability of our set of alternative models $M_{k}$. Consider three such models, $k=1,2,3$. Then, the model-averaged posterior is calculated as

\begin{eqnarray}\label{average}
p(\theta_i|d)&=&\sum_{k=1,2,3}p(\theta_i|d,M_k)p(M_k|d)\\\nonumber
&=& p(M_1|d)\sum_{k=1,2,3}BF_{k1} p(\theta_i|d,M_k),
\end{eqnarray}
where in the second equality we have adopted $M_1$ as the reference model and replaced the models' posterior probabilities by the Bayes factors, $BF_{k1}$, with respect to the reference model. Obviously, $BF_{11}=1$. The posterior for the reference model, $M_1$, can be calculated by considering that the sum of the probabilities for all three models must be unity, thus

\begin{equation}
p(M_1|d)=\frac{1}{1+\sum_{k=2,3}B_{k1}}.
\end{equation}
The obtained result is independent on the model chosen to be the reference.

\subsection{Hierarchical Bayesian inference}\label{hierarchicalbayes}

Bayesian inference using Equation~(\ref{bayes}) is the inference tool when a single analysis,  using a particular event, is to be carried out. Now suppose we have multiple observations of different events of the same kind (e.g.,\ kink oscillations observed in different active regions at different times), each one giving independent results from each other. We might wish to combine them in a single global analysis. This can be achieved by using hierarchical models which provide a framework to combine these multiple observations in a single ensemble in which each analysis is improved by the information from the others. The overall information is summarised in an empirical prior distribution.

To construct such a hierarchical multi-level model let us gather the set of unknown parameters from $N$ observations in the vector
$\Thetabold$. Their posterior probability distribution is given by

\begin{equation}
p(\Thetabold| \mathbf{D}) = 
\frac{p(\mathbf{D}|\Thetabold) p(\Thetabold)}{p(\mathbf{D})},
\label{eq:bayes_theorem_simple}
\end{equation}

\noindent
where $\mathbf{D}$ refers to the measured observable for all the observations, the function $p(\mathbf{D}|\Thetabold)$ is the likelihood function, and $p(\Thetabold)$ is the prior distribution.

The idea behind hierarchical models is to impose parametric shapes for the priors $p(\Thetabold)$ for the parameters of interest to learn their value from the large set of observations. They are made dependent on a set of hyperparameters, or parameters of the priors, $\Omegabold$, which will have their corresponding hyperpriors, $p(\Omegabold)$, that are included in the inference scheme. The final posterior will be given by

\begin{equation}
p(\Thetabold,\Omegabold| \mathbf{D}) = 
\frac{p(\mathbf{D}|\Thetabold) p(\Thetabold|\Omegabold) p(\Omegabold)}{p(\mathbf{D})}.
\label{eq:bayes_theorem_final}
\end{equation}

\noindent
Under the assumption that there is not any correlation between any two events from the set of $N$ observations, the likelihood and the priors can be factorised, and the posterior simplifies to 

\begin{equation}
p(\Thetabold,\Omegabold| \mathbf{D}) = \frac{1}{p(\mathbf{D})} \prod_{i=1}^N p(D_i|\thetabold_i) p(\thetabold_i|\Omegabold) p(\Omegabold).
\label{eq:final_posterior}
\end{equation}

\noindent
As a final step, in order to compute the statistical properties of the hyperparameters, $\Omegabold$, all the individual physical parameters $\Thetabold$ have to be integrated out from the posterior \citep[e.g.,][]{gregory05}

\begin{equation}
p(\Omegabold| \mathbf{D}) = \frac{p(\Omegabold)}{p(\mathbf{D})} \prod_{i=1}^N \int d\thetabold_i p(D_i|\thetabold_i) p(\thetabold_i|\Omegabold).
\label{eq:marginal_hyperparameters}
\end{equation}
\noindent
This integration propagates information from all individual events simultaneously to the hyperparameters. The obtained priors inferred from the data summarise the global information one currently has about the unknown parameters. This scheme was applied by \cite{asensioramos13} in a study that combines information from multiple coronal loop oscillations and is discussed in Section~\ref{andresresults}.

\subsection{Computation of posteriors and marginal likelihoods}

The computation of posterior distributions and marginal likelihoods using Equations (\ref{marginalposterior}) and (\ref{marginallikelihood}) requires the calculation of integrals over the parameter space. On rare occasions the probability distribution functions will have analytical solutions. In general, one needs to numerically evaluate the distributions. When the dimension of the parameter space is low, a brute force approach consisting on the direct numerical integration over a grid of points is still a feasible option. This approach is followed by e.g., \cite{arregui13a,arregui13b,arregui14,arregui15b,arregui15c} in the studies discussed in Sections~\ref{magneticfield}, \ref{gaussexp}, and \ref{densityacross}.

As the complexity of the theoretical model and the dimension of the parameter space increase, one has to resort to alternatives such as 
the Markov Chain Monte Carlo (MCMC) sampling or Monte Carlo integration to numerically evaluate the posterior distributions and marginal likelihoods, respectively \citep[see e.g.,][for a recent review]{sharma17}. These algorithms enable us to construct a sequence of points in parameter space whose density is proportional to the posterior distribution function therefore providing a method for sampling the posterior distribution up to a multiplicative constant. After the sampling, the results can be given as a histogram of samples simulated from the posterior distribution.

Although MCMC techniques exist since more than 50 years ago \citep{metro53}, it was the increase in computational power which made them feasible and increasingly popular. This approach is followed in the studies by \cite{arregui11b,asensioramos13,pascoe17a,pascoe17b} discussed in Sections~\ref{seisdamp}, \ref{andresresults}, and \ref{pascoeresults}. 

%%%%%%%%%%%%%%%%%%%%%%%%%%%%%%%%%%%%%%%%%%%%%%%%%%%%%%%%%%%%%%%%%%%%%
\section{Applications and results}\label{apps}
%%%%%%%%%%%%%%%%%%%%%%%%%%%%%%%%%%%%%%%%%%%%%%%%%%%%%%%%%%%%%%%%%%%%%

This section presents a selection of recent applications of the Bayesian methodology to seismology of coronal loops and prominence fine structures.

%%%%%%%%%%%%%%%%%%%%%%%%%%%%%%%%%%%%%%%%%%%%%%%%%%%%%%%%%%%%%%%%%%%%%
\subsection{Seismology of damped transverse coronal loop oscillations}\label{seisdamp}
%%%%%%%%%%%%%%%%%%%%%%%%%%%%%%%%%%%%%%%%%%%%%%%%%%%%%%%%%%%%%%%%%%%%%

Our first example deals with the determination of physical parameters in oscillating coronal waveguides. Soon after the first TRACE observations of transverse loop oscillations by \cite{nakariakov99} and \cite{aschwanden99}, their use on the determination of the magnetic field strength was  demonstrated by \cite{nakariakov01}, by making use of observed wave periods and loop lengths. The mechanism of resonant absorption, a transfer of wave motions from large to small scales at the boundaries of the waveguide \citep{hollweg88}, was proposed to explain the observed rapid damping \citep{goossens02a,ruderman02}. It was soon realised that the damping offers an additional seismological tool by relating observed damping time scales with unobservable physical quantities, such as the cross-field density variation length scales \citep{goossens02a}. 

In coronal loops, assuming resonant damping and under the thin tube and thin boundary (TTTB) approximations, the forward problem is reduced to the solution of two algebraic equations for the period and damping ratio of resonantly damped kink oscillations in 1D magnetic flux tubes models

\begin{equation}\label{analyticaltttb}
P=\tau_{\rm Ai}\sqrt{2}\left(\frac{\zeta+1}{\zeta}\right)^{1/2} \mbox{\hspace{0.1cm}} \mbox{and}\mbox{\hspace{0.2cm}} \frac{\tau_{\rm d}}{P}=\frac{2}{\pi}\frac{\zeta+1}{\zeta-1}\frac{1}{l/R}.
\end{equation}
These equations express the period, $P$, and the damping time, $\tau_{\rm d}$, which are observable quantities in terms of the internal Alfv\'en travel time, $\tau_{\rm Ai}$, the density contrast, $\zeta= \rho_{\rm i}/\rho_{\rm e}$, and the transverse inhomogeneity length scale, $l/R$, in units of the loop radius.  The two observable wave properties are therefore functions of three unknown parameters: density contrast, transverse inhomogeneity length-scale, and internal Alfv\'en travel time. The classic inversion technique simply consists of imposing these two functions to be equal to the observed periods and damping times. As we have two observables and three unknowns, there is an infinite number of equally valid equilibrium models that explain observations. However, these solutions must follow a particular 1D solution curve in the 3D parameter space. 

\begin{figure*}
\centering
\includegraphics[scale=0.35]{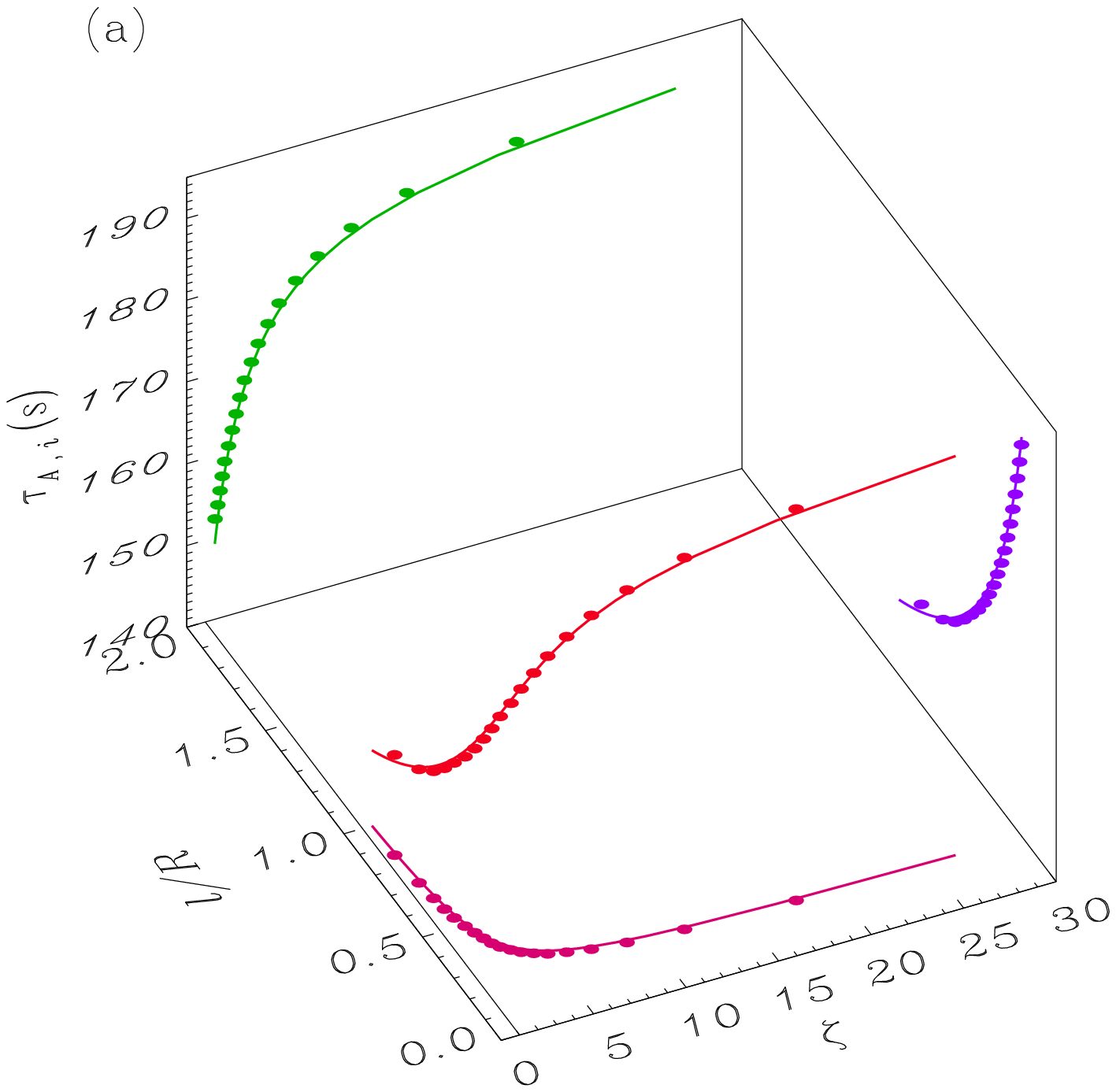}
\includegraphics[scale=0.35]{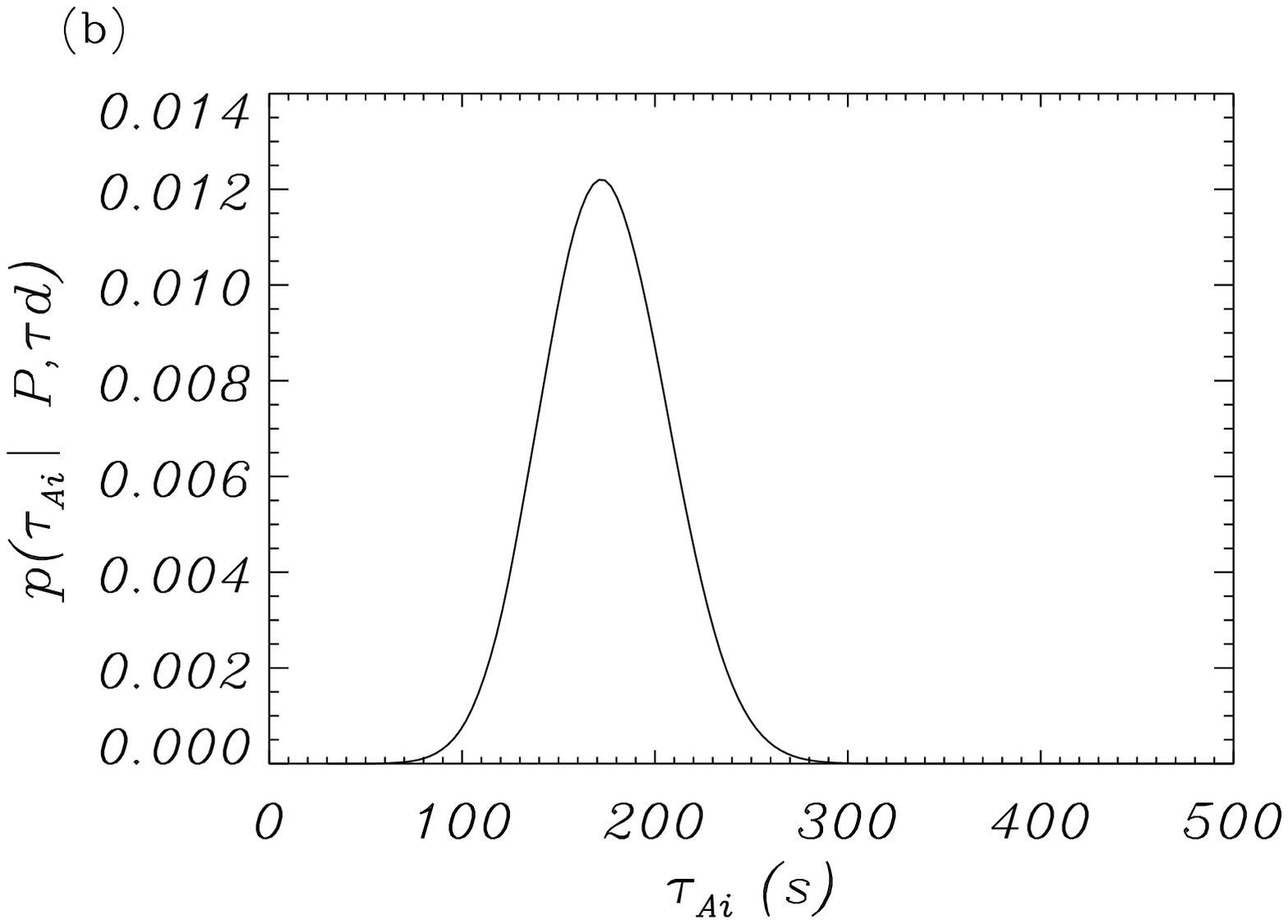}\\
\includegraphics[scale=0.35]{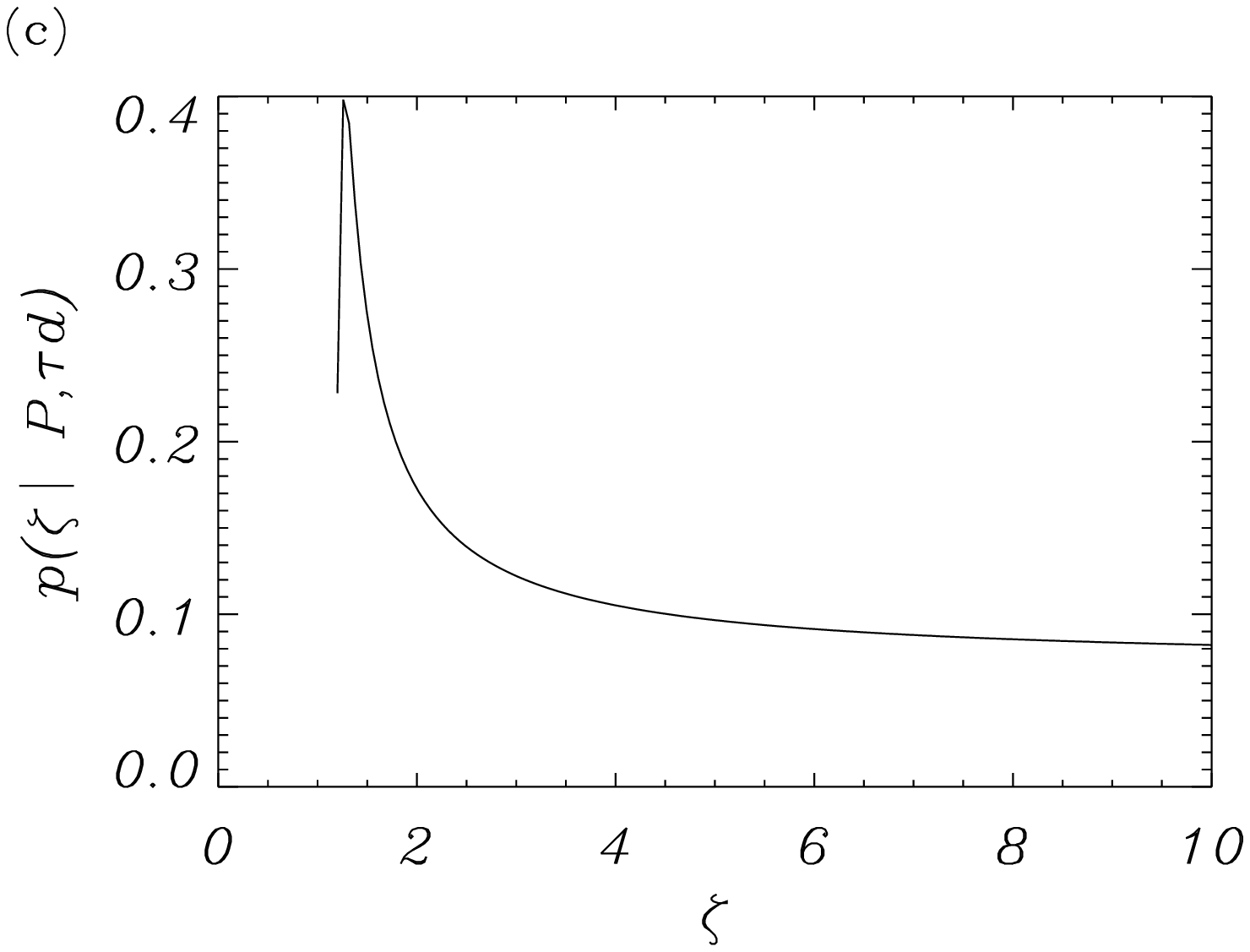}
\includegraphics[scale=0.35]{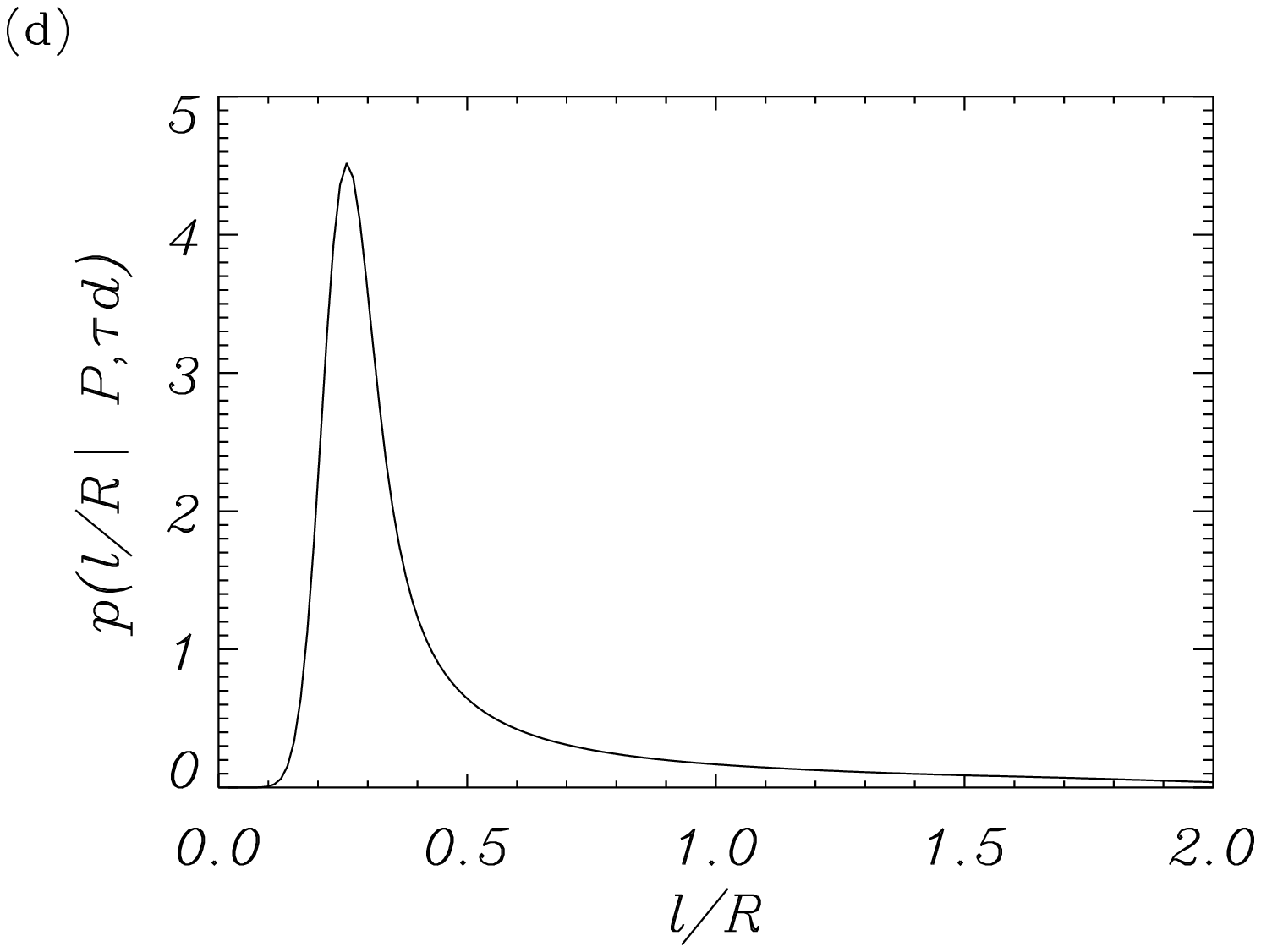}
\caption{(a) Three-dimensional view of the 1D solution curve representing the valid equilibrium models that reproduce observed periods and damping rates in the parameter space ($\zeta$, $l/R$, $\tau_{\rm Ai}$) for a loop oscillation event with $P=272$ sec and $\tau_{\rm d}=849$ sec, using the numerical and analytical inversion schemes by \cite{arregui07a} and \cite{goossens08a}. The solutions are also projected onto the different two-dimensional parameter planes. The dotted line in the horizontal plane represents the minimum density contrast ($\zeta=1.5$) considered in the numerical solutions. (b)-(d) Marginal posteriors for the three unknowns obtained from the Bayesian solution to the same inversion problem with a Gaussian likelihood, uniform priors, and a 10\% uncertainty on the measured period and damping time, following the scheme developed by \cite{arregui11b}.}
\label{3dcurvesposteriors}
\end{figure*}

The solution to the inverse problem, shown in Figure~\ref{3dcurvesposteriors}, was first obtained numerically by \cite{arregui07a} and by following an analytical procedure by \cite{goossens08a}. The figure shows the valid equilibrium models that reproduce observed period and damping rates in the three-dimensional parameter space of density contrast, transverse inhomogeneity, and Alfv\'en travel time. Although there are in principle infinite possibilities, the Alfv\'en travel time is found to be constrained to a rather narrow range. Also, as soon as information on one of the three parameters becomes available, the remaining two can be further constrained. However, the main drawback of this method lies in its inability to produce estimates for the uncertainty on the inferred parameters. Considering this,  \cite{arregui11b} solved the same problem using Bayesian parameter inference. The remaining panels in Figure~\ref{3dcurvesposteriors} display marginal posteriors distributions for the three parameters of interest for the same observed wave properties and assuming a given uncertainty on period and damping time. The magnitude of each posterior at a particular parameter value is a measure of the grade of belief on the parameter adopting that particular value. The resulting posteriors indicate how this degree of belief is distributed among the different parameter values. In contrast to the classic inversion curves obtained by \cite{arregui07a} and \cite{goossens08a} not all equilibrium models are equally probable. The degree of plausibility of different parameter values is given by the marginal posterior density. The shape of the posteriors indicates how well they can be constrained. For instance, the Alfv\'en travel time and the transverse inhomogeneity length-scale show well constrained posteriors. The same is not true with the density contrast that displays a long tail towards large values.  Although one could summarise the posteriors by giving e.g., the median and errors at a given credible interval, the full solution to the inference problem consists of the posteriors themselves, which are the result of a process in which the theoretical predictions and the data with their uncertainty  are combined in a consistent way. In the process, the uncertainties are correctly propagated from observed data to inferred parameters. 

\begin{figure*}
\centering
\includegraphics[width=0.405\textwidth]{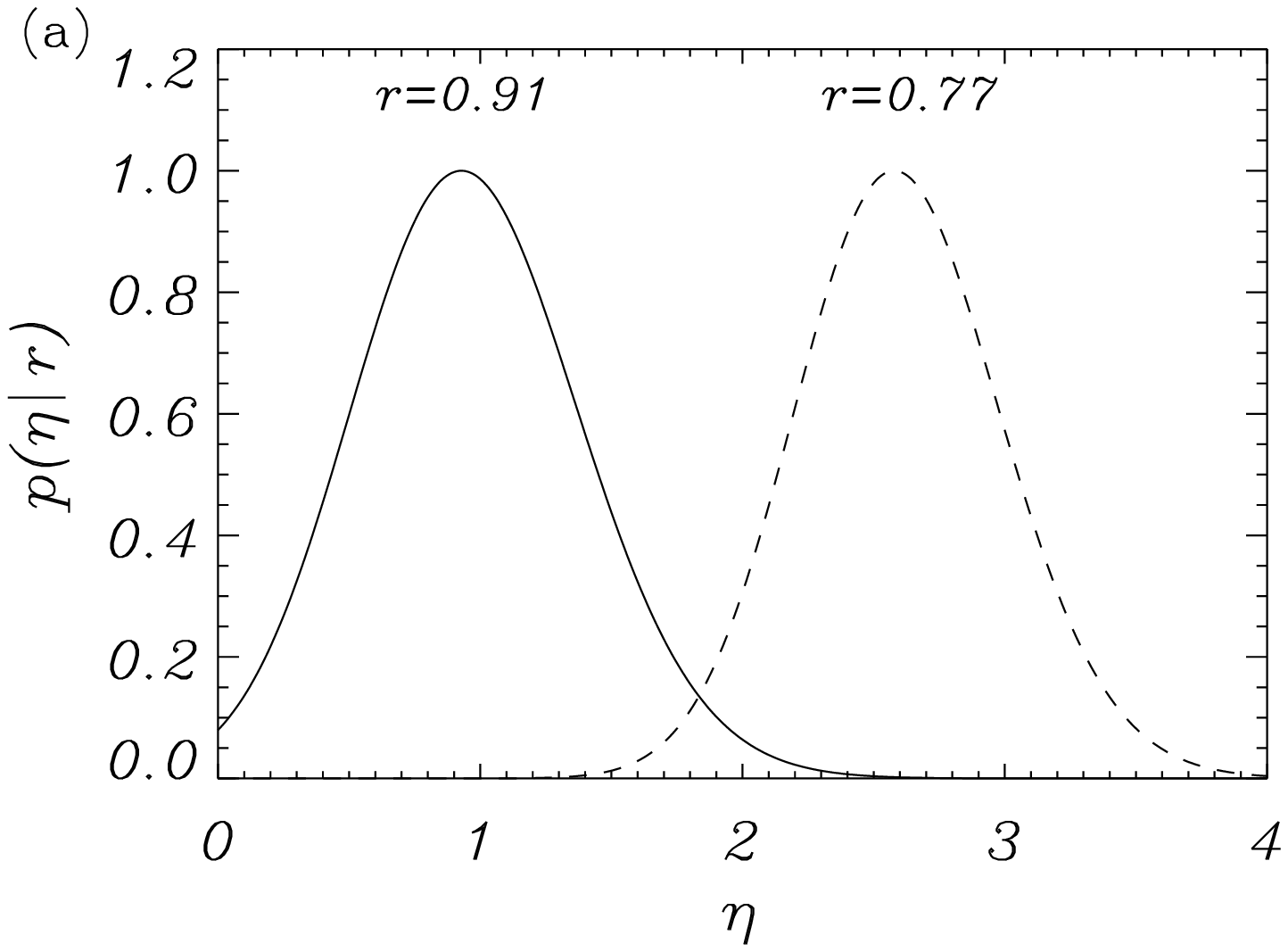}
\includegraphics[width=0.405\textwidth]{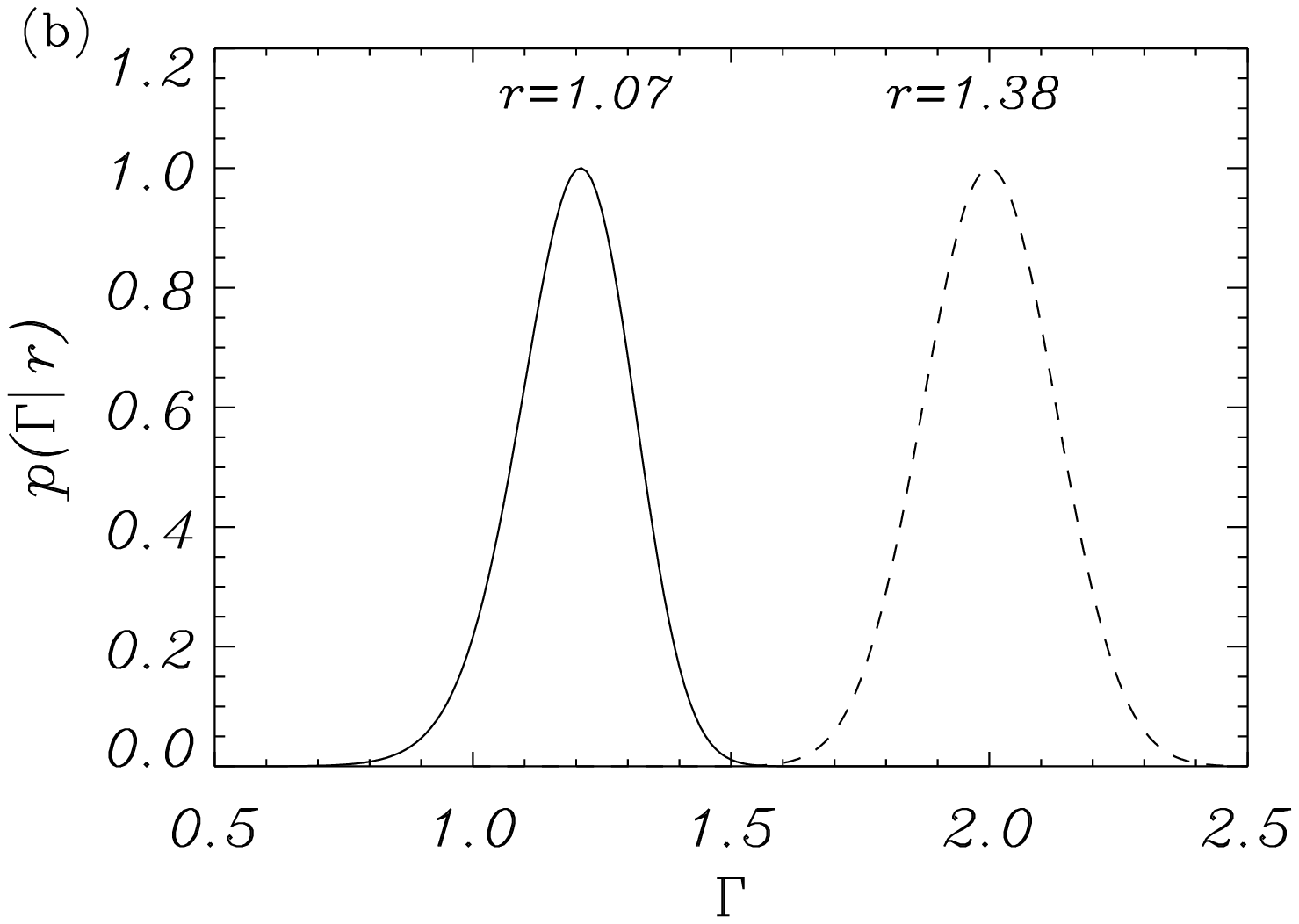}
\caption{(a) Posterior distributions for $\eta=L/\pi H$, under the density stratification model, for two values of the period ratio. (b) Posterior distributions for the magnetic tube expansion, $\Gamma=r_a/r_f$, under the magnetic expansion model, for two values of the period ratio. The measured period ratio and the inferred median of the distribution, with uncertainties given at the 68\% credible interval are (a) $r = 0.91 \pm 0.04$, $\eta = 1.26^{+0.65}_{-0.59}$ ; $r = 0.77 \pm 0.03$, $\eta = 3.39^{+0.72}_{-0.64}$ and (b) $r = 1.07 \pm 0.04$, 
$\Gamma = 1.20^{+0.10}_{-0.12}$ ; $r = 1.38 \pm 0.04$, $\Gamma = 1.87^{+0.07}_{-0.07}$. Adapted from \cite{arregui13a}.}
\label{p1p2}
\end{figure*}

%%%%%%%%%%%%%%%%%%%%%%%%%%%%%%%%%%%%%%%%%%%%%%%%%%%%%%%%%%%%%%%%%%%%%
\subsection{Magnetic field and density structuring along coronal and prominence waveguides}\label{magneticfield}
%%%%%%%%%%%%%%%%%%%%%%%%%%%%%%%%%%%%%%%%%%%%%%%%%%%%%%%%%%%%%%%%%%%%%

One of the most successful concepts developed in coronal seismology has been the suggestion by \cite{andries05a} to use the simultaneous presence of multiple harmonic oscillations in the same structure, first detected in coronal loops by \cite{verwichte04},  to obtain information on the coronal density scale height. This was followed by a similar suggestion by  \cite{verth08b} about the loop expansion rate, see \cite{andries09b} for a review.
The presence of multiple harmonic oscillations in coronal loops has been interpreted as a manifestation of the simultaneous presence of two longitudinal harmonics, the fundamental mode with period $P_1$ and the first overtone with period $P_2$. The ratio between these two periods offers information that can be related to the magnetic and density structuring along coronal waveguides. The measured period ratios, $P_1/2P_2$, deviate from the value of one that would correspond to uniform plasma and field along the loops. Models with longitudinally stratified density predict period ratios below one \citep{andries05a}. Models with magnetic flux tube expansion predict period ratios above one \citep{verth08a}.  By using the measured period ratios, inversion methods can be applied to infer the coronal density stratification and the magnetic flux tube expansion, as shown by \cite{andries05a} and \cite{verth08b}.

The Bayesian solution to these two inference problems was presented by \cite{arregui13a}. These authors considered separate models for the influence of density stratification and magnetic field expansion on the period ratio. In the model by \cite{andries05b}  coronal density stratification produces a decrease of the period ratio. The analytical expression for the period ratio as a function of the density scale height derived by \cite{safari07} is

\begin{equation}\label{forward1}
r_1=\frac{P_1}{2P_2}=1-\frac{4}{5}\left(\frac{\eta}{\eta+3\pi^2}\right),
\end{equation}

\noindent
with  $\eta=L/\pi H$ the ratio of the loop height at the apex to the density scale height $H$, and  $L$ the loop length.  On the other hand, in the model by \cite{verth08a} magnetic field expansion produces an increase of the period ratio. An analytical expression for their dependency is

\begin{equation}\label{forward2}
r_2=\frac{P_1}{2P_2}=1+\frac{3(\Gamma^2-1)}{2\pi^2},
\end{equation}
with the expansion defined as $\Gamma=r_a/r_f$, where $r_a$ is the radius at the apex and $r_f$ is the radius at the footpoint.  
Both models have forward models that relate one observed quantity, the period ratio $P_1/2P_2$, to one physical quantity to be inferred, either $\eta$ or $\Gamma$. 

To perform the inference using Bayes Theorem (\ref{bayes}),  \cite{arregui13a} used period ratio measurements reported in observations by \cite{verwichte04} and additional values discussed in the review by \cite{andries09b}. Figure~\ref{p1p2}a shows posterior probability distributions for $\eta$ computed using Eq.~(\ref{bayes}) for two period ratio measurements by \cite{verwichte04}. Well constrained distributions are obtained.  For the measured period ratios  $r\sim0.77$  and $r\sim0.91$,  the inversion leads to density scale heights of $H\sim21$ Mm and $H\sim56$ Mm, respectively, for a loop with a height at the apex of  $L/\pi=70$ Mm. Figure~\ref{p1p2}b shows posterior probability distributions for $\Gamma$ for two period ratio measurements.  Again, well constrained distributions are obtained.  \cite{andries09b} discuss period ratio measurements in Table 1 by \cite{demoortel07}. Assuming that either the most power is in the fundamental mode or in the first overtone, mean values for the period ratio of  $r\sim1.07$  and $r\sim1.38$ are obtained. For those values, the inversion leads to tube expansion factors that are compatible with previous estimates by \cite{klimchuk00} and \cite{watko00}.  Note however that, according to Figure~\ref{p1p2}b, a period ratio of $r\sim 1.38$ requires an expansion of the tube by a factor of $\Gamma\sim 1.85$, while observations by \cite{watko00}  seem to indicate that in only very few cases does this parameter approach or exceed a value of $2$. Despite the solution to the two inversion problems using the algebraic forward predictions given by Eqs.~(\ref{forward1}) and (\ref{forward2}) seems straightforward, the Bayesian framework makes use of all the available information in a consistent manner and enables us to consistently propagate errors from observations to inferred parameters.

\begin{figure*}
\centering
\includegraphics[width=0.375\textwidth]{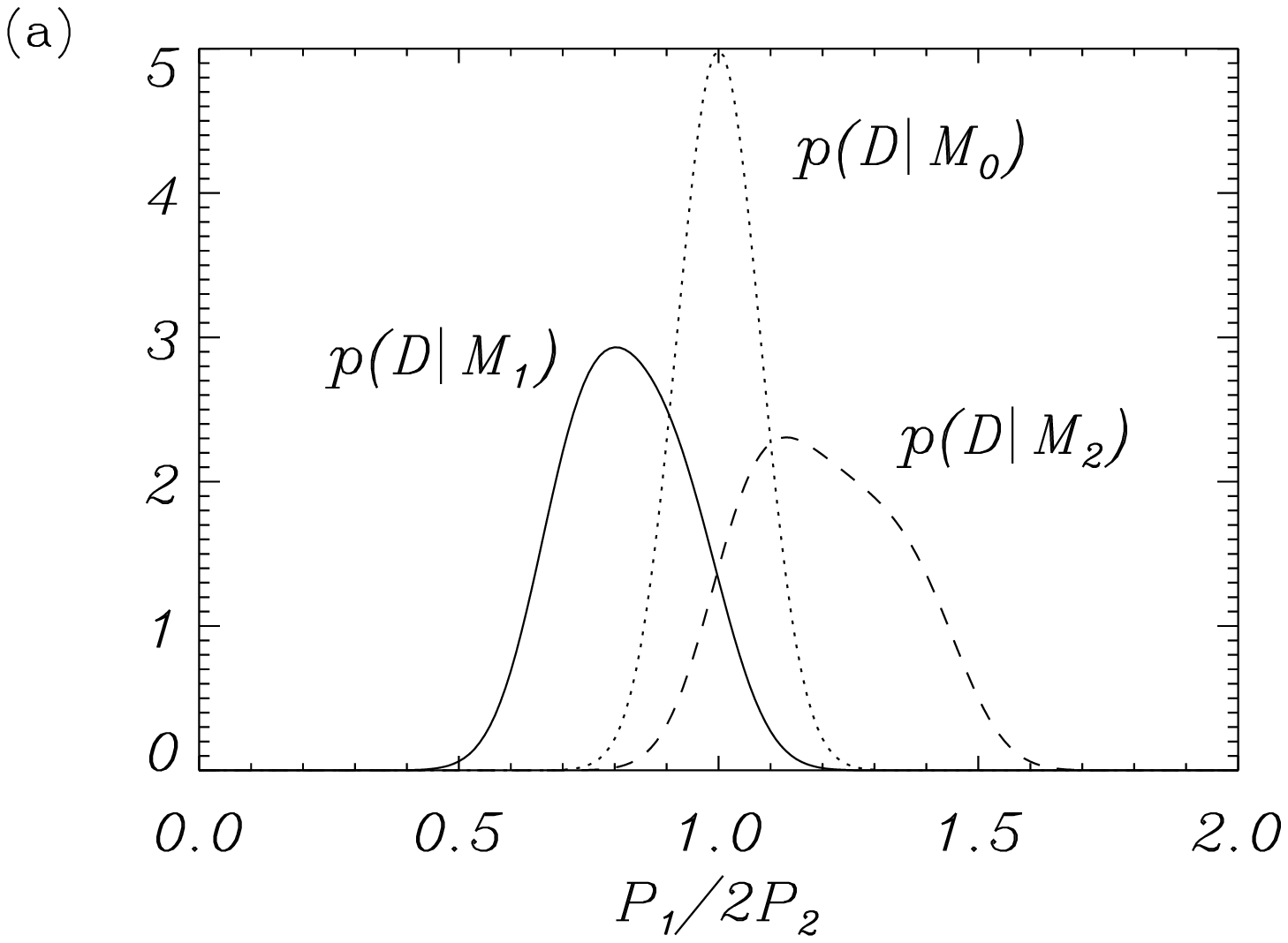}
\includegraphics[width=0.375\textwidth]{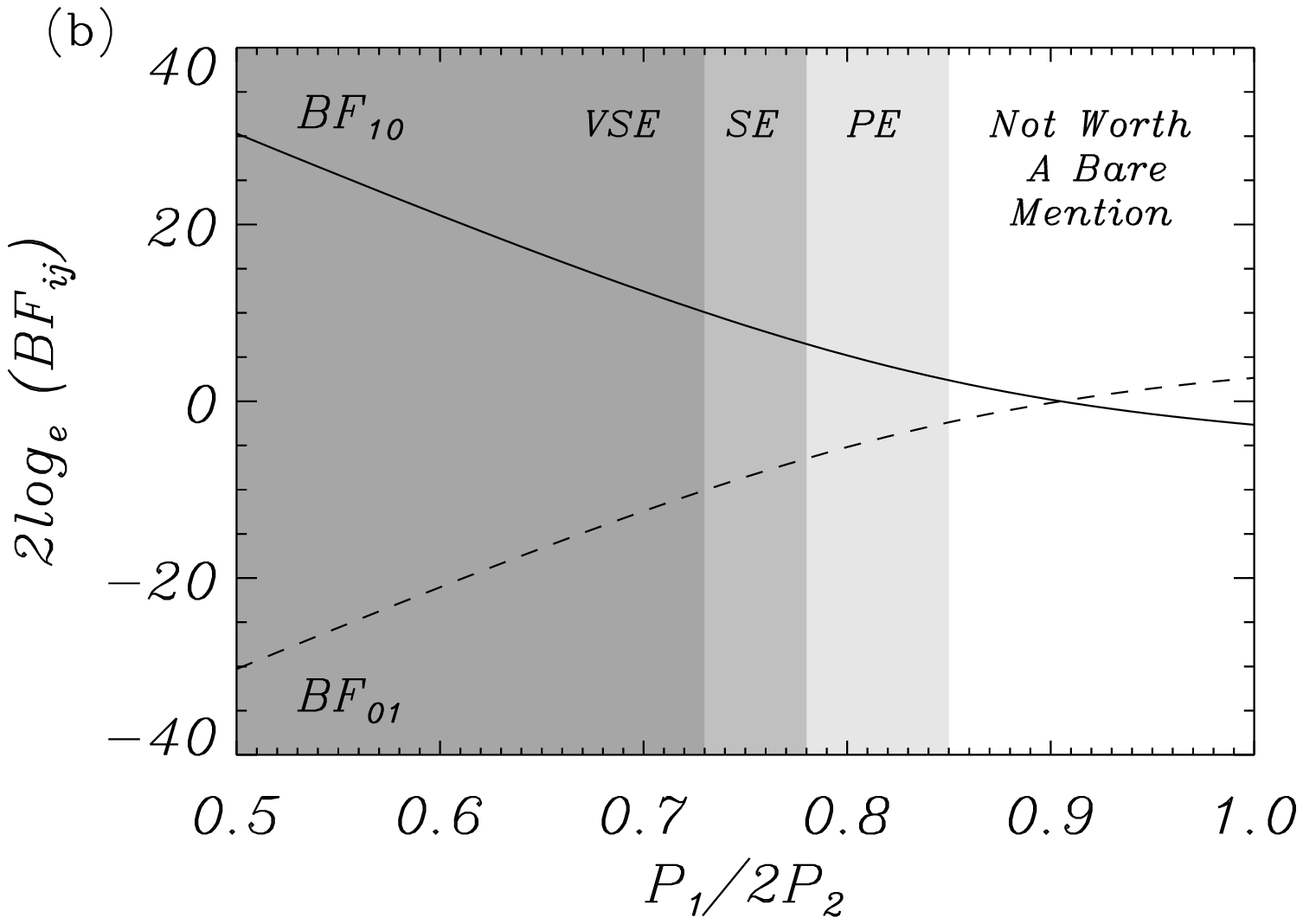}\\
\includegraphics[width=0.375\textwidth]{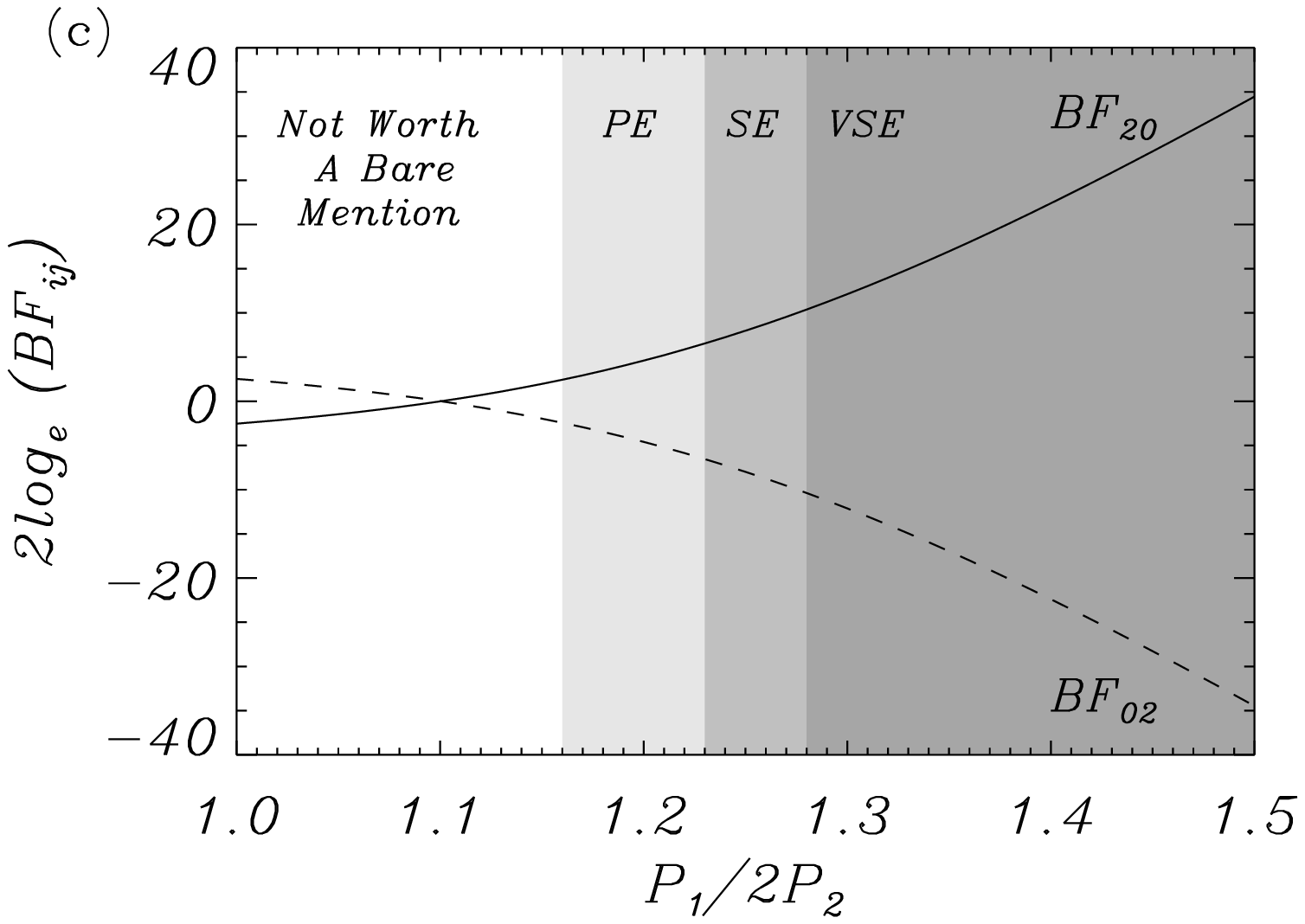}
\includegraphics[width=0.375\textwidth]{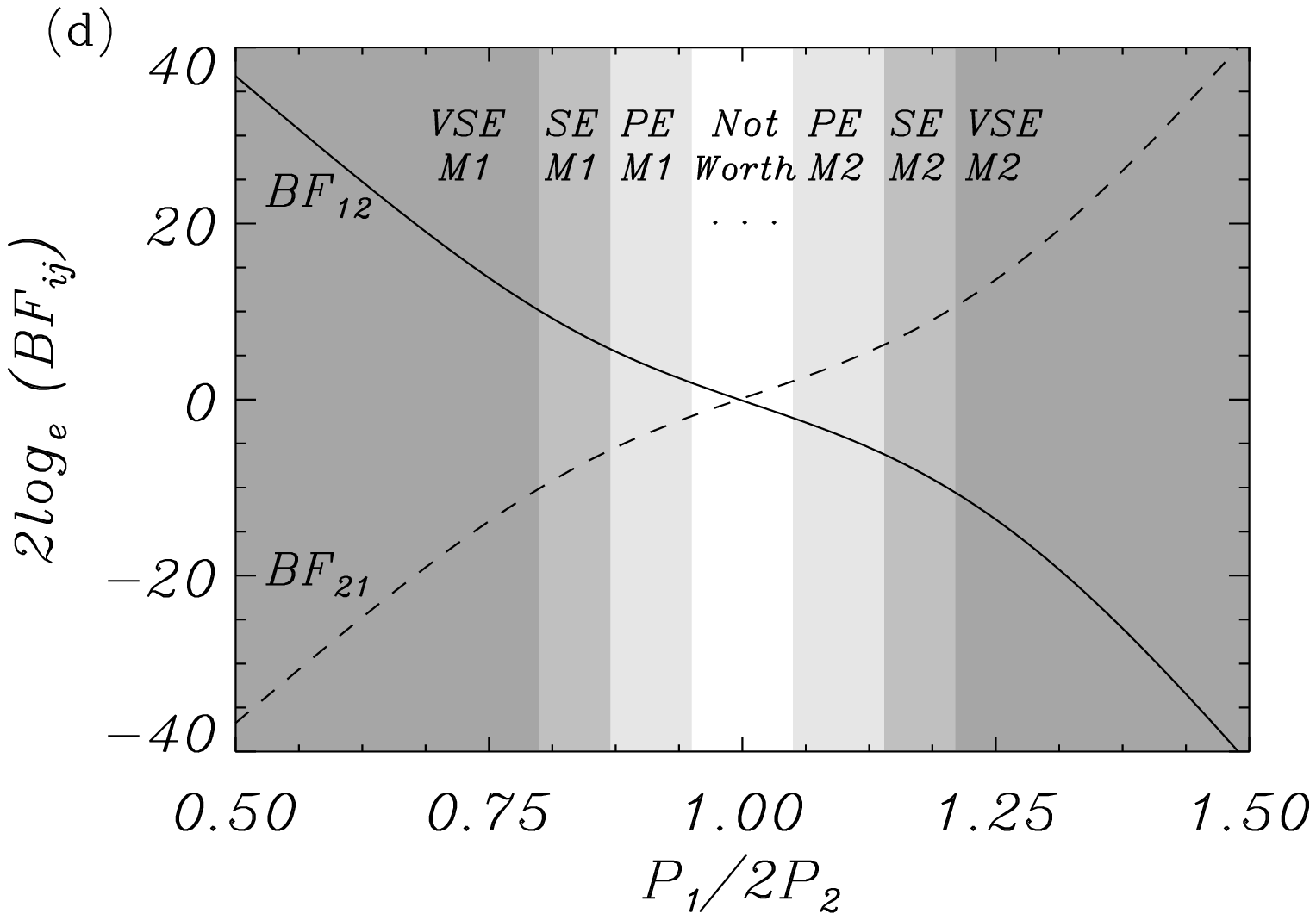}
\caption{(a) Marginal likelihoods computed using Eq.~(\ref{marginallikelihood}) for models M$_0$, M$_1$ and M$_2$ as a function of the data. (b)--(d) Bayes factors computed using Eq.~(\ref{bayesfactor}) for the model comparisons between M$_1$ over M$_0$, M$_2$ over M$_0$, and M$_1$ over M$_2$. In (b)--(d), white regions indicate IE: inconclusive evidence ($2 log_{e}BF\in [0,2]$); PE: positive evidence ($2 log_{e}BF \in [2,6]$); SE: strong evidence ($2 log_{e}BF \in [6,10]$); VSE: very strong evidence ($2 log_{e}BF > 10$). Uniform priors in the ranges $\eta\in [0, 8]$ and $\Gamma\in[1, 2.5]$ have been taken. In all figures $\sigma = 0.08$. Adapted from \cite{arregui13a}.}
\label{ml}
\end{figure*}

The potential of Bayesian methods becomes more visible when we move to the second level of Bayesian inference, namely model comparison. The key question here is what is the amount of evidence that we have in support of one model (the density structured model) in front of the other (the magnetically structured model), or even in support of any of them in front of the null hypothesis, in this case that the waveguides are uniform along the field. The solution to the inference problem is unable to answer these questions, since any inference is conditional to the assumption that a particular model under consideration is true.  To answer these questions we need to apply Bayesian model comparison. 

For this reason, \cite{arregui13a} considered the relative performance of three models in explaining observed period ratio values. In the following,  M$_0$, M$_1$, and M$_2$ stand for the uniform, density stratified, and expanding loop models, respectively. Upon computation of the marginal likelihood for each of them, by making use of Eq.~(\ref{marginallikelihood}) and displayed in Figure~\ref{ml}a, one can appreciate  that model M$_0$ is likely for period ratios around one, M$_1$ for period ratios shorter than one and M$_2$ for period ratios larger than one. The lateral extent of the three marginal likelihoods depends on the particular value assumed for the uncertainty on the measured period ratio. By assuming that the three models are equally probable a priori, the posterior ratio is equal to the Bayes factor, which expresses how well observed data are predicted by one model against the other. A quantitative model comparison can then be performed by computing the Bayes factor and by using the empirical evidence classification scale in Table~\ref{kasstable} to assign a level of evidence to one model in front of the alternative. 

\begin{figure*}
\centering
\includegraphics[width=0.375\textwidth,angle=0]{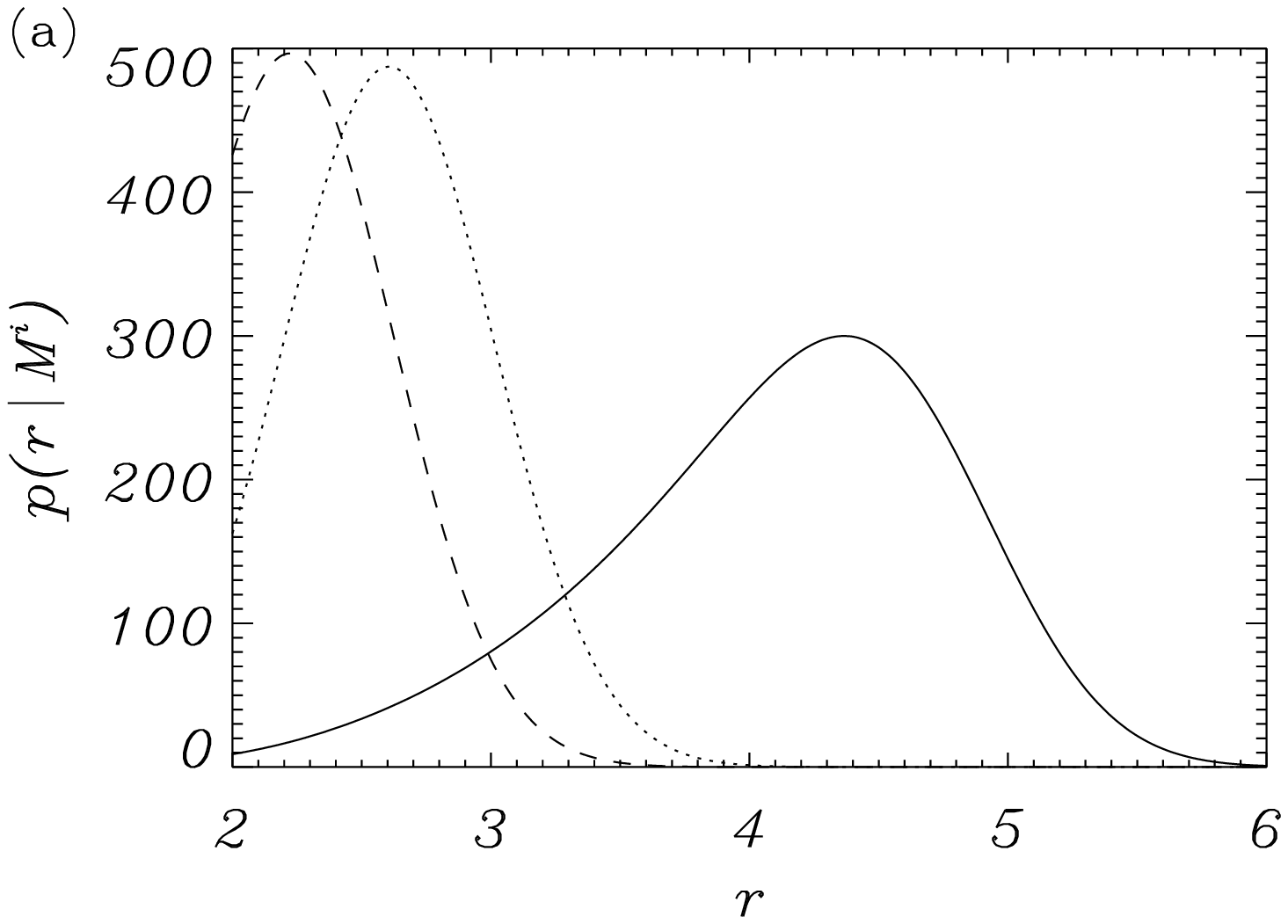}
\includegraphics[width=0.375\textwidth,angle=0]{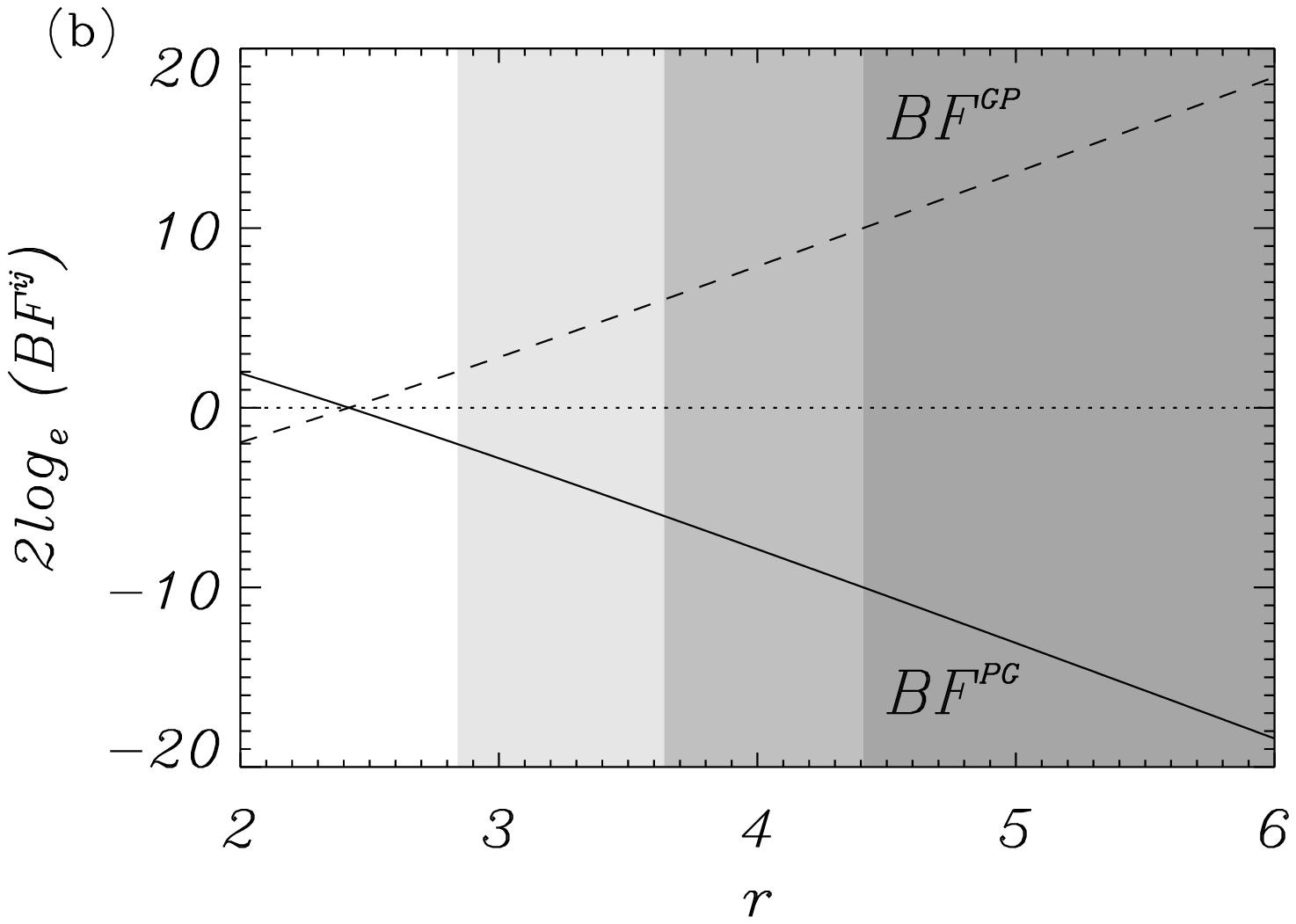}
\includegraphics[width=0.375\textwidth,angle=0]{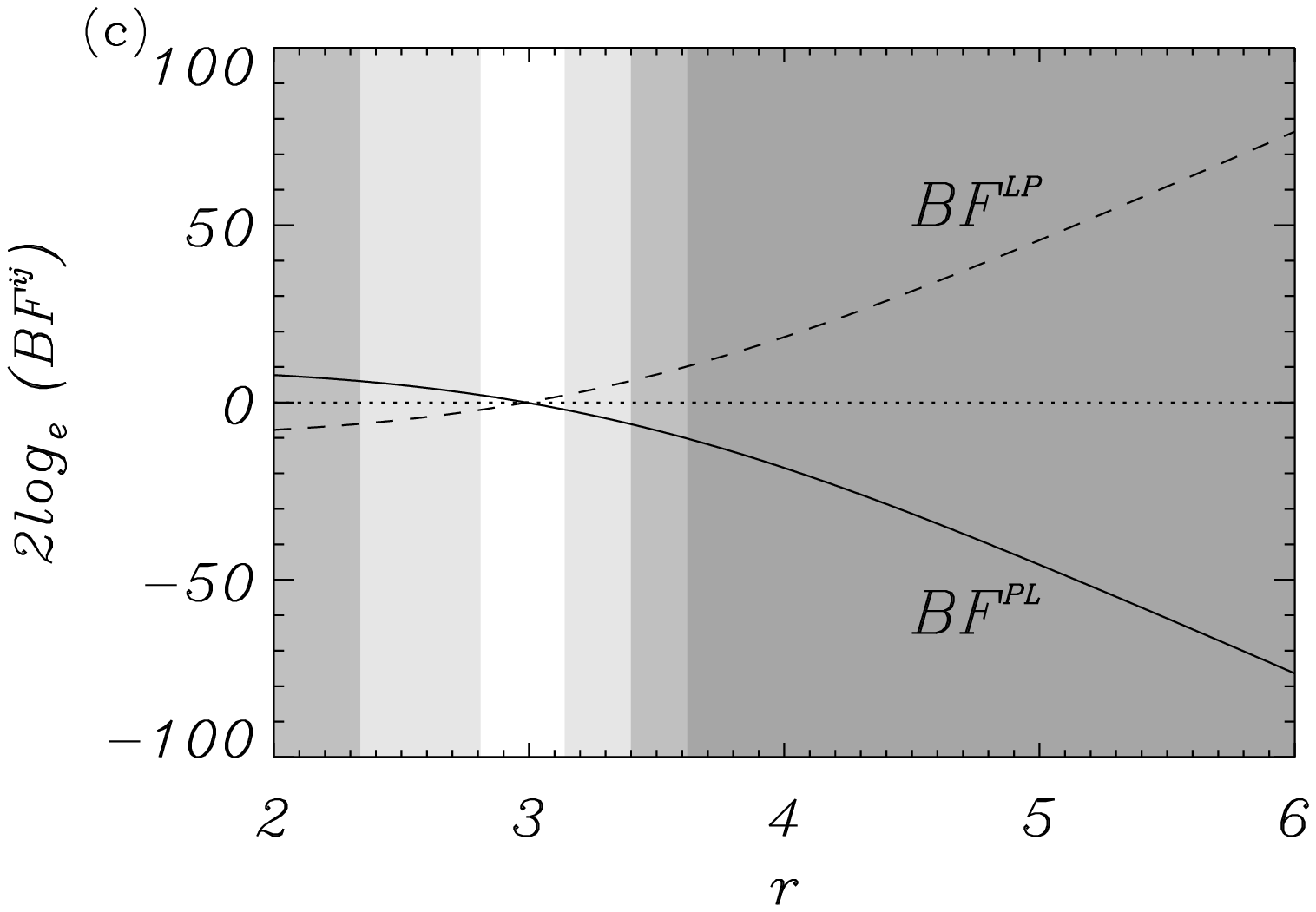}
\includegraphics[width=0.375\textwidth,angle=0]{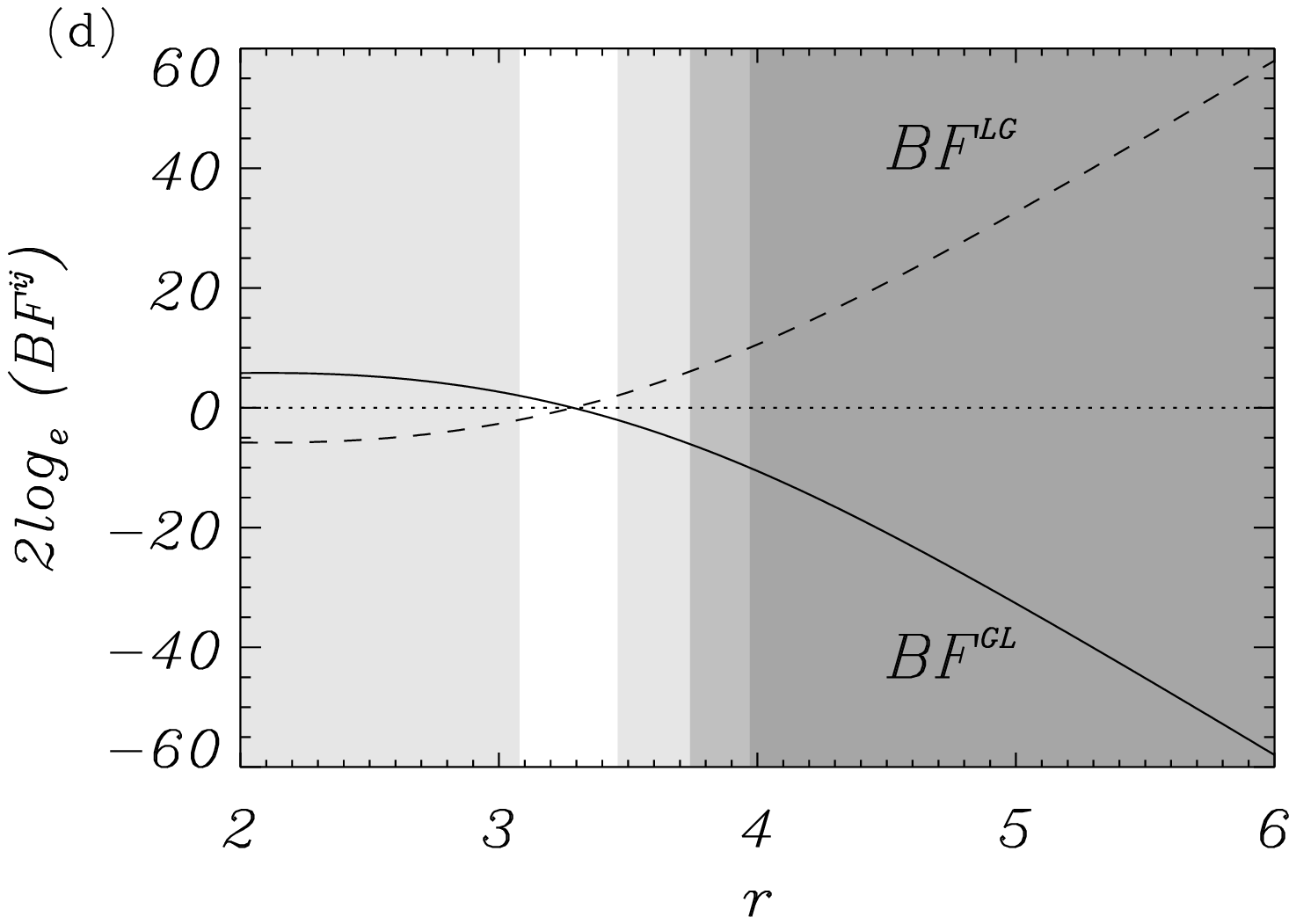}
\caption{(a) Marginal likelihoods for three density models considered by \cite{arregui15b} as a function of the observable period ratio, $r$: $p(r|M^{\rm P})$ (dashed line), $p(r|M^{\rm G})$ (dotted line) and $p(r|M^{\rm L})$ (solid line).  (b)--(d) Bayes factors for the model comparisons between (b) the parabolic and the Gaussian density models; (c)  the parabolic and the Lorentzian density models; and (d) the Gaussian and the Lorentzian density models, as a function of the period ratio. Adapted from \cite{arregui15b}.}
\label{prom}
\end{figure*}

Consider first the comparison between model M$_1$ for stratified loops with model M$_0$ corresponding to a uniform loop. Figure~\ref{ml}b shows the Bayes factors BF$_{10}$ and BF$_{01}$  as a function of the data for a given error. In the figure, different shades of grey limit ranges in period ratio with different levels of evidence, depending on the magnitude of the Bayes factors following the empirical Table~\ref{kasstable}. White regions indicate values of period ratio for which there is minimal evidence for any of the models that are being compared. Then, different shades of grey indicate regions with positive, strong and very strong evidence, with the level of evidence being larger for darker regions. 
The lower the measured period ratio, the more plausible is the stratified model in front of the uniform model. However, a period ratio shorter than one does not automatically imply evidence for a stratified loop model. For the assumed uncertainty of period ratio, only for period ratio measurements below 0.87 we have Bayes factors that indicate positive evidence for the stratified model. Strong evidence is obtained below 0.78 and very strong evidence for period ratios below 0.71. A similar exercise can be performed by comparing the magnetic expansion and uniform models (see Figure~\ref{ml}c). A period ratio larger than one does not automatically translate into evidence for an expanding loop model. Positive, strong, and very strong evidence for the expanded model are obtained for period ratio measurements above 1.16, 1.23, and 1.28, respectively.
Finally, we can compare the stratified and magnetic expansion models. Figure~\ref{ml}d shows that  different levels of evidence for one model or the alternative are obtained depending on the measured period ratio and the uncertainty for period ratio measurements to the left or to the right of the value of $P_1/2P_2=1$ which would correspond to a uniform plasma and a uniform field along an infinitely thin loop. Around this value, there is a region, with  $0.96<r<1.06$, in which no firm conclusion can be established. Positive, strong, and very strong evidence for $M_1$ occur below 0.96, 0.87, and 0.80, respectively. Positive, strong, and very strong evidence for $M_2$ occur above 1.06, 1.15, and 1.21, respectively. 

In the model comparison example here explained, a value $\sigma=0.08$ has been selected for the uncertainty on the period ratio measurement, so as to clearly show the different regimes for the evidence. An increase (decrease) of $\sigma$ produces a decrease (increase) of any evidence. The inference in Figure~\ref{p1p2}a for $r=0.91$ with uncertainty of $\sigma=0.04$ falls into the region of positive evidence for hypothesis $M_1$. The inference in Figure~\ref{p1p2}b for $r=1.07$ should have an uncertainty of $\sigma=0.03$ (close to the reported error) to conclude that there exists positive evidence for hypothesis $M_2$.

Seismology inversion techniques akin to those employed with coronal magnetic and plasma structures can also offer information about the physical conditions of prominence plasmas. When observed with high resolution instrumentation, solar prominences are seen to be composed of a myriad of fine threads that also display transverse oscillations \citep{lin05,lin09}. These threads are clearly structured in density since the cool and dense plasma occupies only a fraction of a larger magnetic flux tube. The observation of multiple periods in transverse thread oscillations also offers information on the structuring of the plasma along the magnetic field. \cite{soler15} have shown that theoretical predictions for the period ratio depend on the particular profile that is adopted to model field-aligned density variations. As the exact profile is unknown, this poses a problem since inference results depend on the theoretical model that has been assumed.
To shed some light into the most plausible density profile along prominence threads, \cite{arregui15b} considered three alternative density profiles: parabolic (M$^P$), Gaussian (M$^G$), and Lorentzian (M$^L$) and their theoretical predictions for the period ratio to perform parameter inference and model comparison in the Bayesian framework.

Figure~\ref{prom} shows the results of the model comparison between the three adopted profiles, a similar exercise to the one described before for coronal loop oscillations. The marginal likelihoods displayed in Figure~\ref{prom}a show that the parabolic and Gaussian profiles are likely to produce period ratios in the lower half of the considered period ratios range,  from 2 to 4. Beyond that, their likelihood decreases significantly.  Hence, if our observed period ratio is in the  range between 2 and 4, it can be difficult to obtain significant evidence for one model to be preferred over the other.  The Lorentzian profile is more likely to reproduce values of the period ratio larger than those that can be reproduced by the parabolic and the Gaussian profiles. The distribution peaks at about 4.4, but is rather extended and covers almost all values of the considered range for the period ratio.

By computing Bayes factors, the one-to-one relative plausibility between the three models was assessed. The results displayed in Figures~\ref{prom}b-d indicate that a Lorentzian density profile with plasma density concentrated around the centre of the tube offers the most plausible density distribution. The model comparison results indicate that the evidence points to the Gaussian and parabolic profiles for period ratios in between 2 and 3, while the Lorentzian profile is preferred for higher period ratio values. This model comparison technique could be used to obtain information on the plasma structure along threads, provided period ratio measurements become widely available.

\begin{figure*}
\centering
 \includegraphics[width=0.375\textwidth]{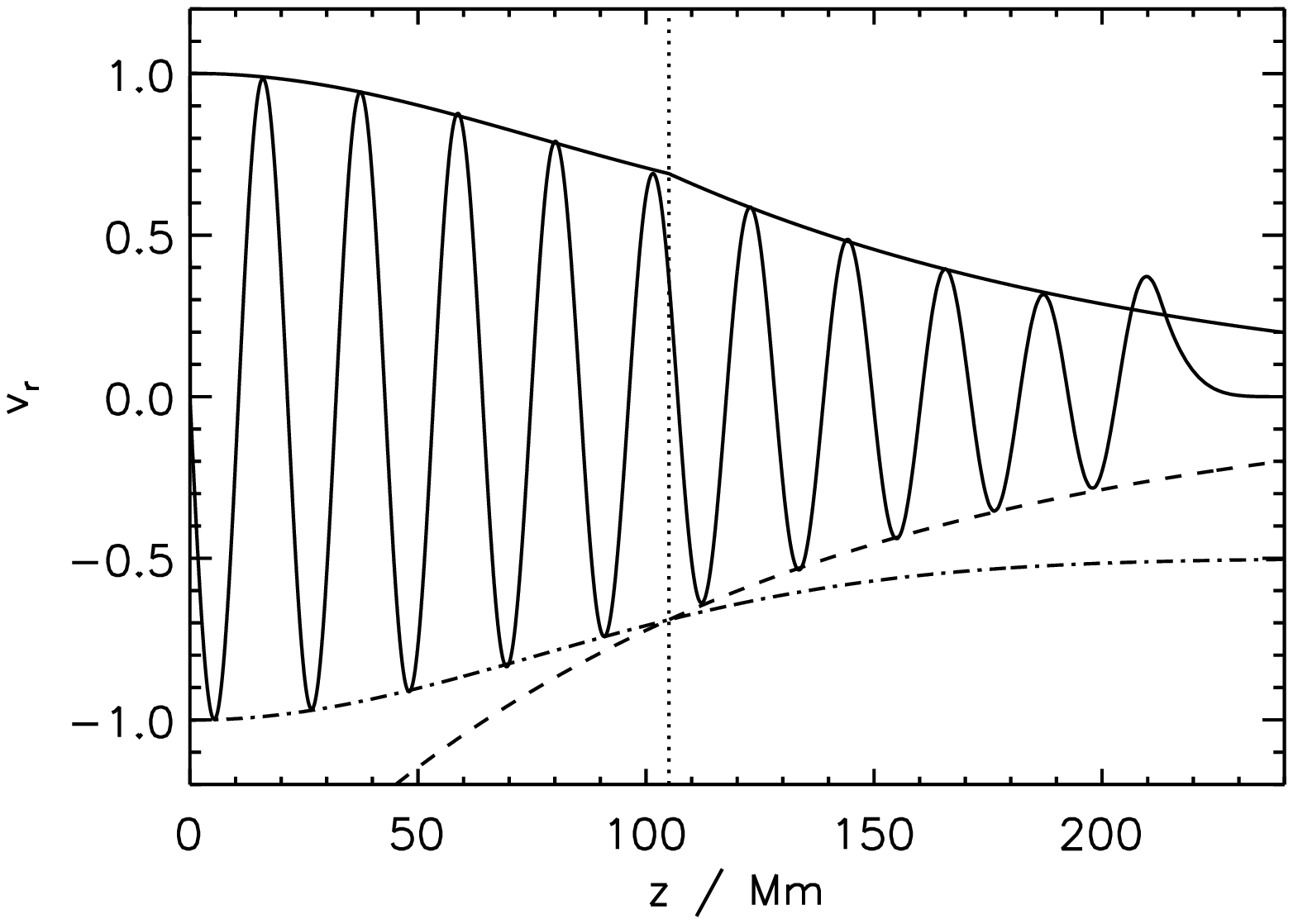}
  \includegraphics[width=0.375\textwidth]{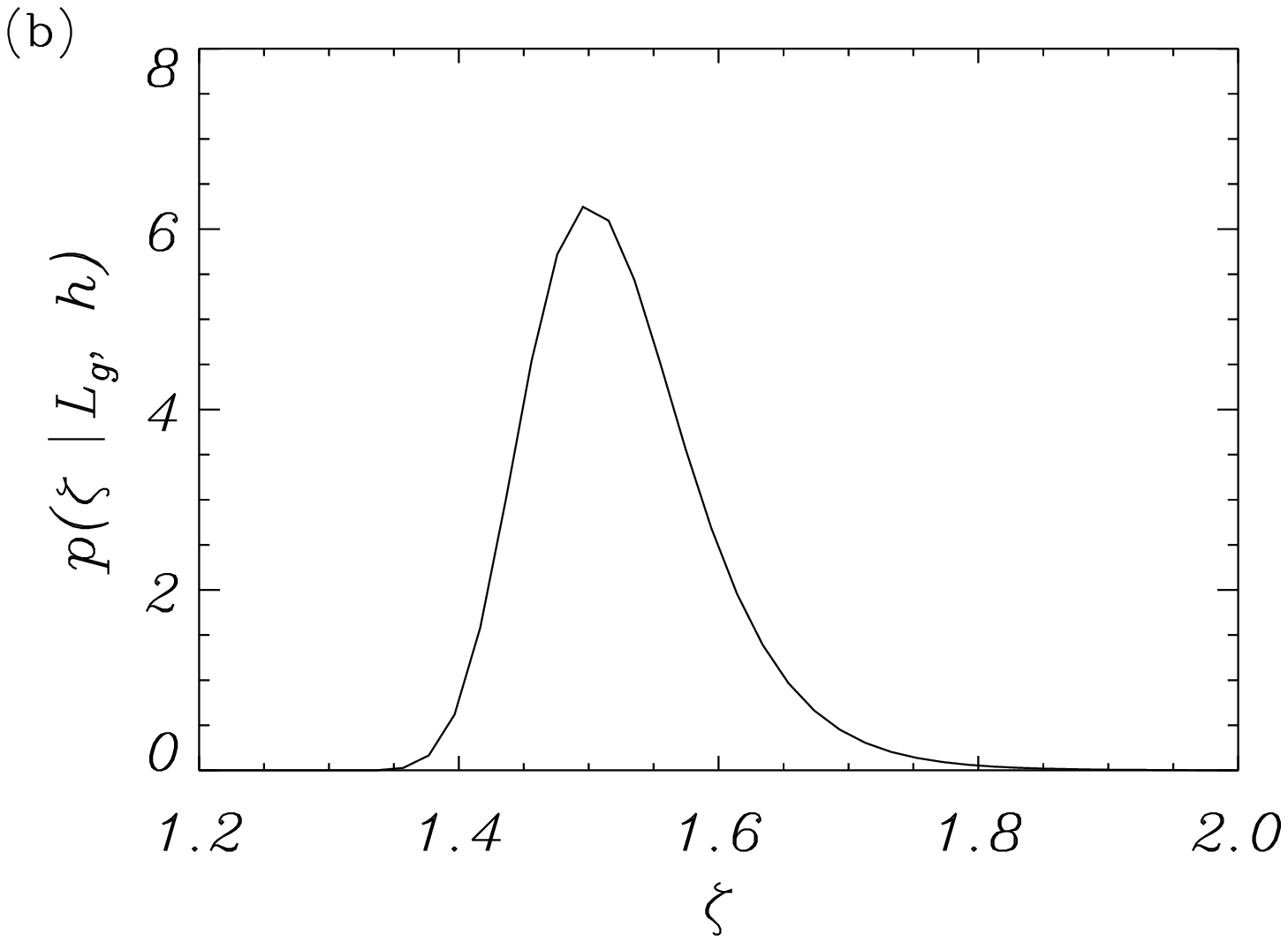}\\
 \includegraphics[width=0.375\textwidth]{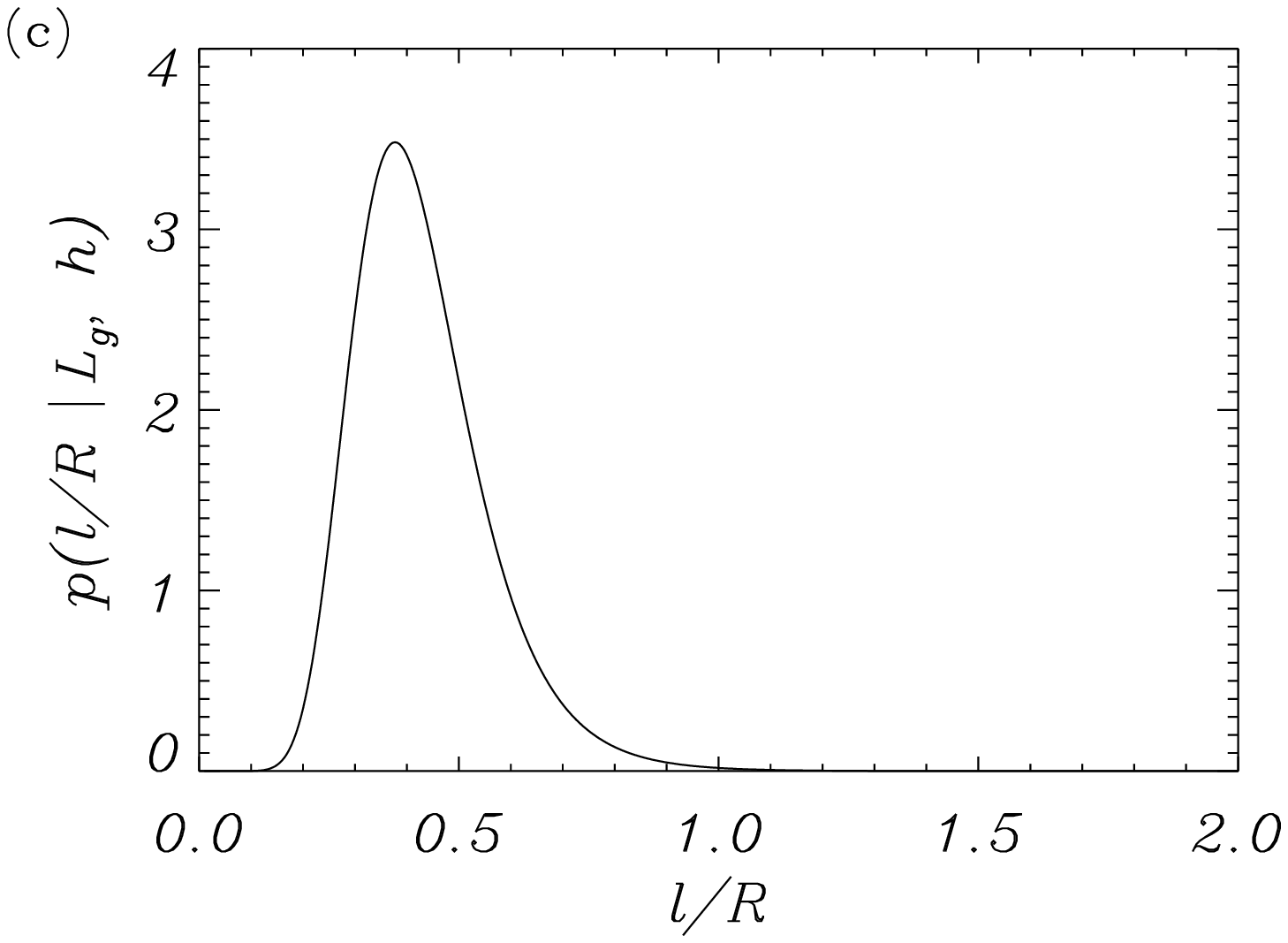}
  \includegraphics[width=0.375\textwidth]{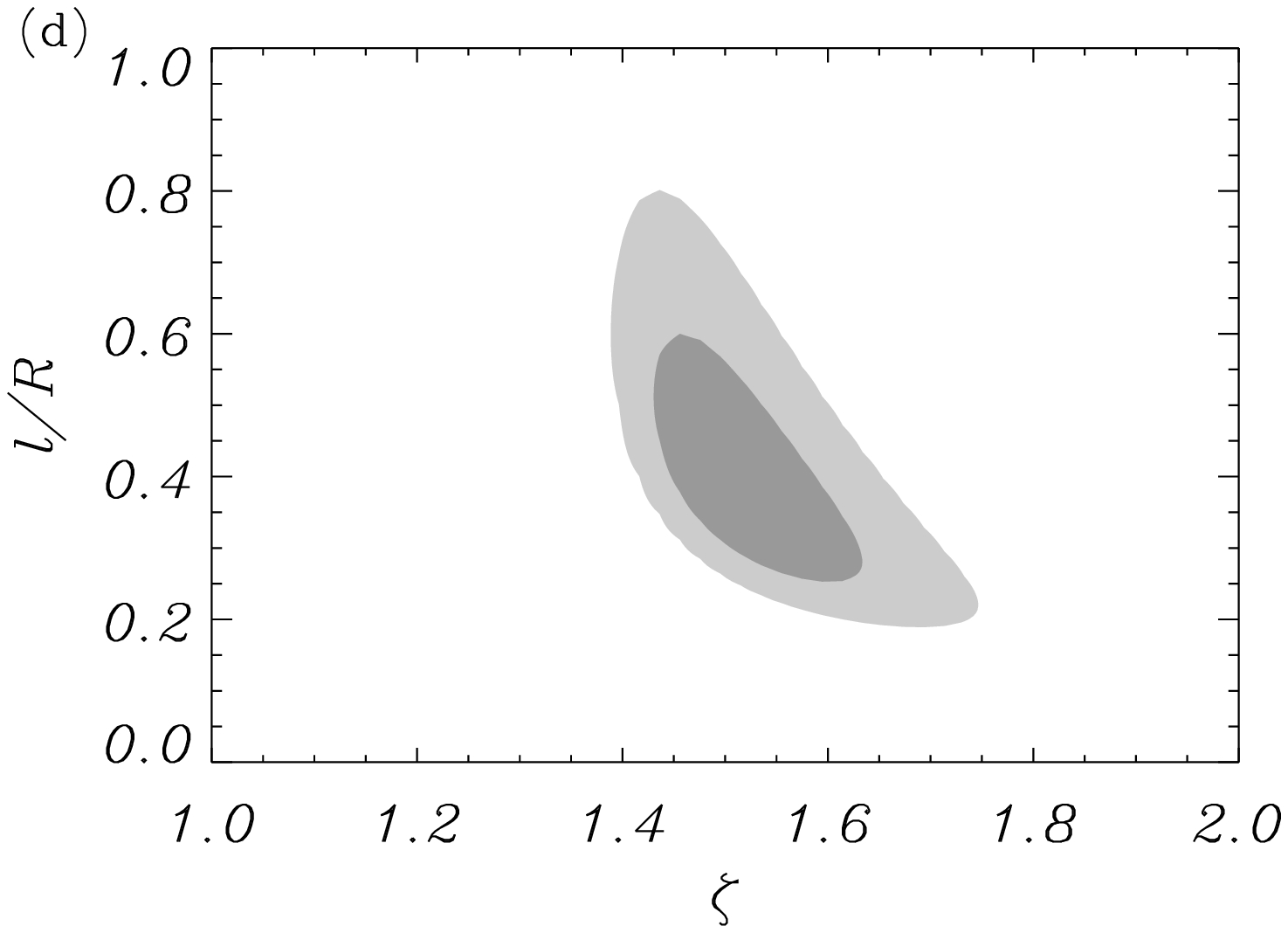}
\caption{Transverse velocity component as a function of the height at the axis of the tube for propagating kink waves in a numerical simulation with $\zeta = 1.5$ and $l/R = 0.4$. One-dimensional marginal posterior distributions for the density contrast (b) and the transverse inhomogeneity length scale (c) corresponding to the inversion of a spatially damped transverse oscillation with $L_{\rm g}/\lambda = 5.0 \pm 0.1$ and $h/\lambda = 4.9\pm0.1$. (d) Joint two-dimensional posterior distribution. The light and dark gray shaded regions cover the 95\% and 68\% credible regions, respectively. Adapted from \cite{arregui13b}.}
\label{figpascoe}
\end{figure*}

The two examples discussed in this subsection show that Bayesian analysis techniques offer useful tools to discriminate between competing models. The marginal likelihoods and Bayes factors enable us to quantify the relative performance of alternative physical models in explaining observed data, taking into account their uncertainty.

%%%%%%%%%%%%%%%%%%%%%%%%%%%%%%%%%%%%%%%%%%%%%%%%%%%%%%%%%%%%%%%%%%%%%
\subsection{Seismology using Gaussian and exponential damping regimes}\label{gaussexp}
%%%%%%%%%%%%%%%%%%%%%%%%%%%%%%%%%%%%%%%%%%%%%%%%%%%%%%%%%%%%%%%%%%%%%

Theoretical studies of damping by resonant absorption of standing and propagating waves have predominantly focused on the modelling of the decay of the oscillations using exponential profiles \citep{ruderman02,goossens02a}. Subsequent studies have shown that this exponential profile does not cover the initial phase of the damping unless an eigenmode is excited right from the start.  Analytical and numerical studies have shown that the full nonlinear damping profile is better approximated by an initial Gaussian profile for short distances/times followed by the usual exponential profile for long distances/times \citep{pascoe12,hood13}.  As shown by \cite{pascoe13}, the Gaussian damping regime is important for low density contrasts and for short time spans.

\begin{figure*}
\centering
\includegraphics[width=0.40\textwidth]{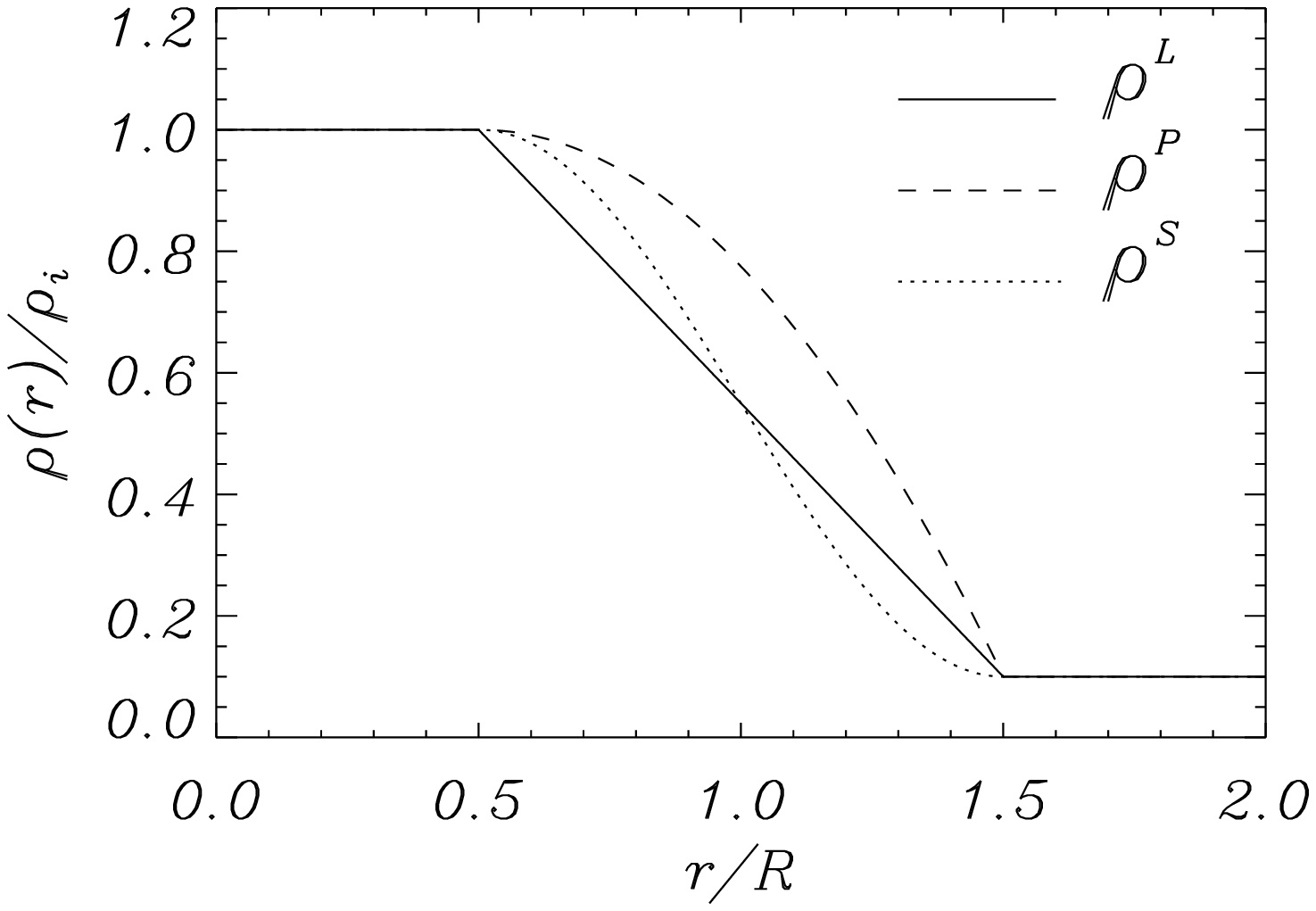}
 \includegraphics[width=0.375\textwidth]{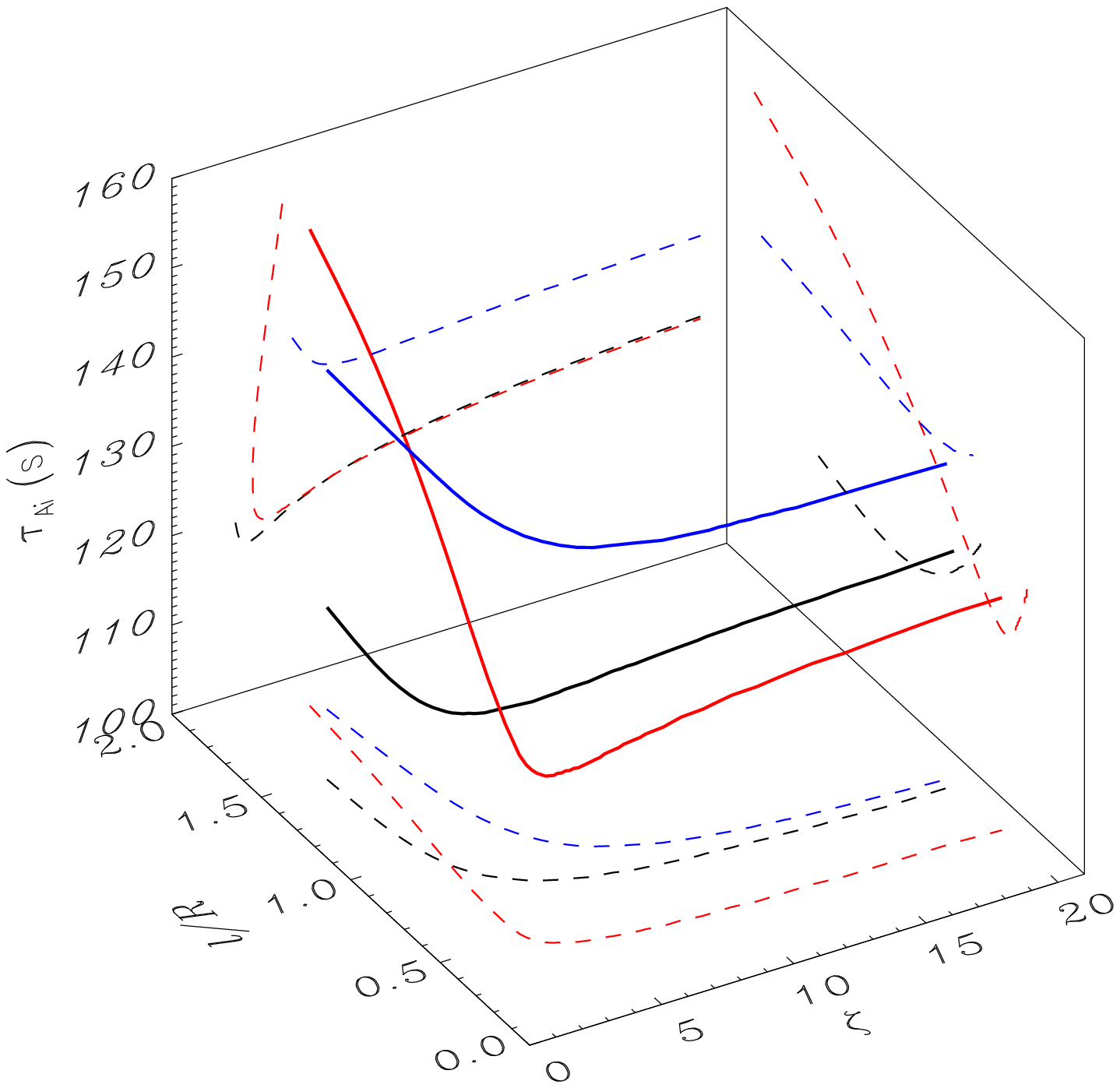}
\caption{Spatial variation of the equilibrium mass density across the magnetic flux tube for waveguide models with a linear profile (solid line), a sinusoidal profile (dotted line), and a parabolic profile (dashed line). The ratio of internal to external density is $\rho_{\rm i}/\rho_{\rm e}=10$ and the thickness of the nonuniform layer is $l/R = 1$. Classic seismological inversion result of the event with strong damping ($P = 185$ [sec] 
and $\tau_{\rm d}=200$ [sec]) with the one-dimensional solution curves in the three-dimensional parameter space corresponding to the sinusoidal (black), linear (read), and parabolic (blue) density models. The dashed lines are the projections of the inversion curves to the various planes. Adapted from \cite{arregui15c}.}
\label{crossf1}
\end{figure*}

The use of a combined damping profile has a marked advantage from the seismology point of view. Consider, following \cite{pascoe13}, a damped propagating transverse wave in a magnetic flux tube.  An example of the spatial dependence of the velocity amplitude from numerical simulations is shown in Figure~\ref{figpascoe}. The amplitude is fitted to a double profile. The initial Gaussian stage, with a damping length scale $L_{\rm g}$, can be approximated as

\begin{equation}\label{gaussdamp}
\frac{L_{\rm g}}{\lambda}= \left(\frac{2}{\pi}\right) \left(\frac{R}{l}\right)^{1/2}\left(\frac{\zeta+1}{\zeta-1}\right),
\end{equation}
\noindent
with $\lambda$ the wavelength and $\zeta$ and $l/R$ the two parameters that define the variation of density at the non-uniform layer. For the ensuing exponential stage, the classic expression for the damping length-scale is

\begin{equation}\label{expdamp}
\frac{L_{\rm d}}{\lambda}=  \left(\frac{2}{\pi}\right)^2\left(\frac{R}{l}\right)\left(\frac{\zeta+1}{\zeta-1}\right).
\end{equation}

\noindent
Notice that both damping lengths are functions of the same two physical parameters that we aim to infer. The additional information comes from the height, $h$, at which the damping regime changes from Gaussian to exponential. This is given by \citep[see][]{pascoe13}

\begin{equation}\label{ratio}
h=\frac{L^2_{\rm g}}{L_{\rm d}}=\lambda\left(\frac{\zeta+1}{\zeta-1}\right).
\end{equation}
As first proposed by \cite{pascoe13}, the double profile fitting of such a general profile leaves us with two observables (e.g., one damping length and the height of regime change) and two unknowns, $\zeta$ and $l/R$, thus making feasible a proper seismology inversion. On the top left panel of Figure~\ref{figpascoe}  is the general spatial damping profile given by the solid line. The transition between Gaussian and exponential damping is given by the vertical dotted line. At the bottom, the general profile is split into its two components: Gaussian (dot-dashed) and exponential (dashed). 

Using analytical expressions for the Gaussian and exponential damping length scales and the position at which the damping regime changes, \cite{arregui13b} showed how Bayesian inference can be used to obtain the density contrast and the transverse inhomogeneity length scale. 
In this particular example, a hypothetical value for $h$ was used. In a recent study of standing loop oscillations, \cite{pascoe16} obtain the transition time from Gaussian to exponential damping with an accuracy of a few minutes.
The results for a particular choice of observed damping length-scale are shown in Figures~\ref{figpascoe}b-d. The figures display the marginal posteriors for the density contrast and the transverse inhomogeneity length-scale and their joint two-dimensional posterior, indicating the 68\% and 95 \% credible regions. The inference fully constrains the two parameters that define the cross-field density structure of the waveguide. 
We note that the inference results in Figure~\ref{figpascoe} were obtained for a rather large value of $l/R=0.4$ in the numerical simulation, while the inference was performed by making use of expressions (\ref{gaussdamp})--(\ref{ratio}) which are in principle only valid under the TTTB approximations.

%%%%%%%%%%%%%%%%%%%%%%%%%%%%%%%%%%%%%%%%%%%%%%%%%%%%%%%%%%%%%%%%%%%%%
\subsection{Density structure across coronal waveguides}\label{densityacross}
%%%%%%%%%%%%%%%%%%%%%%%%%%%%%%%%%%%%%%%%%%%%%%%%%%%%%%%%%%%%%%%%%%%%%

In Sections \ref{seisdamp} and \ref{gaussexp} we have shown that obtaining information about the cross-field density structuring of coronal loops has motivated several studies in this field. In all cases, particular assumptions are made about the exact profile of the density variation at the non-uniform layers for resonantly damped transverse oscillations. For instance, in expressions (\ref{gaussdamp}) and (\ref{expdamp}), the numerical factors in front of the functions of the density contrast and transverse inhomogeneity length-scale depend on the particular profile that is adopted for the density variation at the layer, which is unknown and not directly measurable. In a recent study, \cite{arregui15c} have presented the application of the three levels of Bayesian inference (parameter inference - model comparison - model averaging) to the problem of inferring the cross-field density variation in transversely oscillating coronal waveguides.

Figure~\ref{crossf1} (left panel) displays the spatial variation of three commonly used density models at the boundary of the waveguide: the linear, sinusoidal, and parabolic profiles. Although they lead to a priori not very different numerical factors in the expression for the damping rate ($2/\pi\sim0.64$ for the sinusoidal profile, $4/\pi^2\sim0.41$ for the linear profile, and $4\sqrt{2}/\pi^2\sim0.57$ for the parabolic profile), the solutions to the three inversion problems, following the classical methods by \cite{arregui07a} and \cite{goossens08a} and computed by \cite{soler14a} display marked differences (see right panel in Figure~\ref{crossf1}). One could conclude that the inverse solutions for resonantly damped transverse oscillations strongly depend on the assumed density profile at the tube boundary. 

\begin{figure*}
   \centering
   \includegraphics[width=0.32\textwidth]{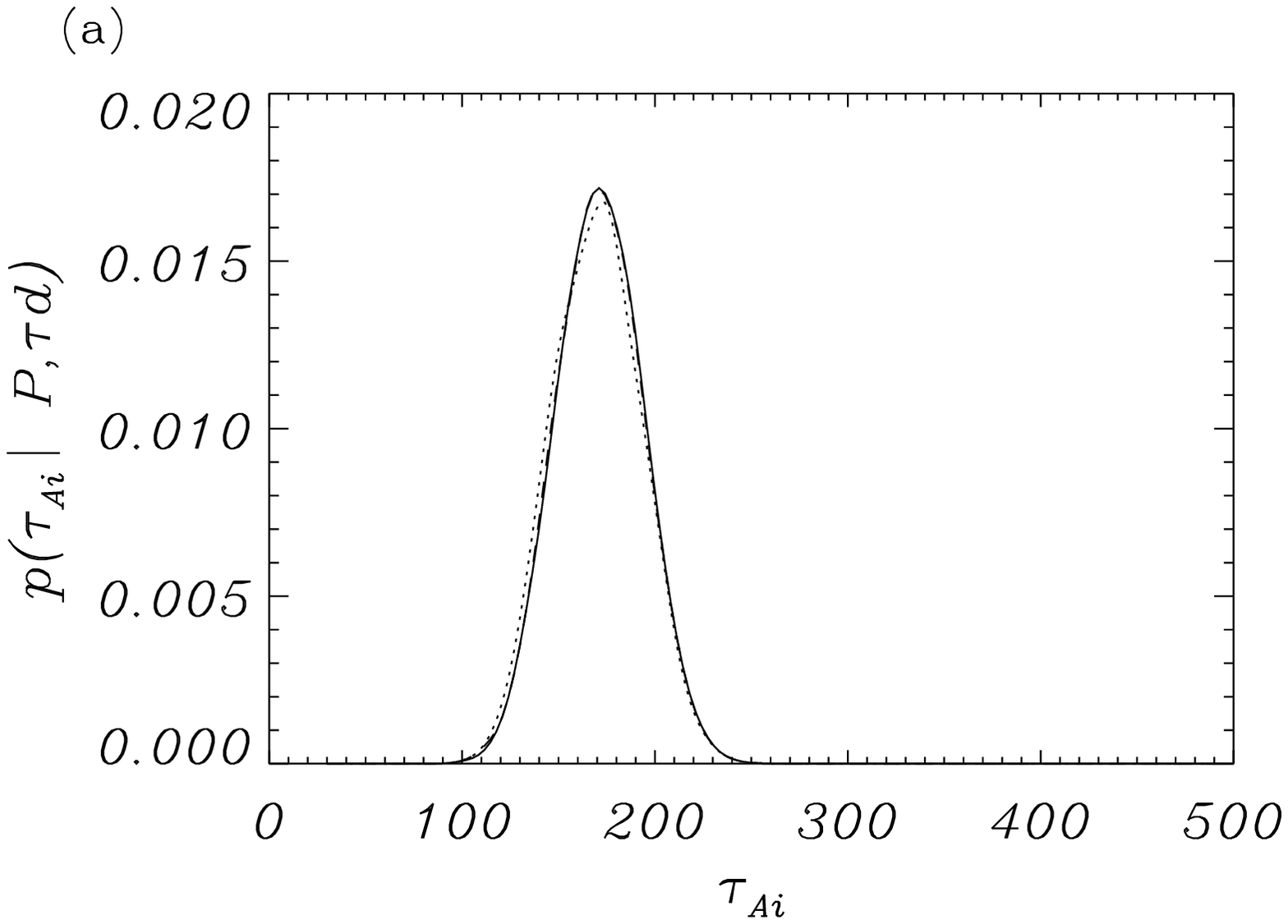}
   \includegraphics[width=0.32\textwidth]{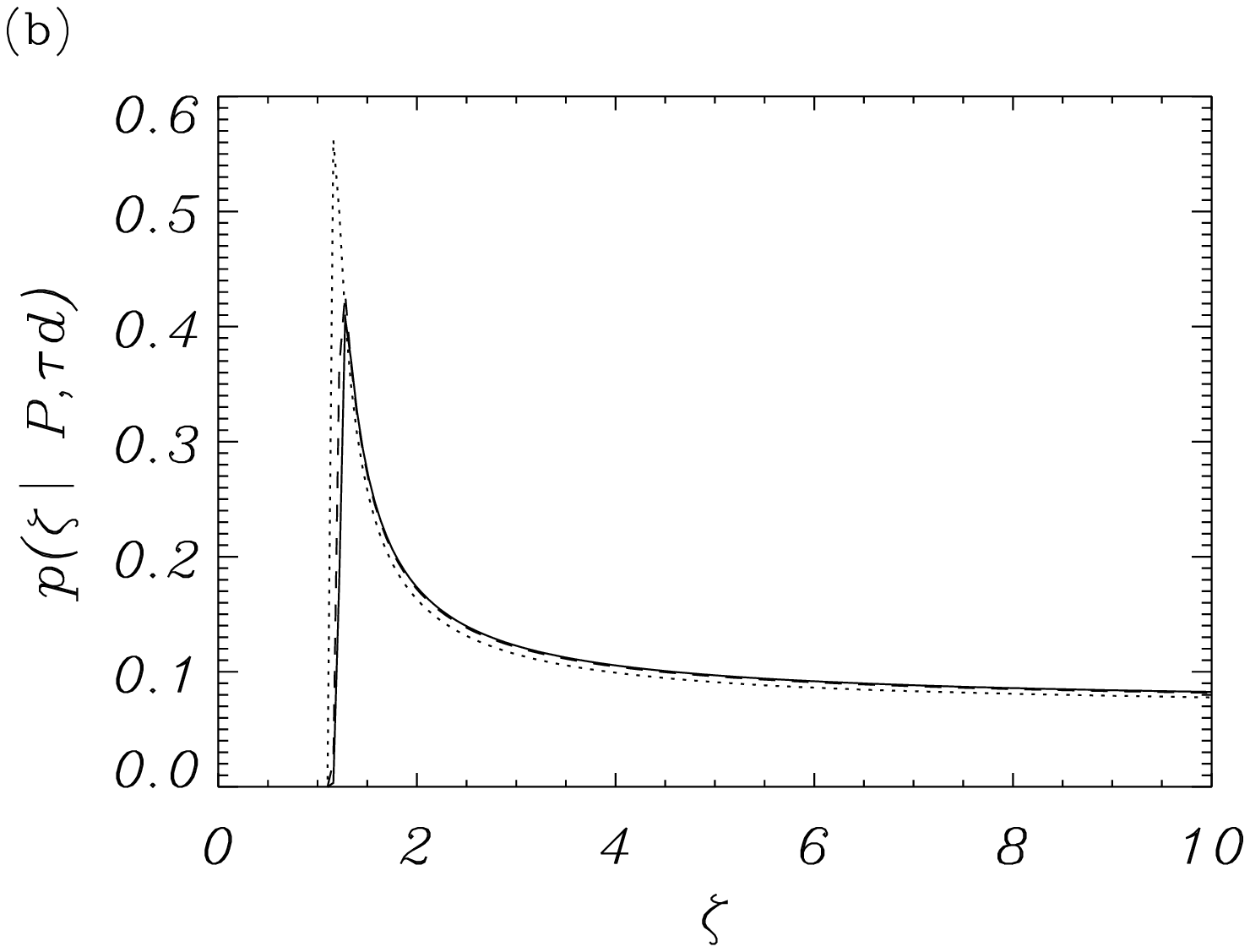}
   \includegraphics[width=0.32\textwidth]{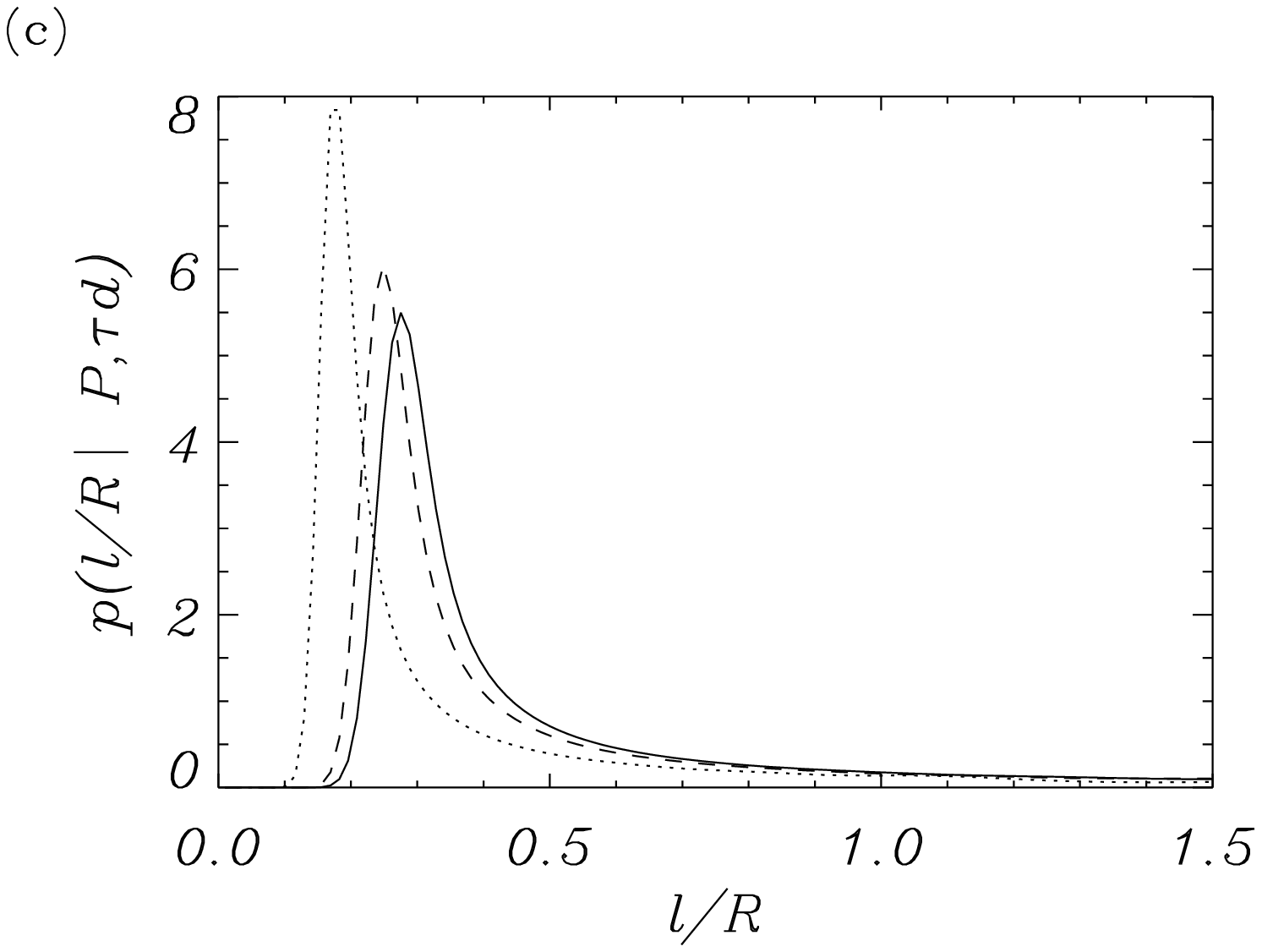}\\
   \includegraphics[width=0.32\textwidth]{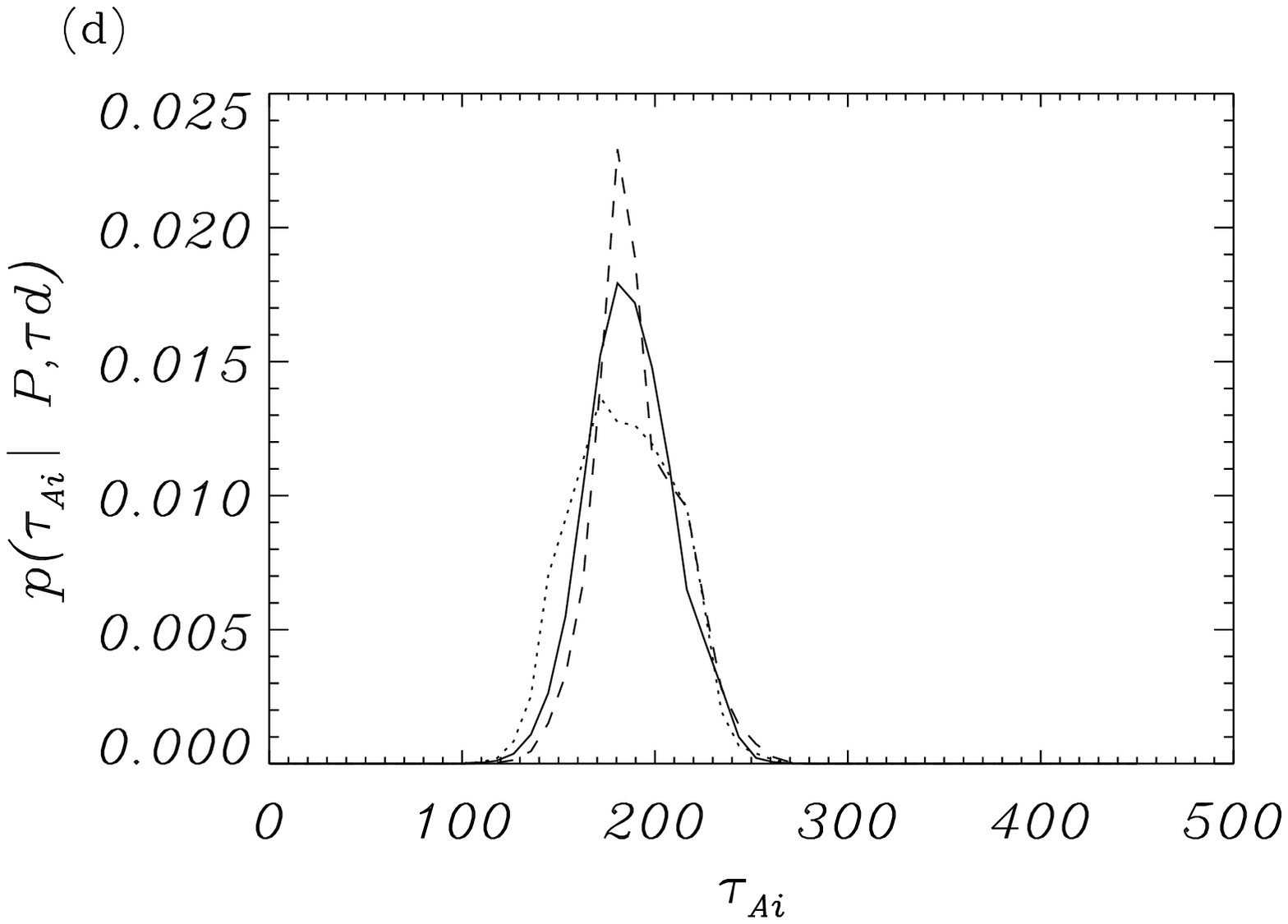}
   \includegraphics[width=0.32\textwidth]{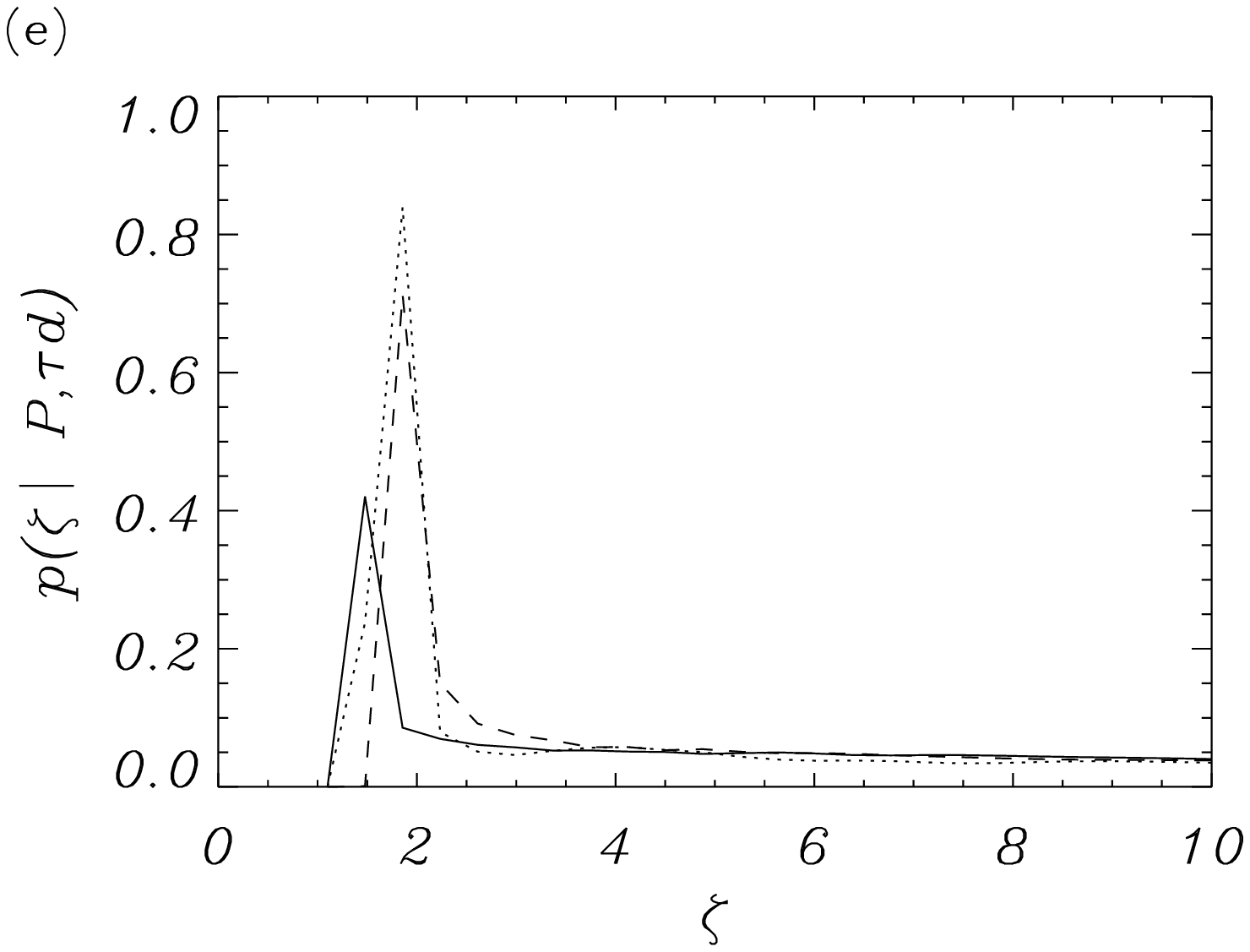}
   \includegraphics[width=0.32\textwidth]{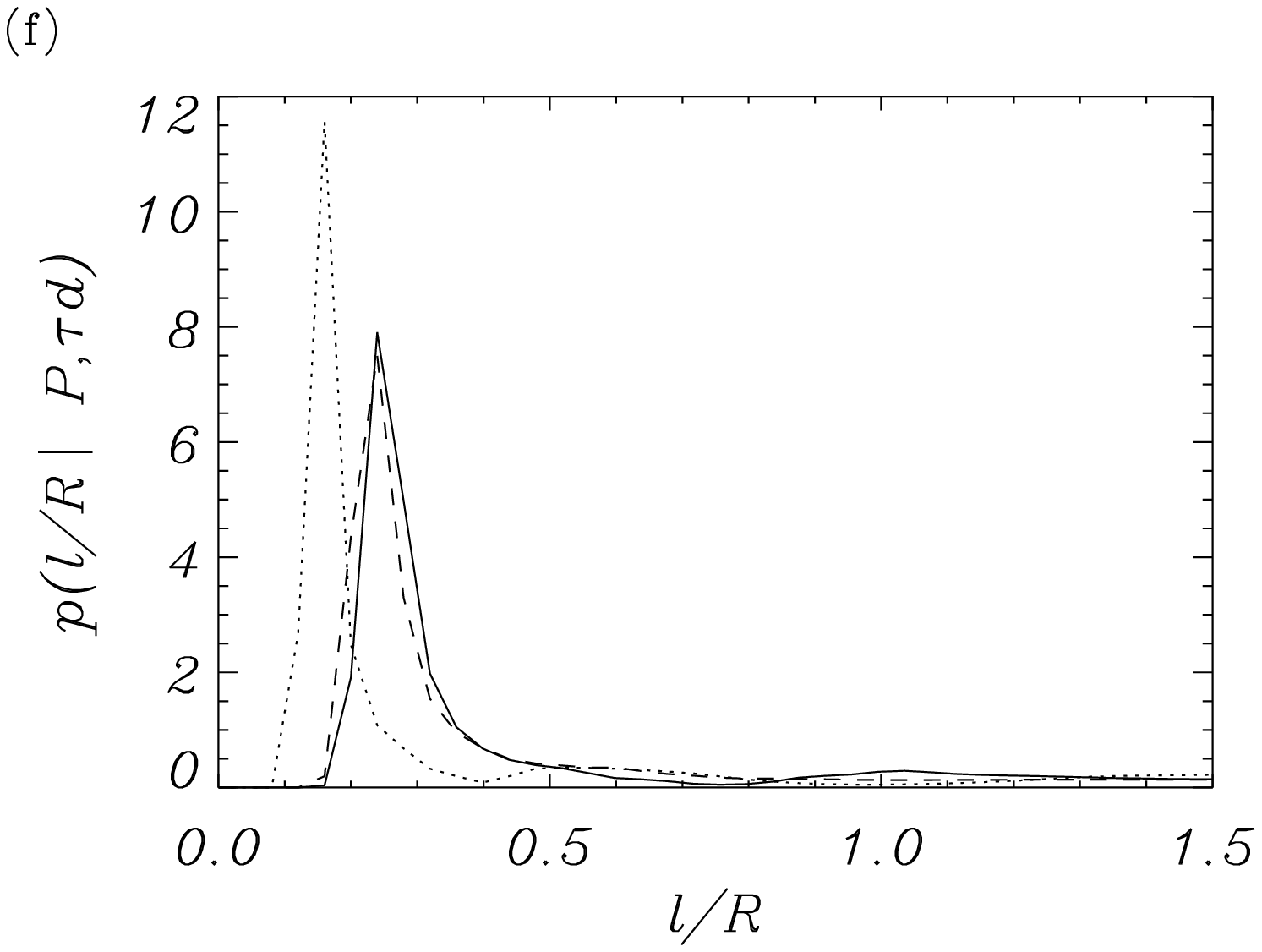}
   \caption{Marginal posteriors for ($\tau_{\rm Ai}$, $\zeta$, $l/R$) for the moderate damping case ($P=272$ [sec]; $\tau_{\rm d}=849$ [sec]; $\sigma_{\rm P}=\sigma_{\rm \tau_{\rm d}}= 30$ [sec)]) and for the three alternative density models: sinusoidal (solid); linear (dotted); parabolic (dashed). Top row: TTTB results; bottom row: numerical results. Adapted from \cite{arregui15c}.}
              \label{inf1f2}%
    \end{figure*}
\begin{figure*}
   \centering
   \includegraphics[width=0.32\textwidth]{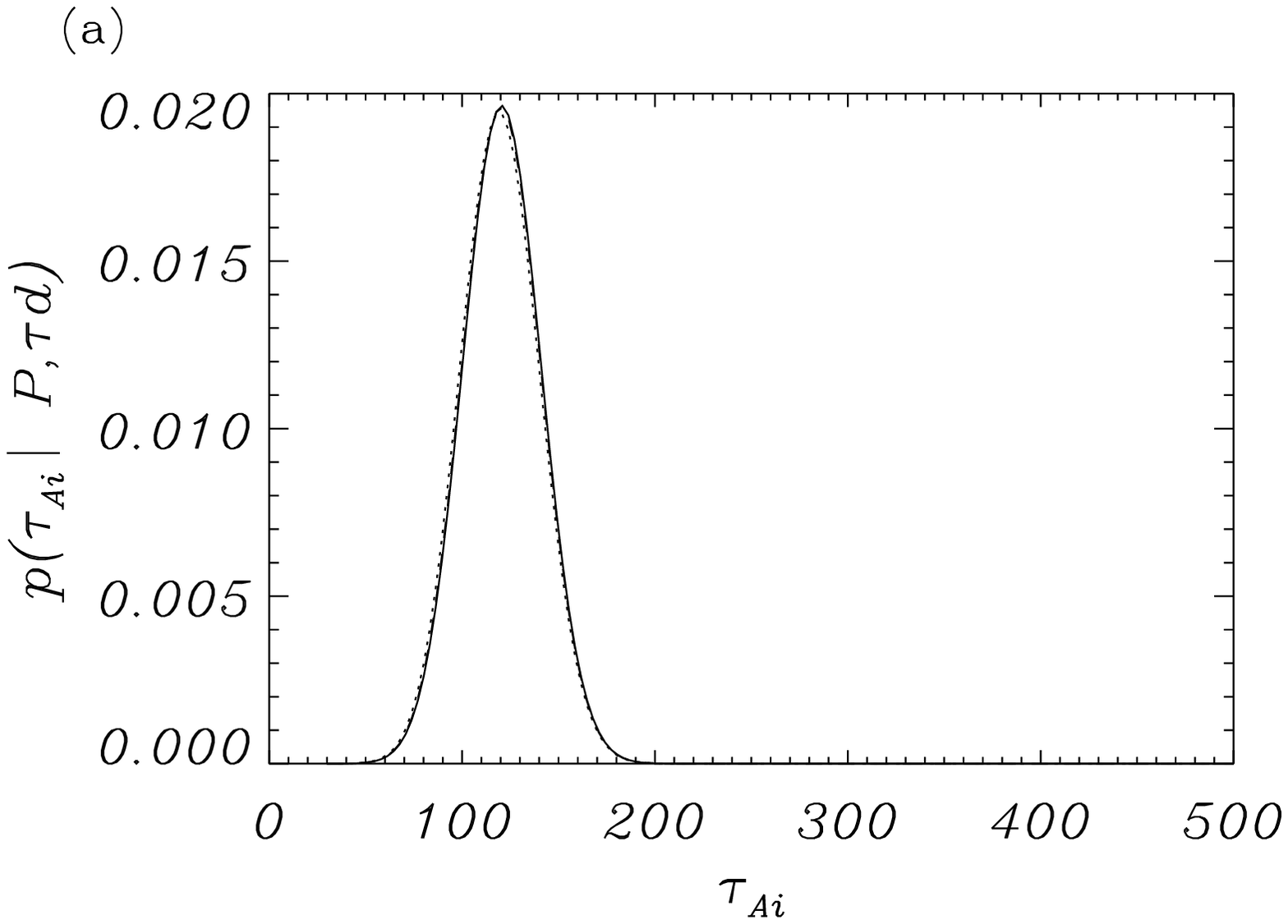}
   \includegraphics[width=0.32\textwidth]{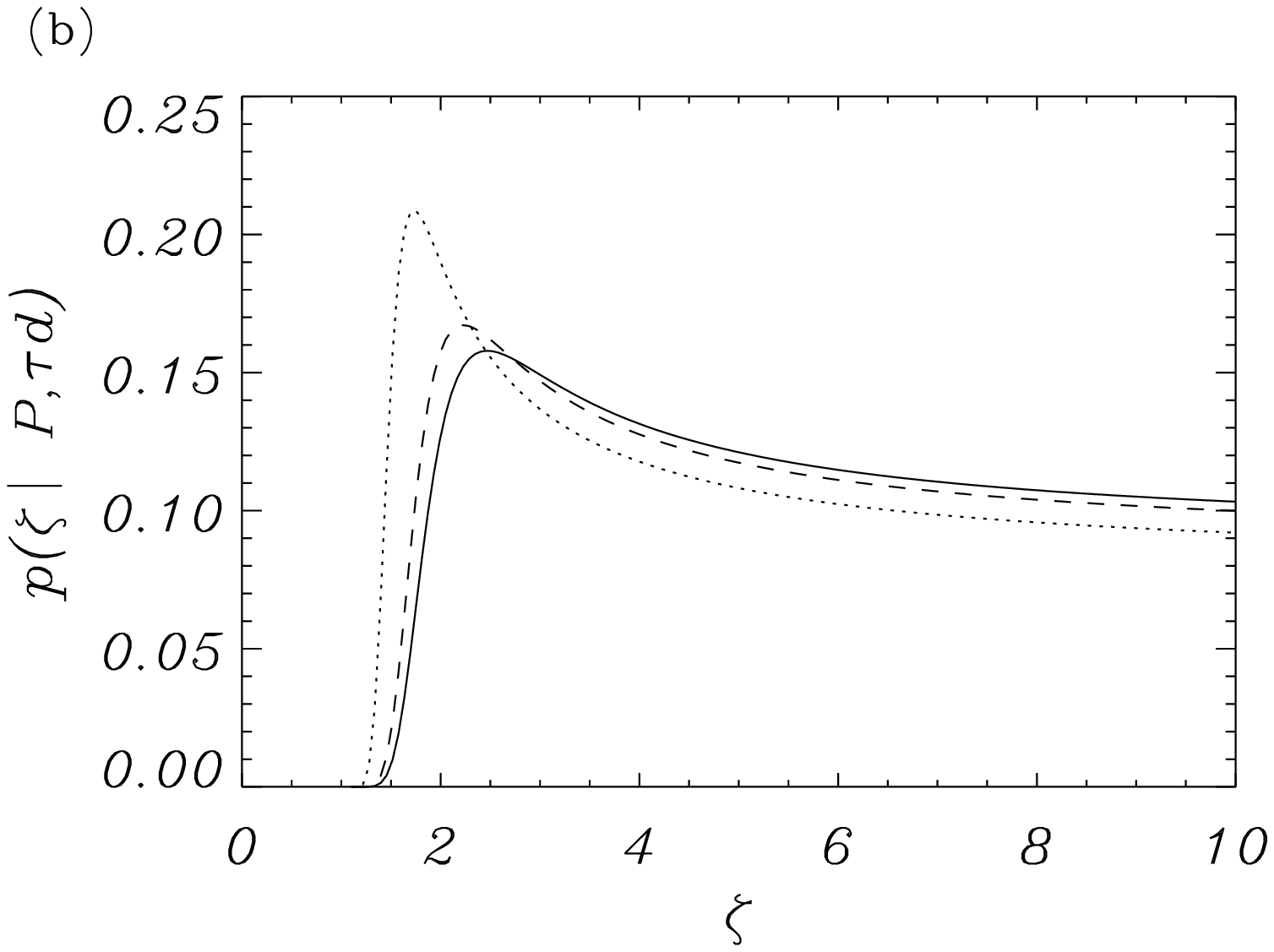}
   \includegraphics[width=0.32\textwidth]{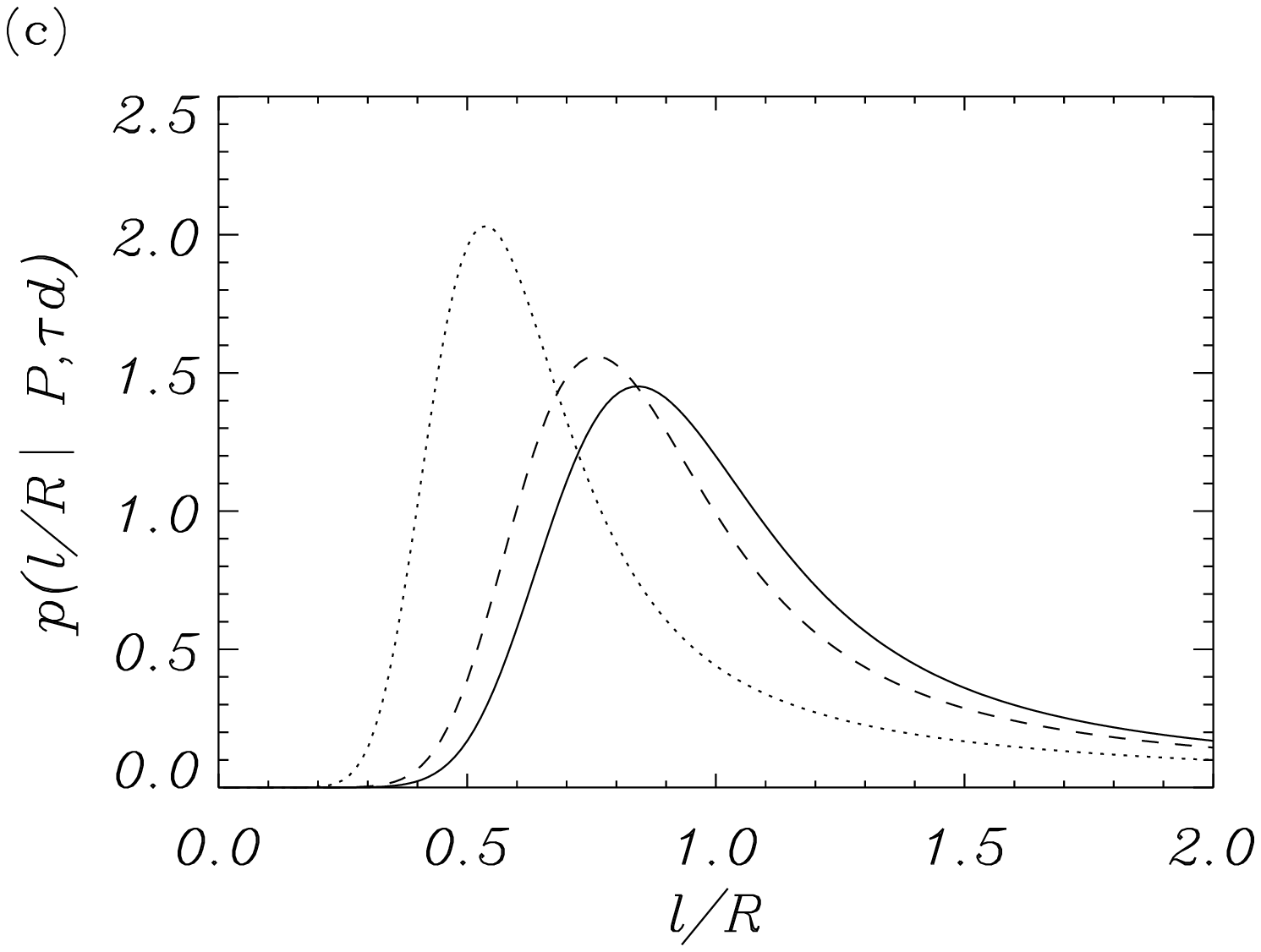}\\
   \includegraphics[width=0.32\textwidth]{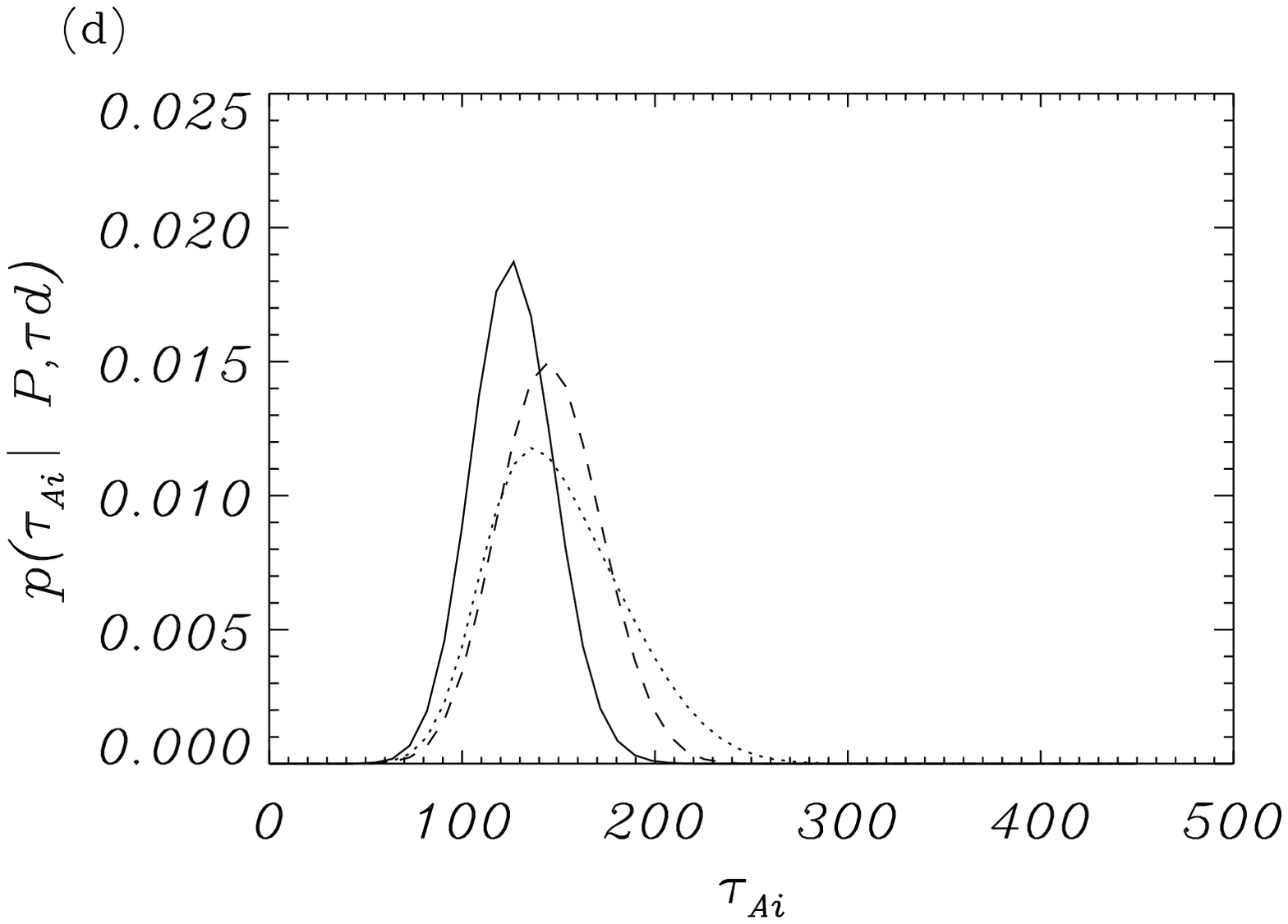}
   \includegraphics[width=0.32\textwidth]{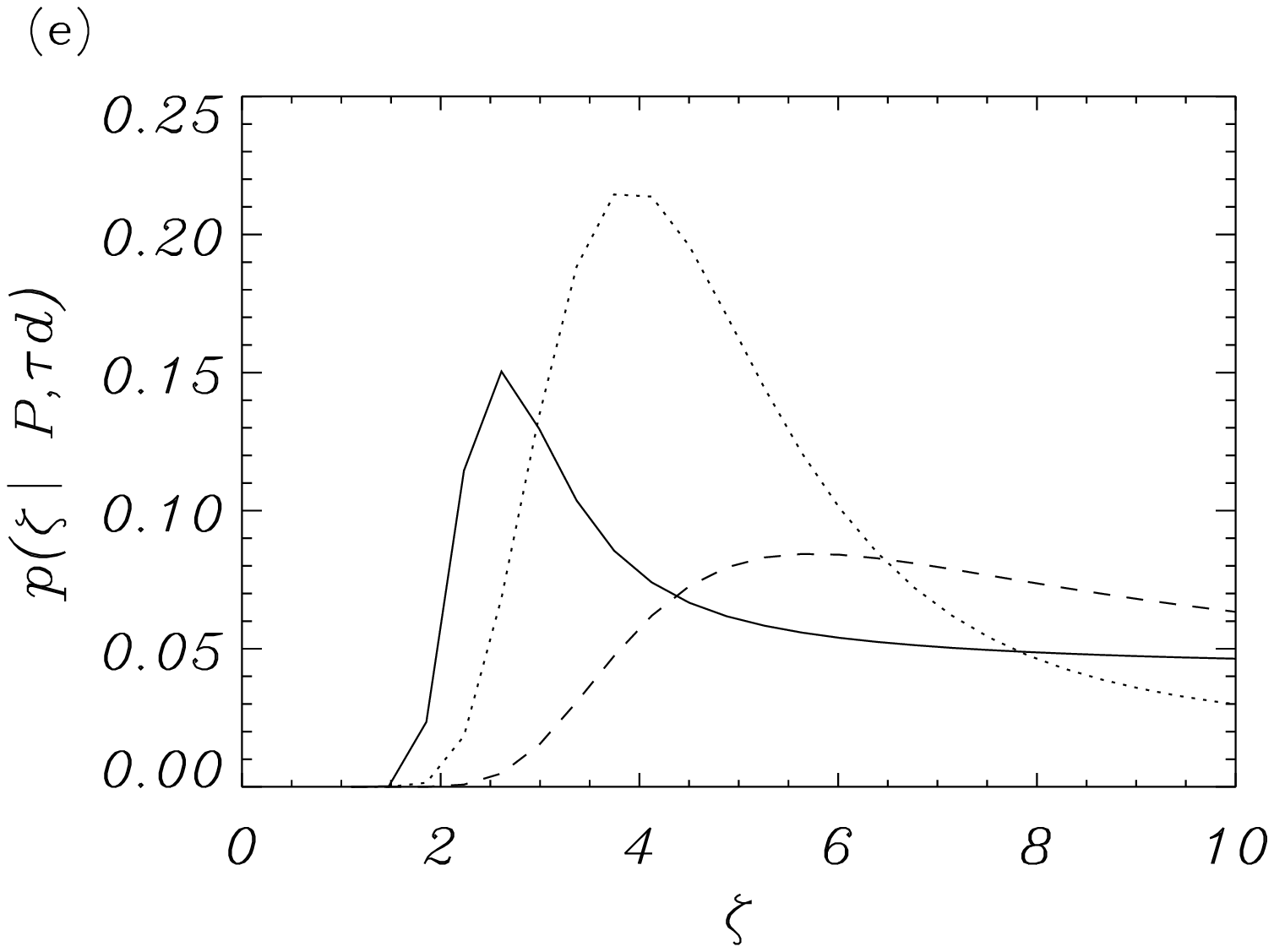}
   \includegraphics[width=0.32\textwidth]{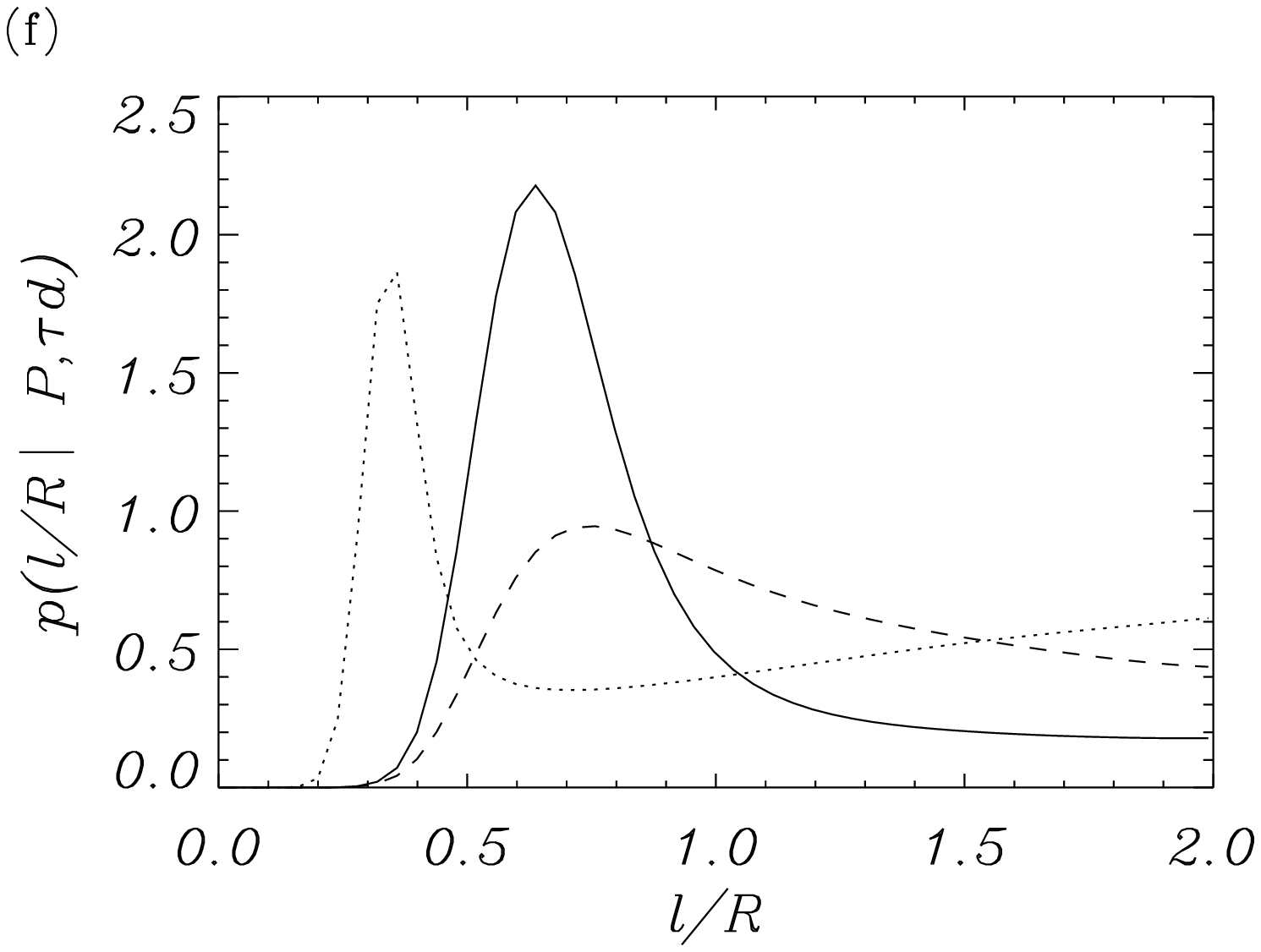}
   \caption{Marginal posteriors ($\tau_{\rm Ai}$, $\zeta$, $l/R$) for the strong damping case ($P=185$ [sec]; $\tau_{\rm d}=200$ [sec]; $\sigma_{\rm P}=\sigma_{\rm \tau_{\rm d}}= 30$ [sec)]) and for the three alternative density models: sinusoidal (solid); linear (dotted); parabolic (dashed). Top row: TTTB results; bottom row: numerical results. Adapted from \cite{arregui15c}.}
              \label{inf2f2}%
    \end{figure*}

The problems of determining how different the solutions are, when considering the uncertainty on the inversion, and of assessing which one of the density models better explains data have been considered by \cite{arregui15c}. First, the inversion problem was solved using Bayesian analysis. The analysis was applied to two cases from a set of observations by \cite{ofman02b}, one representing observations with moderate damping ($P=272$ sec,  $\tau_{\rm d}=849$ sec, $\tau_{\rm d}/P=3.12$) and another representing observations with strong damping ($P=185$ sec, $\tau_{\rm d}=200$ sec, $\tau_{\rm d}/P=1.08$). A timescale of 30 sec was taken as  the uncertainty on measured period and damping time, in line with current observational capabilities. 
\begin{figure*}
   \centering
   \includegraphics[width=0.325\textwidth]{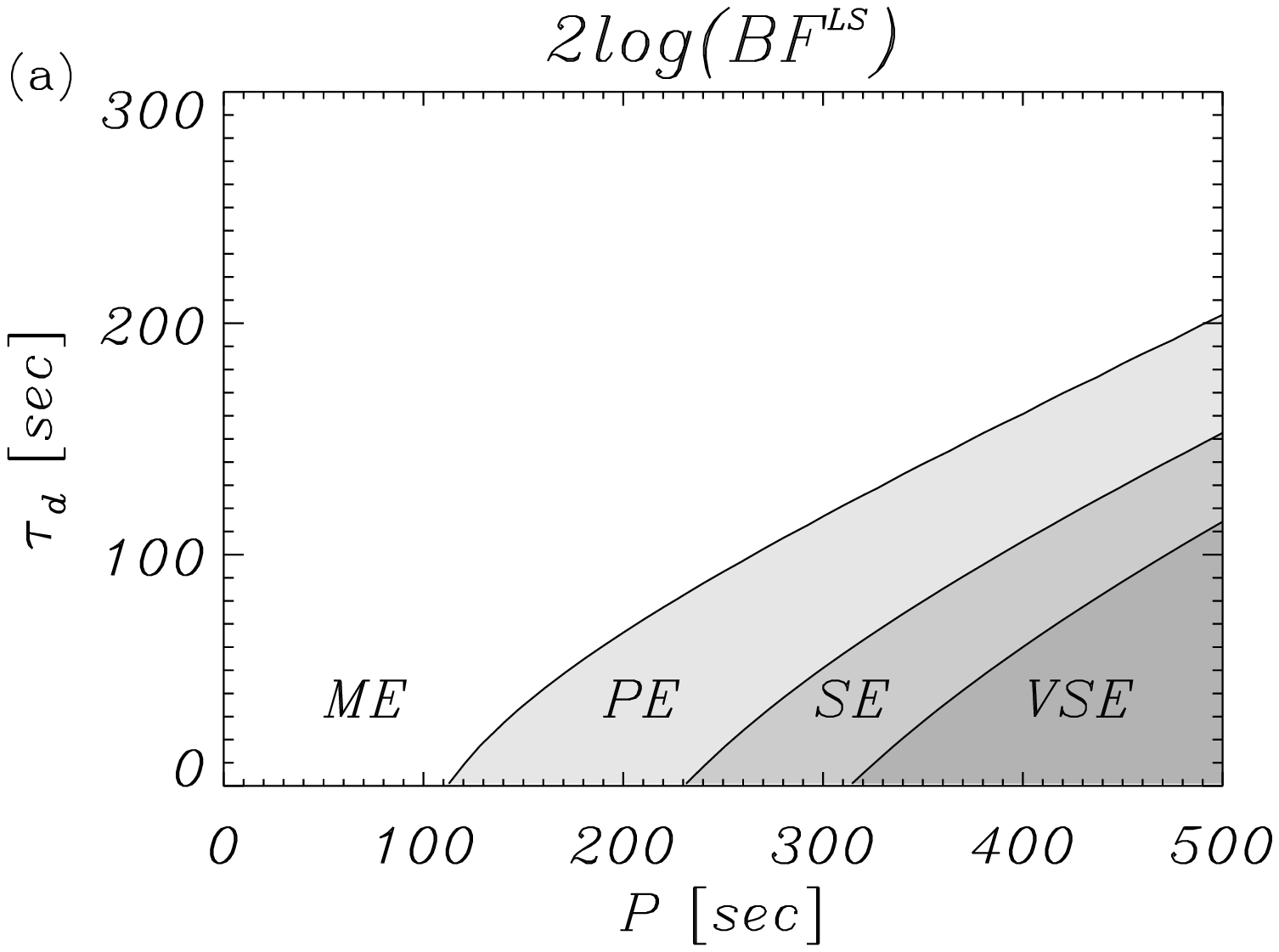}
   \includegraphics[width=0.325\textwidth]{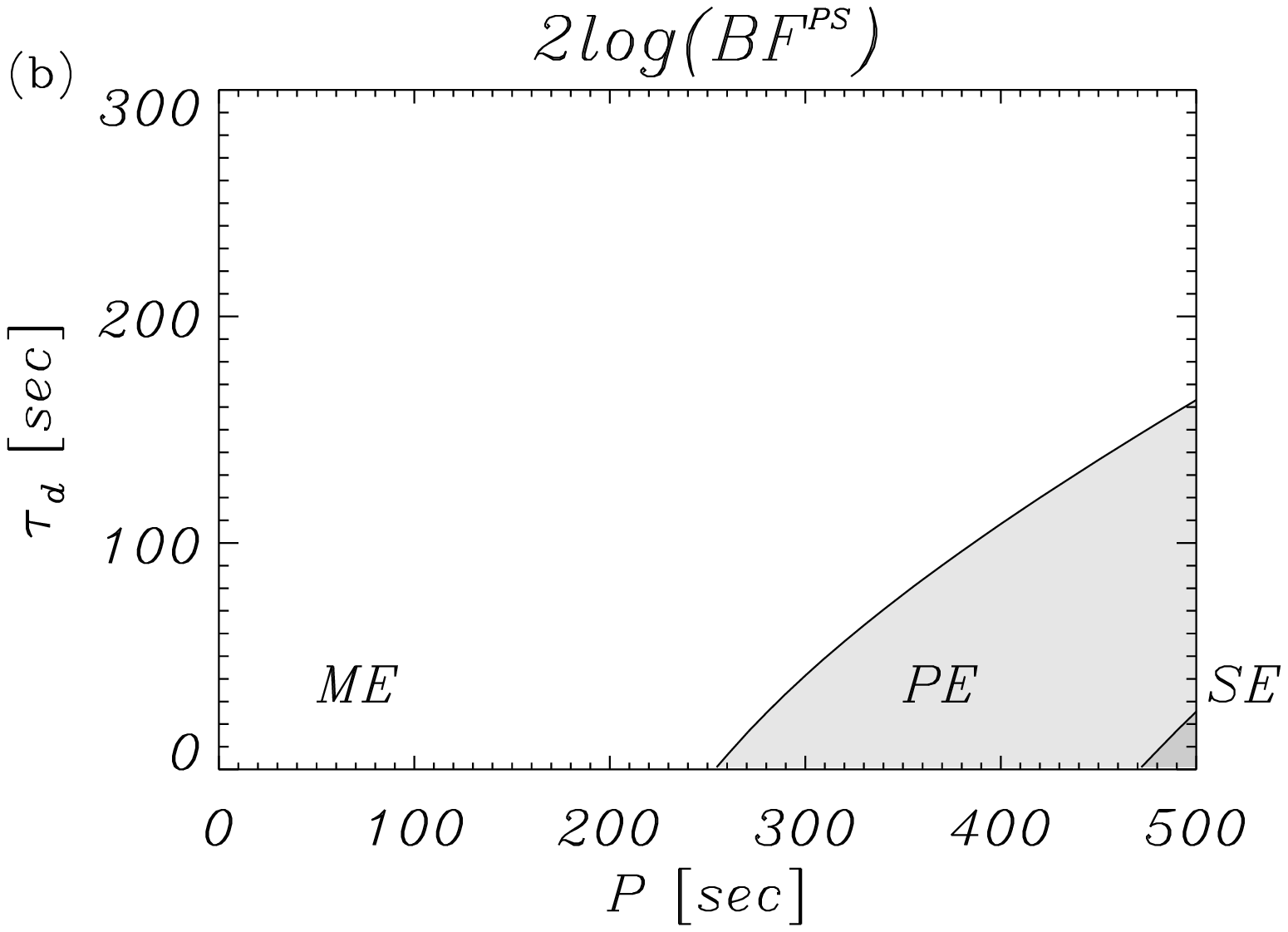}
   \includegraphics[width=0.325\textwidth]{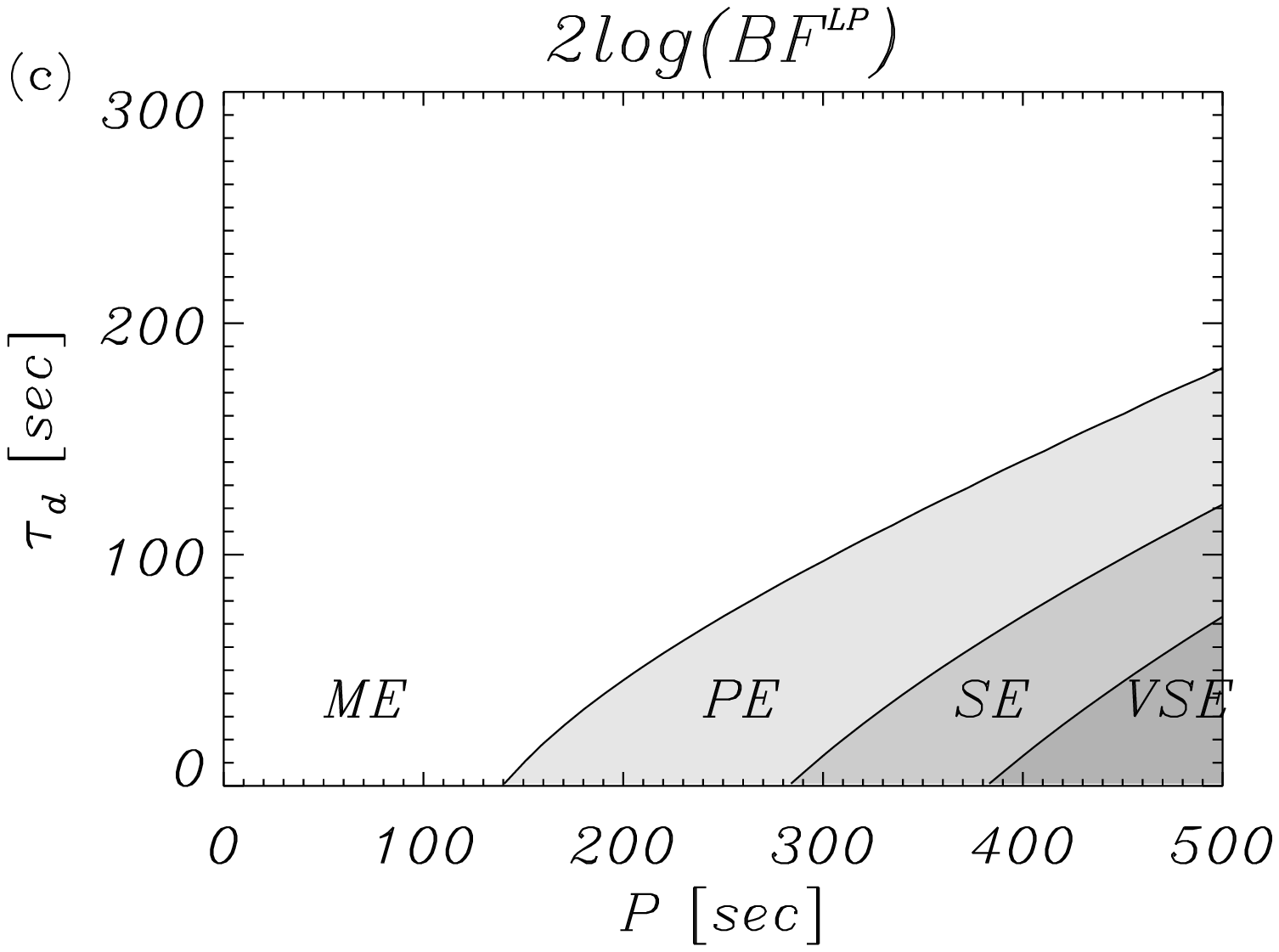}\\
   \includegraphics[width=0.325\textwidth]{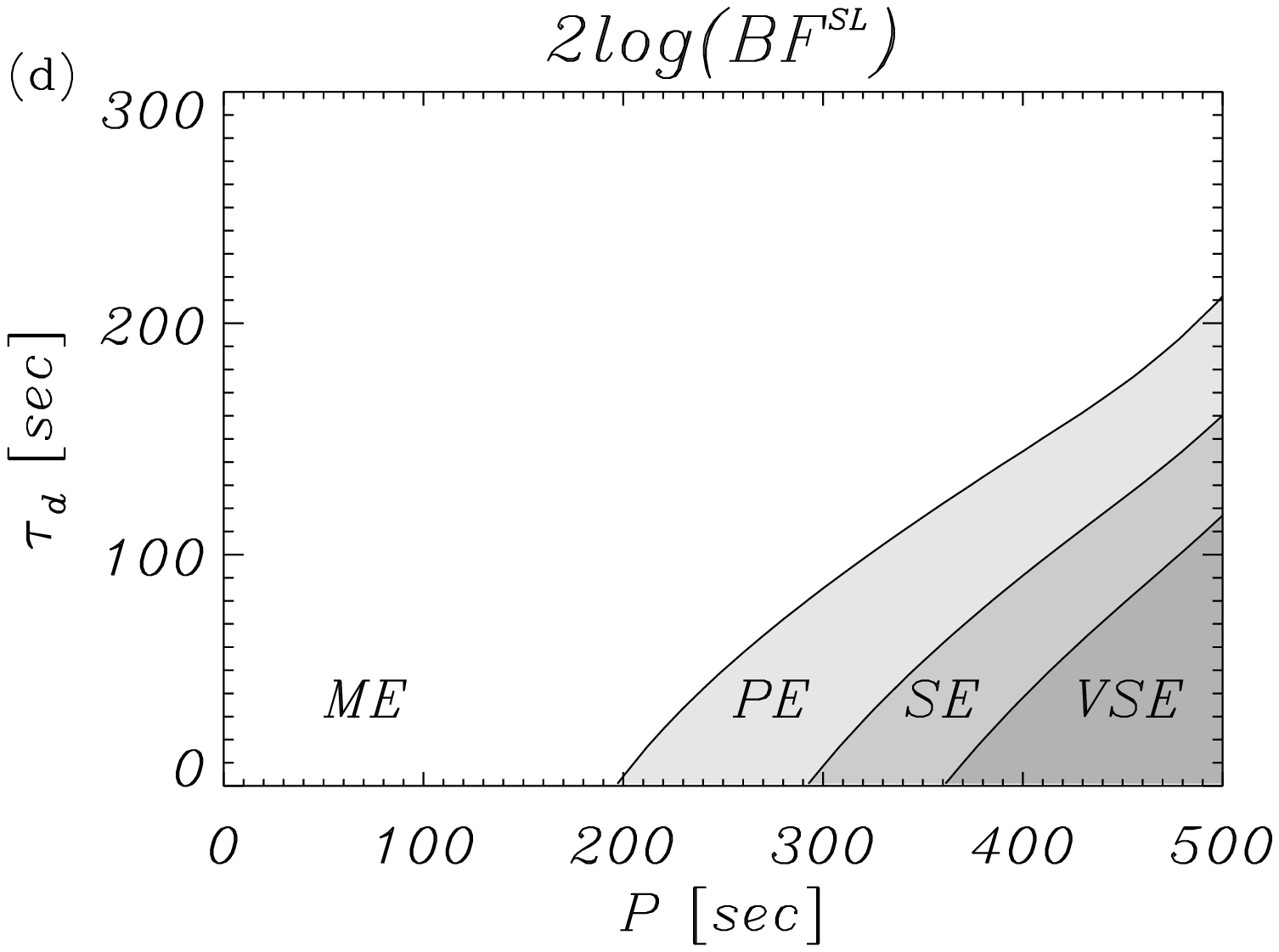}
   \includegraphics[width=0.325\textwidth]{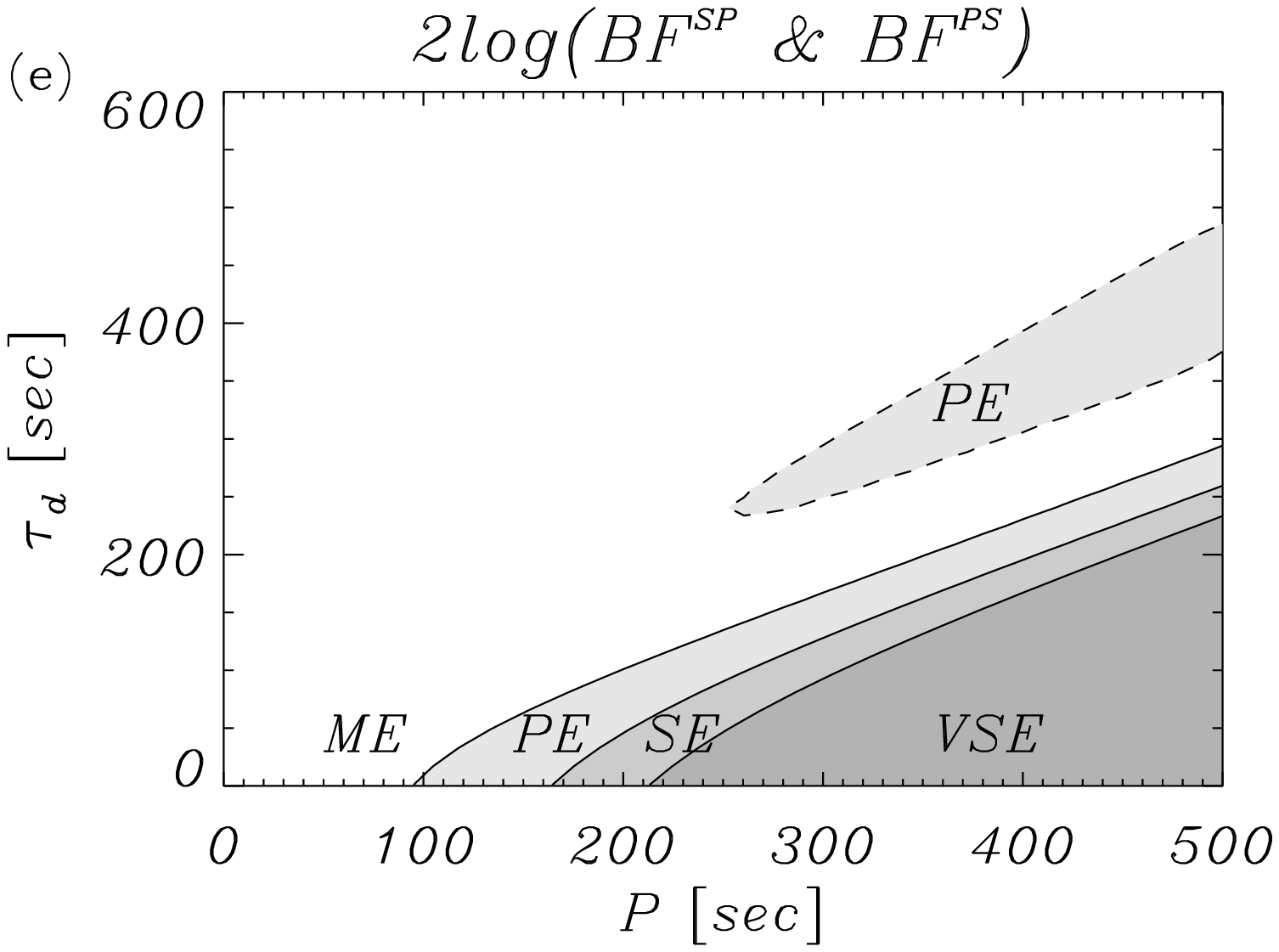}
   \includegraphics[width=0.325\textwidth]{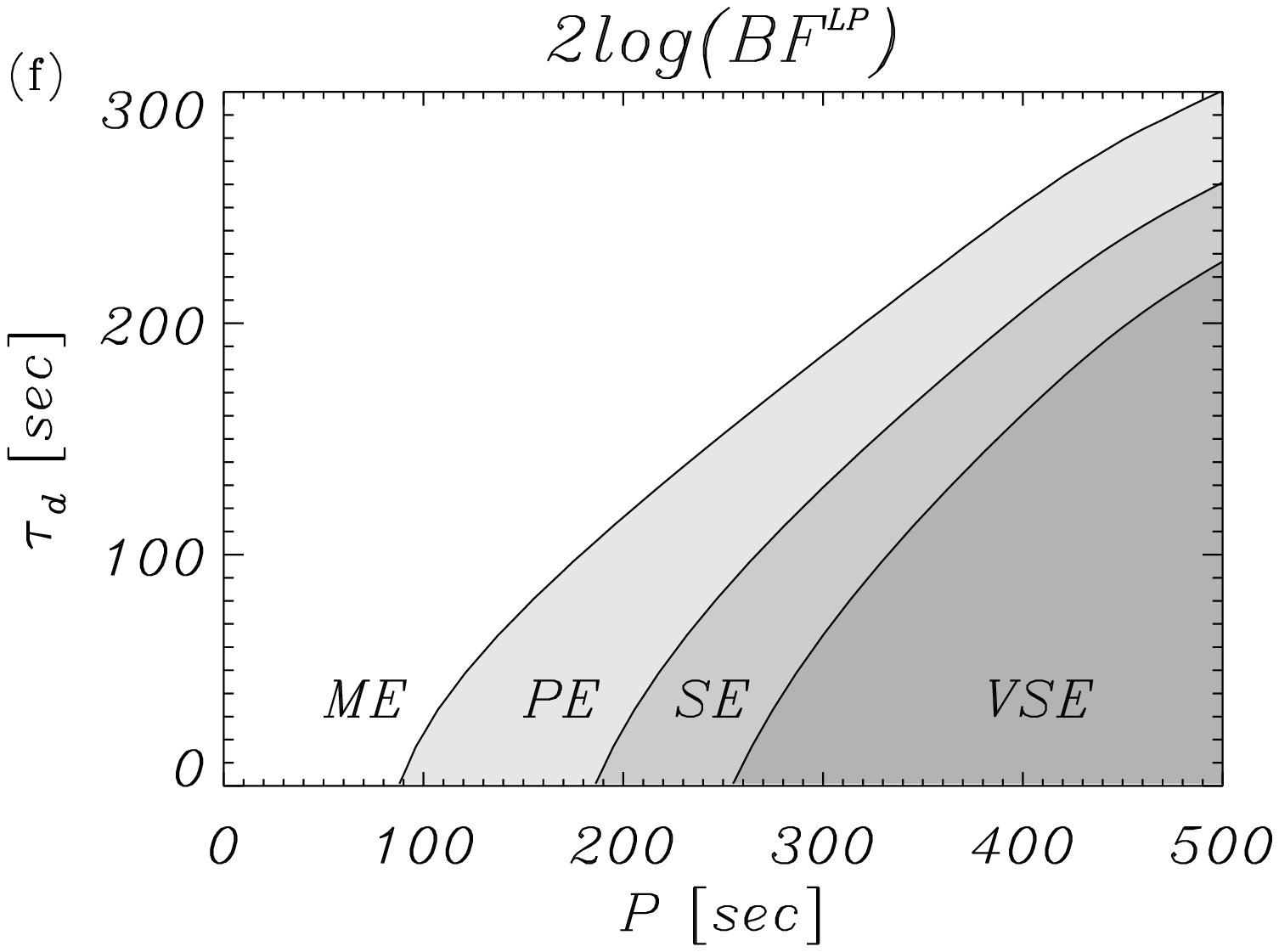}  
   \caption{Contour plots of the regions with different levels of evidence according to Bayes factor values from the comparison between alternative density models in the two-dimensional observational parameter space ($P$, $\tau_{\rm d}$). ME: minimal (inconclusive) evidence; PE: positive evidence; SE: strong evidence; VSE: very strong evidence. Top row panels correspond to the analysis using TTTB  solutions. Bottom row panels correspond to the analysis using numerical solutions. The density models being compared are: (a) and (d): sinusoidal vs. linear; (b) and (e): sinusoidal vs. parabolic; and (c) and (f): linear vs. parabolic. Adapted from \cite{arregui15c}.} 
              \label{bayescross}%
\end{figure*}

\begin{figure*}
   \centering
   \includegraphics[width=0.3\textwidth]{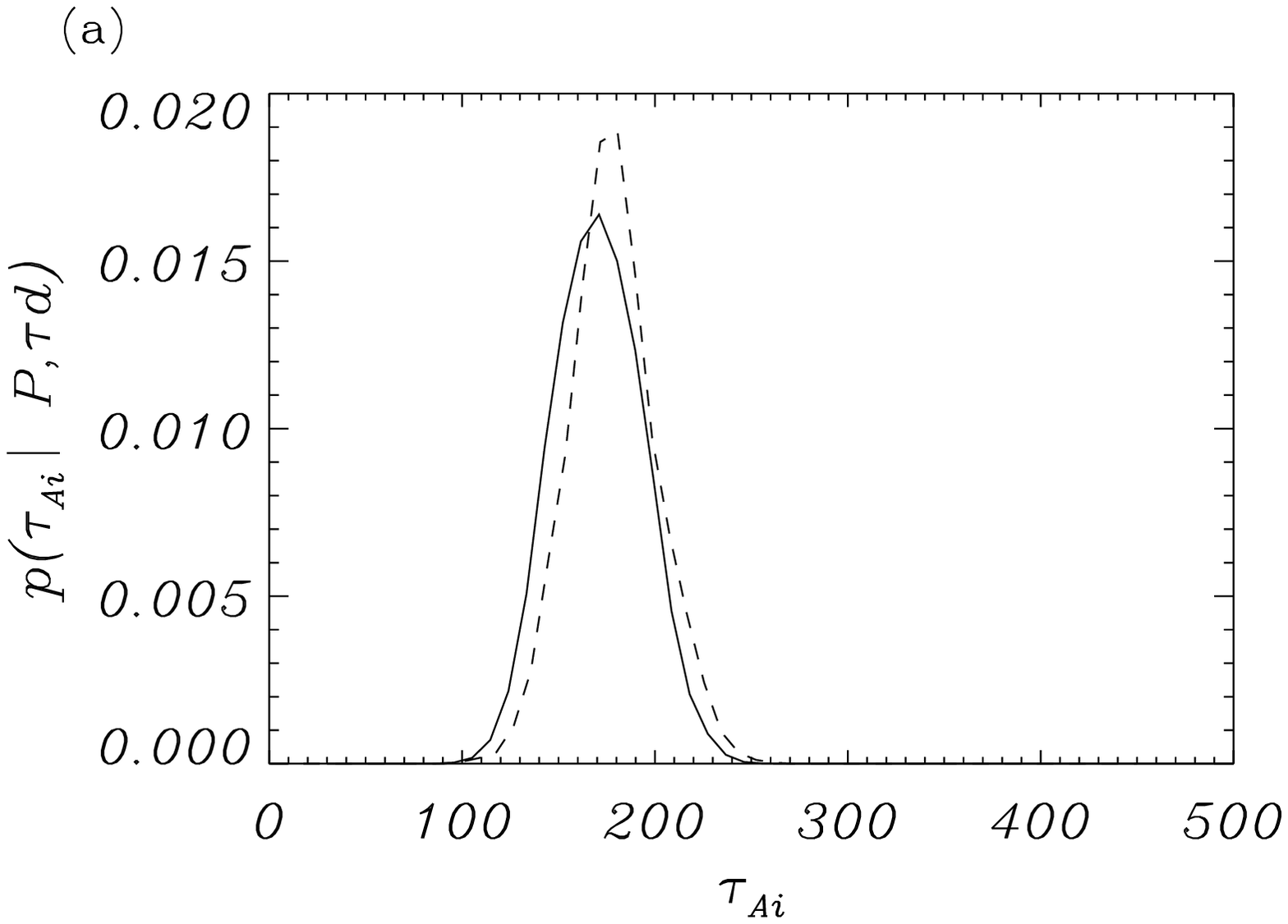}
   \includegraphics[width=0.3\textwidth]{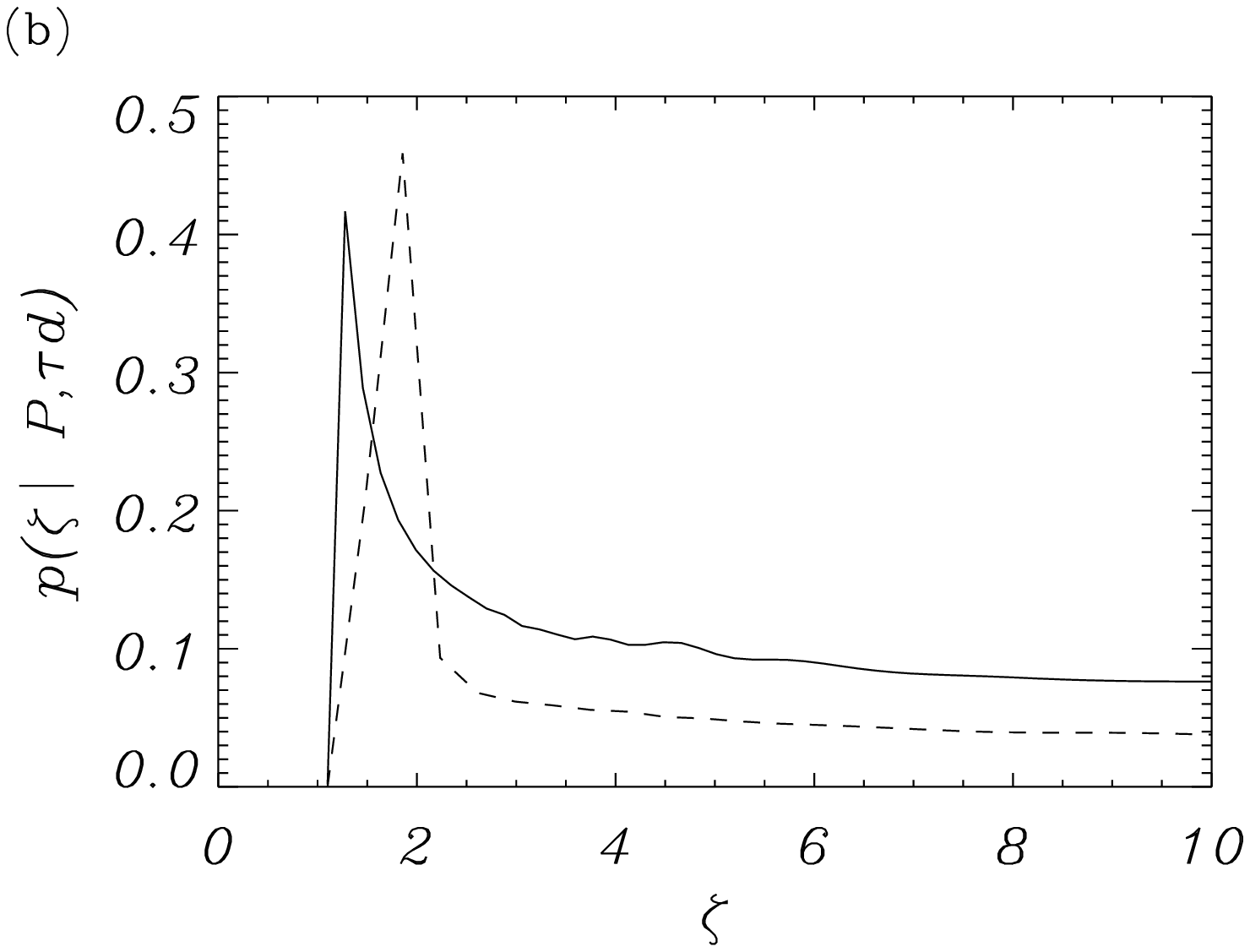}
   \includegraphics[width=0.3\textwidth]{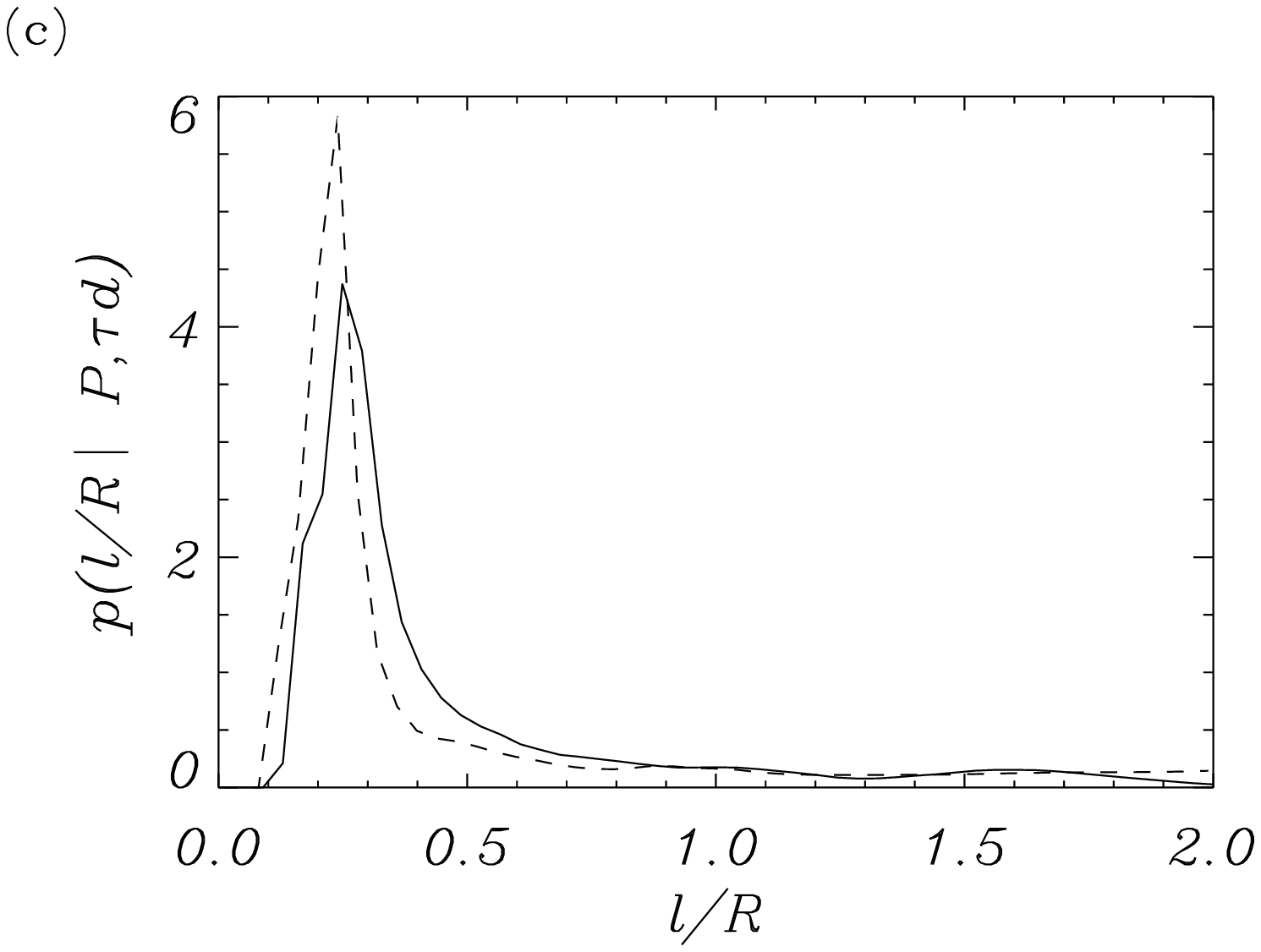}\\
   \includegraphics[width=0.3\textwidth]{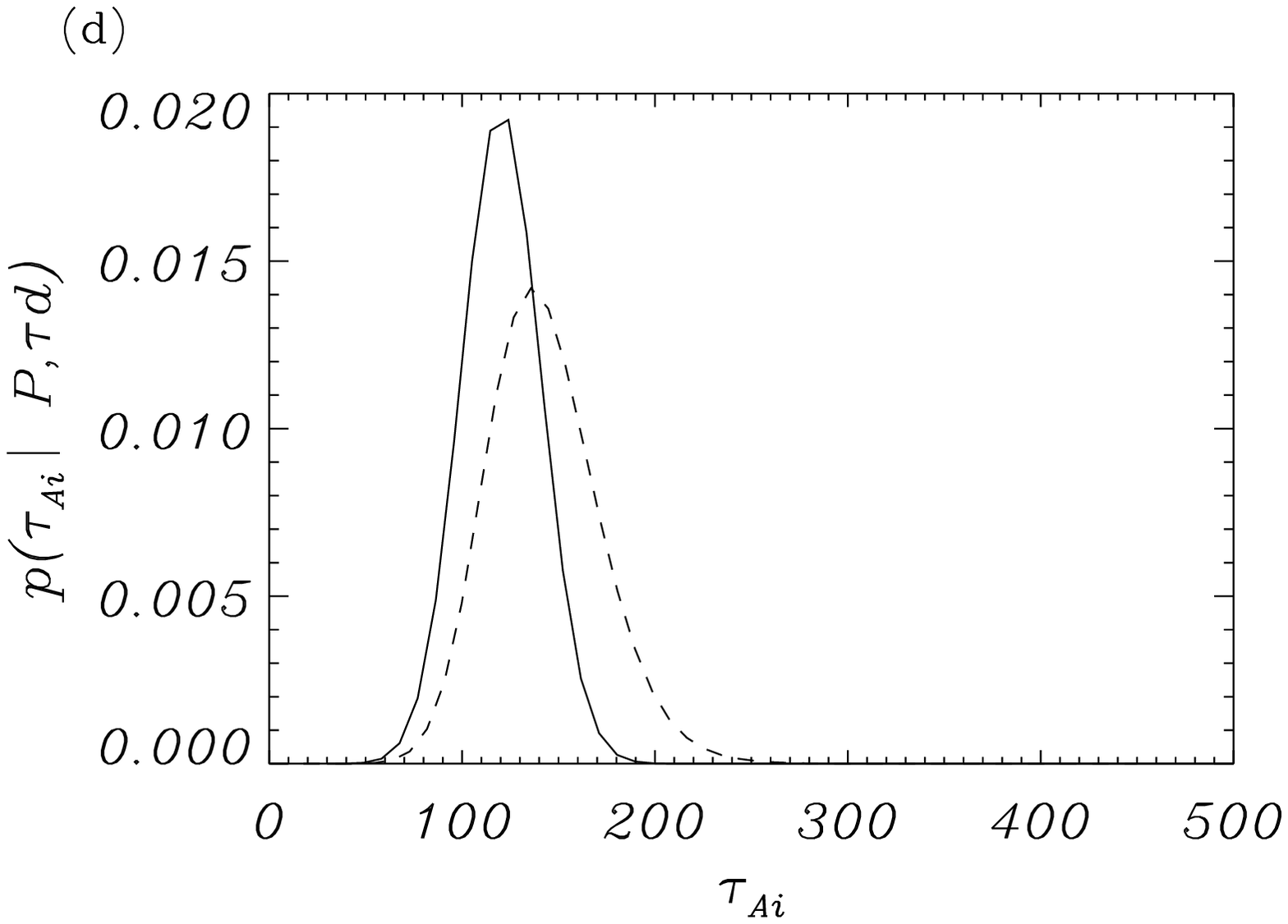}
   \includegraphics[width=0.3\textwidth]{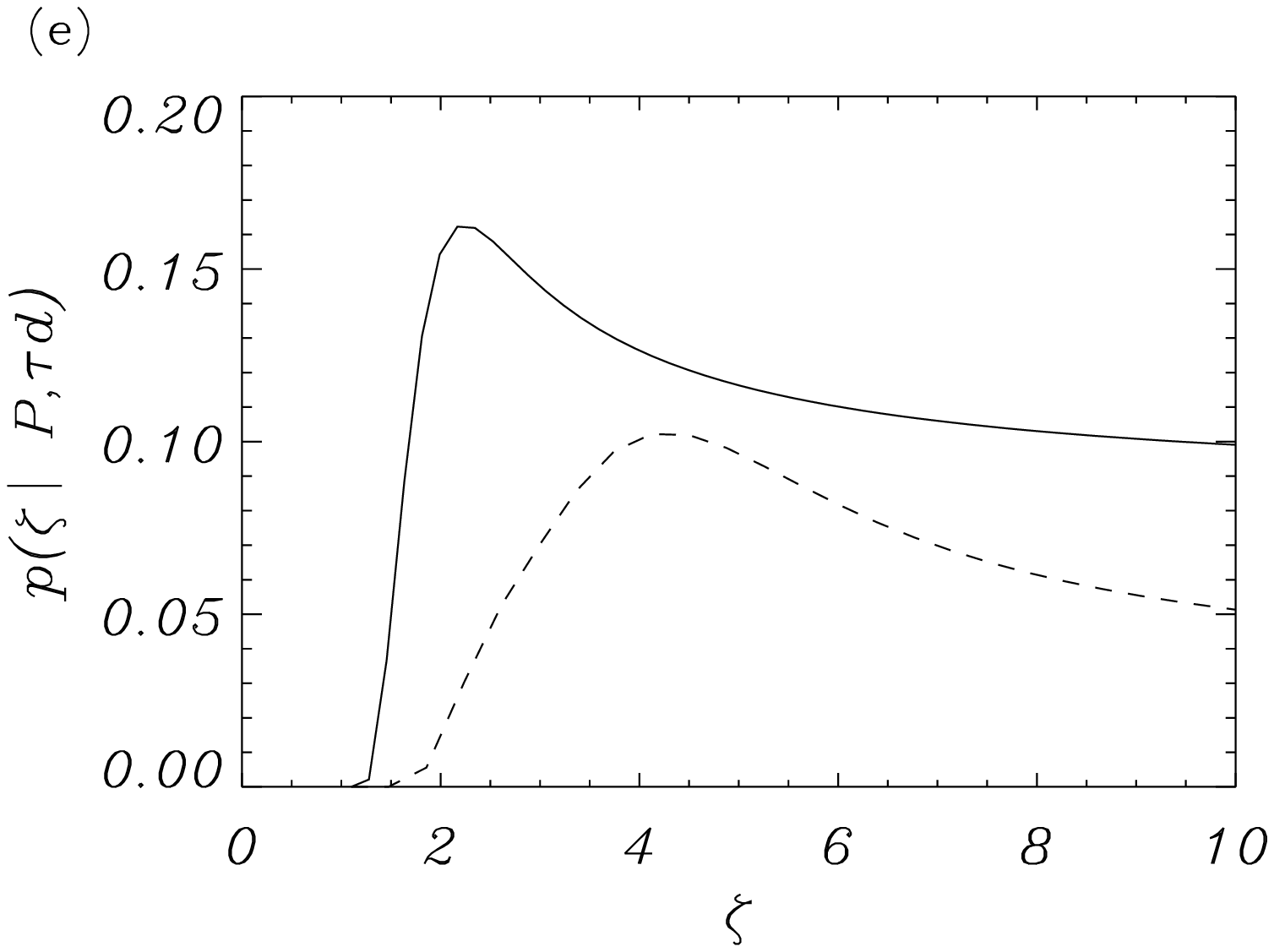}
   \includegraphics[width=0.3\textwidth]{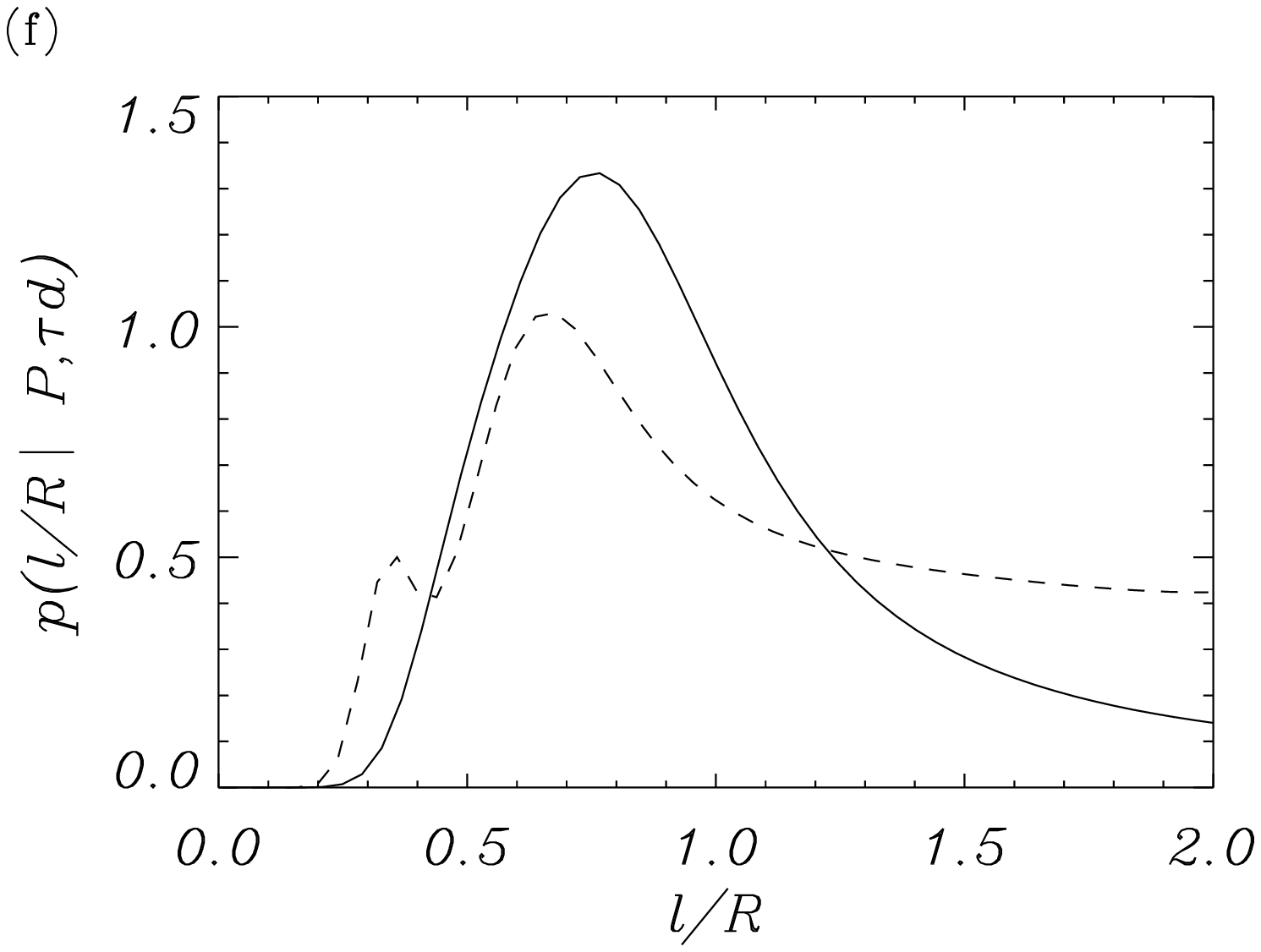}
   \caption{Model-averaged marginal posteriors for the three parameters of interest computed using expression~(\ref{average}). Top and bottom panels correspond to the moderate ($P=272$ [sec]; $\tau_{\rm d}=849$ [sec]) and strong ($P=185$ [sec]; $\tau_{\rm d}=200$ [sec]) damping cases, respectively. In all cases, $\sigma_{\rm P}=\sigma_{\rm \tau_{\rm d}}= 30$ [sec]. On each panel, solid lines represent the inference performed using the TTTB solutions while dashed lines are for the inference performed using the numerical  solutions. Adapted from \cite{arregui15c}.}
              \label{averaged}%
    \end{figure*}

For the case with moderate damping, Figure~\ref{inf1f2} displays the obtained results with the posteriors arranged in such a way that a comparison between the inversions for the three alternative density models can be directly made. The top row shows results employing the analytical thin tube and thin boundary approximations for the period and damping time, the bottom row those employing numerical solutions outside these approximations.  We see that for moderate damping, the adopted density model does not seem to influence the inference result that much. For the case with strong damping, Figure~\ref{inf2f2} displays the obtained results. Again, the top row shows results employing the analytical thin tube and thin boundary approximations for the period and damping time, the bottom row those employing numerical solutions outside these approximations. These figures clearly show that in cases with strong damping, the chosen model has important consequences on the inversion results.

The pertinent question is then how to decide which one among the inferences shown in Figures~\ref{inf1f2} and \ref{inf2f2} for the alternative density models should one belief the most. To try to answer this question, \cite{arregui15c} applied model comparison to assess the plausibility of the alternative models to explain observed data. The panels in Figure~\ref{bayescross} show the two-dimensional distribution of Bayes factors for the comparisons between linear vs. sinusoidal, parabolic vs. sinusoidal, and linear vs. parabolic models. The different grey-shaded regions indicate the level of evidence for one model against the alternative. The results indicate that only for quite strong damping regimes (low values of damping time in comparison to the oscillation period) do we obtain substantial evidence for one model against the alternative. In those regions, the linear model would be the one supported by data.

However, most observed damped loop oscillation events display damping times that are a few times longer than the oscillation period. This means that, for this problem, model comparison does not enable us to make any firm statement about the preferred model for the cross-field density profile at the tube boundary.

Even if the analysis does not give strong enough support for any model, the evidence for each of them is different. The third level of Bayesian inference enables us to perform the most general possible inference by computing a model-averaged posterior distribution, conditional on the observed data, and weighted with the probability of our set of three alternative models. 
Figure~\ref{averaged} shows the model-averaged marginal posteriors for the three parameters of interest computed for each model. In all panels the model-averaged posteriors using the TTTB and the numerical solutions are compared.  In the moderate damping case, very similar model-averaged posteriors are obtained for the three parameters of interest. Differences are more marked in the strong damping case. These model-averaged posteriors offer the most general inference result that can be obtained on the unknown parameter values. They take into account all the available information, i.e., the prior information, the observed data with their uncertainty, the modelling constraints, and the evidence for each model in view of data.

%%%%%%%%%%%%%%%%%%%%%%%%%%%%%%%%%%%%%%%%%%%%%%%%%%%%%%%%%%%%%%%%%%%%%
\subsection{Hierarchical inference of coronal loop oscillations}\label{andresresults}
%%%%%%%%%%%%%%%%%%%%%%%%%%%%%%%%%%%%%%%%%%%%%%%%%%%%%%%%%%%%%%%%%%%%%

The results described so far were based on the inversion of individual loop oscillation properties, thus obtaining the physical parameters of particular events. Also, the wave properties used as observables were restricted to periods and damping times, quantities obtained from the manipulation of primarily measured quantities, such as the displacement as a funcion of time. \cite{asensioramos13} have gone one step forward by using the displacement time series themselves. In addition, instead of using priors constructed from our state of knowledge before considering the data, \cite{asensioramos13} computed the priors using observational information obtained from a global analysis including a large number of events. This was done by performing a fully consistent analysis of a large number of observations of different events using a hierarchical Bayesian framework. In the same way as directly measured properties of transverse loop oscillations can be summarised in the form of histograms,  the hierarchical analysis enables us to obtain similar information for the physical parameters that cannot be directly measured. The results consist of data-favoured distributions for the unknown parameters that can afterwards be used to construct priors based on our current observational information of coronal loops.

The hierarchical scheme by \cite{asensioramos13} considers a generative model for the observed displacement of loop oscillations given by

\begin{equation}
d(t) = d_\mathrm{trend}(t) + d_\mathrm{osc}(t) + \epsilon(t) + b(t).
\label{eq:generative_model}
\end{equation}

\noindent
Here, $\epsilon(t)$ is the uncertainty of the amplitude measurement and $b(t)$ takes into account the presence of any remaining uncertainty produced by non-modelled effects. The oscillatory component is modelled in terms of a sinusoidal with an exponential decay as 

\begin{equation}
d_\mathrm{osc}(t) = A \sin \left[ \frac{2 \pi}{P} (t-t_0) - \phi_0 \right] \exp \left[- \frac{t-t_0}{\tau_d} \right],
\end{equation}

\noindent
with $A$ the amplitude of the oscillatory part, $P$ its period, $t_0$  a reference initial time fixed from the observations,
$\phi_0$ the initial phase of the oscillation and $\tau_d$ the damping time scale. For the detrending
of the oscillatory motion a simple polynomial function is used that absorbs all the unknown effects \citep{aschwanden02}

\begin{equation}
d_\mathrm{trend}(t) = \sum_{i=0}^N a_i (t-t_0)^i,
\end{equation}

\begin{figure*}
\centering
\includegraphics[width=0.98\textwidth,angle=0]{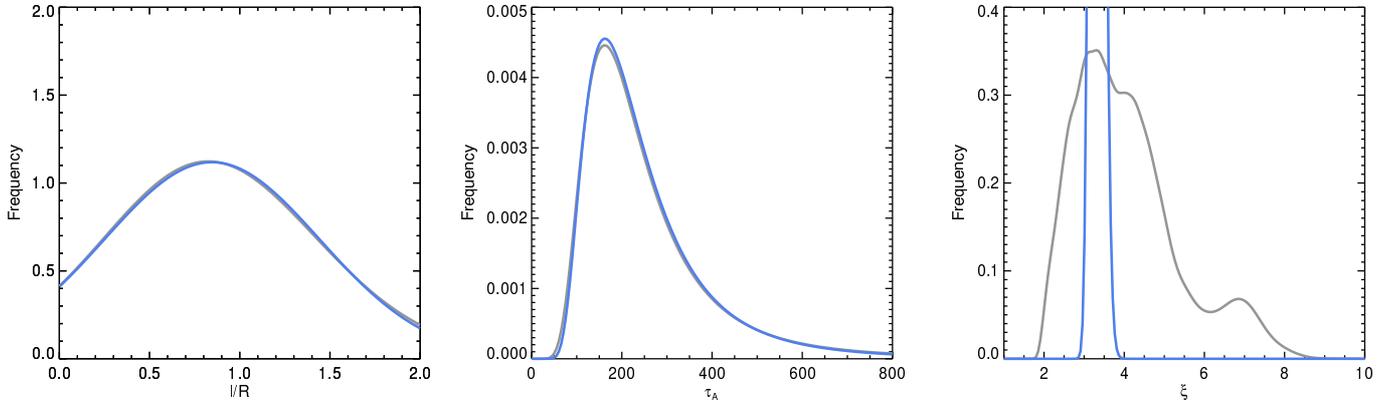}
\caption{Inferred distributions for the transverse inhomogeneity length scale (left panel), the Alfv\'en travel time (central panel), and the density contrast between the tube and the environment (right panel). Grey curves represent the marginalised inferred distribution. Blue lines are the distributions evaluated at the peak of the distributions of the hyperparameters. From \cite{asensioramos13}.}
\label{hyperpriors}
\end{figure*}

\noindent
where the coefficients $a_i$ are obtained for each coronal loop and the order $N$ is adapted to the needed complexity. For simplicity, the values of the $a_i$ coefficients were fixed to those obtained by \cite{aschwanden02}.

The sole use of the generative model does not enable to extract much information. This is complemented with the adoption of a model to explain the observed period and damping. Assuming resonant absorption, these quantities are given by expressions~(\ref{analyticaltttb}).

Following the hierarchical inference scheme described in Section~\ref{hierarchicalbayes},  \cite{asensioramos13} consider that the oscillatory displacement of the $i$-th coronal loop is determined by the set of parameters $\thetabold_i=\{\tau_{A},\xi,l/R,A,\phi_0,\sigma_b\}$.  The vector $\Thetabold=\{\thetabold_1,\thetabold_2,\ldots,\thetabold_n\}$ of length $6N$ then contains all the model parameters for $N$ observed loops.  Once appropriate likelihood and priors are considered, the scheme is applied to a set of 30 observed loop oscillations and a Markov Chain Monte Carlo method is used to sample the posterior in Eq.~(\ref{eq:marginal_hyperparameters}). The marginal posteriors for a sample of 5 among the 30 coronal loops show that the Alfv\'en travel time, the density contrast, and the transverse inhomogeneity length scale  are well constrained.  The hierarchical structure of the model enables to obtain the general properties of coronal loops. To do so, the inferred distributions for the hyperparameters have first to be computed. Once the hyperparameters are known, \cite{asensioramos13} use this information to get the global properties of the physical properties of coronal loops. 

\begin{figure*}
\centering
\includegraphics[width=0.48\textwidth,angle=0]{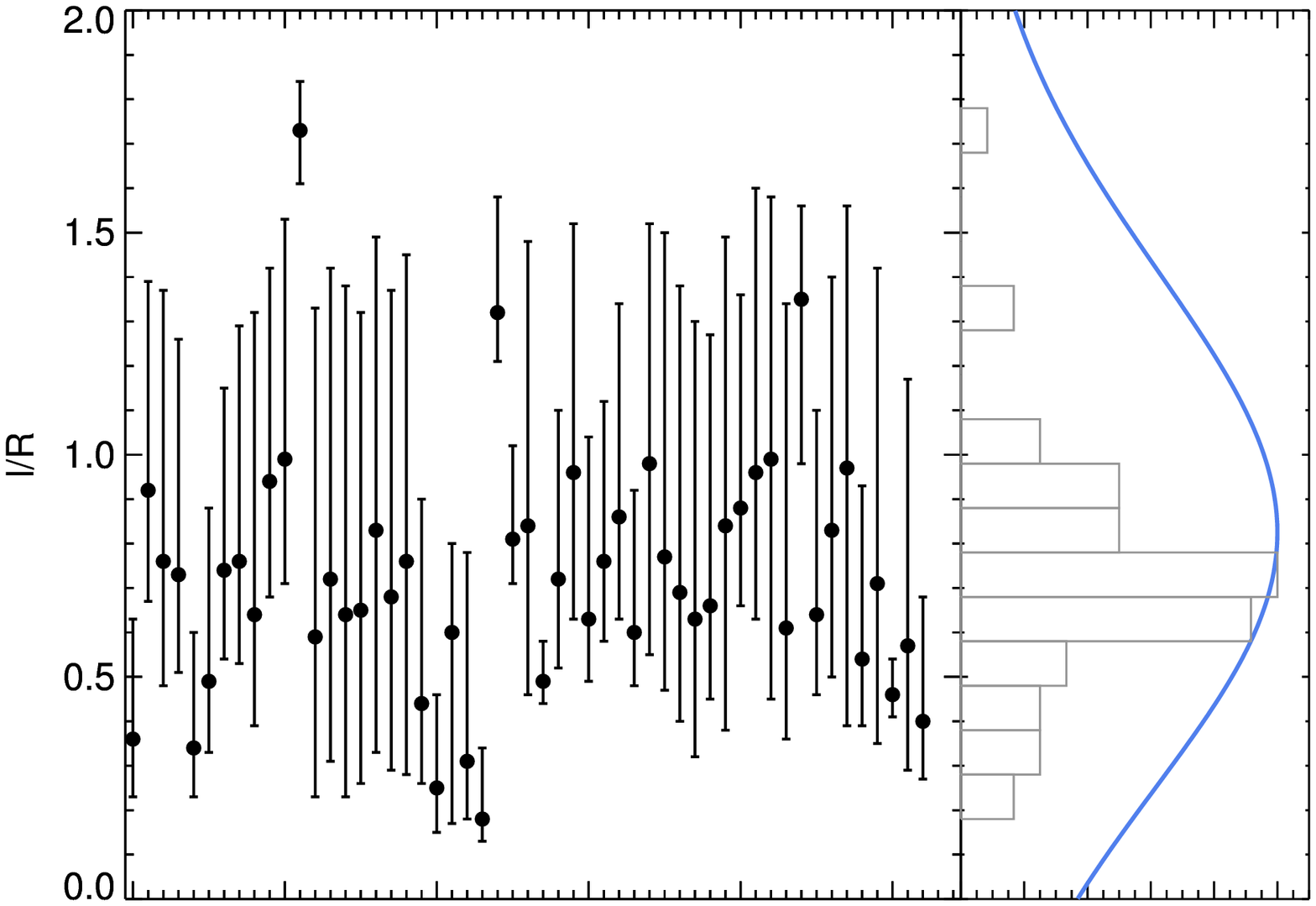}
\includegraphics[width=0.48\textwidth,angle=0]{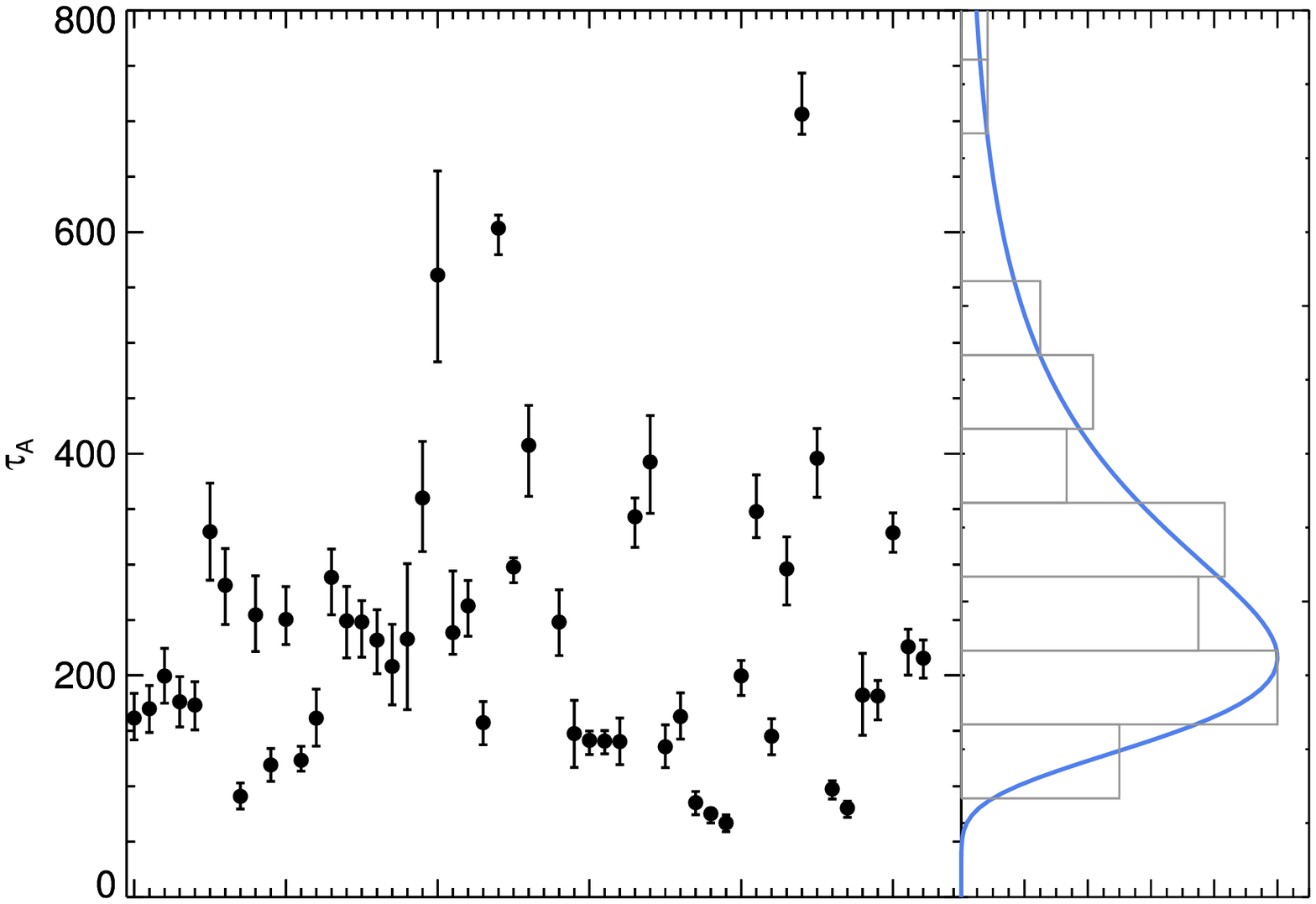}

\caption{Comparison between the inferred distributions shown in Figure~\ref{hyperpriors} and a simple histogram carried out with the inferred values of $\tau_{\rm A}$ and $l/R$ using the formalism of \cite{arregui11b}. From \cite{asensioramos13}.}
\label{comparisonandres}
\end{figure*}

\cite{asensioramos13} first use the so-called type-II maximum likelihood approximation. This consist of evaluating the parametric priors at the most probable values of their parameters, which are obtained from the peaks of the distributions for the hyperparameters. Figure~\ref{hyperpriors} shows the inferred distributions for the unknown parameters as blue lines.  This distributions represent the underlying distribution from which the values of $l/R$, $\tau_{\rm Ai}$, and $\zeta$ have been sampled, under the assumption that this global distribution is shared among all the coronal loops. They can be viewed as  generalised histograms of unobserved quantities. They represent  data-favoured updated priors for the parameters of the model that can be used in any future modelling. The distributions indicate that roughly all allowed values are possible for the transverse inhomogeneity length scale, with a preference for $l/R < 1$. The most probable value for $\tau_{\rm Ai}$ is $\sim$160 s, with a median value of $\sim$212 s.  The density contrast is above 2.3 and below 6.9 with 95\% probability, with a median value of 3.8.

These results are then compared by \cite{asensioramos13} with what one would obtain using a simple histogram with the inferred parameter values using the method by \cite{arregui11b}  to the observations collected in Table 1 of \cite{verwichte13b}. Figure~\ref{comparisonandres} shows that the results are somehow comparable, although the error bars are not taken into account in the histogram. 

In summary, the hierarchical Bayesian analysis by \cite{asensioramos13} offers a tool to obtain global physical information from the analysis of a large set of coronal loops. The resulting distributions contain information that summarises what can be inferred from the analysis of such a large set and that can thereafter be used to construct informative priors in the inversion of individual events.

%%%%%%%%%%%%%%%%%%%%%%%%%%%%%%%%%%%%%%%%%%%%%%%%%%%%%%%%%%%%%%%%%%%%%
\subsection{Coronal loop density profile from measured displacement time-series and forward modelled EUV emission}\label{pascoeresults}
%%%%%%%%%%%%%%%%%%%%%%%%%%%%%%%%%%%%%%%%%%%%%%%%%%%%%%%%%%%%%%%%%%%%%

The results described in Sects.~{\ref{gaussexp} and \ref{densityacross} and those obtained by \cite{pascoe16} have shown that the kink mode damping rate and the use of the Gaussian and exponential damping regimes are powerful diagnostic tools to infer the coronal loop density profile. The use of these tools has been recently extended by including additional physical effects, such as a time-dependent period; the presence of additional longitudinal harmonics; or the decay-less regime of standing kink oscillations by \cite{pascoe17a}. 

In their Bayesian analysis, \cite{pascoe17a} consider a general damping profile to model the measured displacement as a function of time. The model consist of up to four oscillatory components in addition to the background trend. The components enable to take into account: the time-dependent period of oscillations, by means of a time-dependent Alfv\'en transit time; the possible presence of longitudinal harmonics; and a decayless component. In contrast to \cite{arregui11b,arregui13a,arregui13b,arregui14}, who perform the inference from measured periods and damping times, but similarly to \cite{asensioramos13}, the analysis is applied to the time series of a transverse oscillation using the measured loop positions $Y_{\rm i}$ at times $t_{\rm i}$.  The background trend is simultaneously fitted with the harmonic components. 

\begin{figure*}
\centering
\includegraphics[width=0.49\textwidth,angle=0]{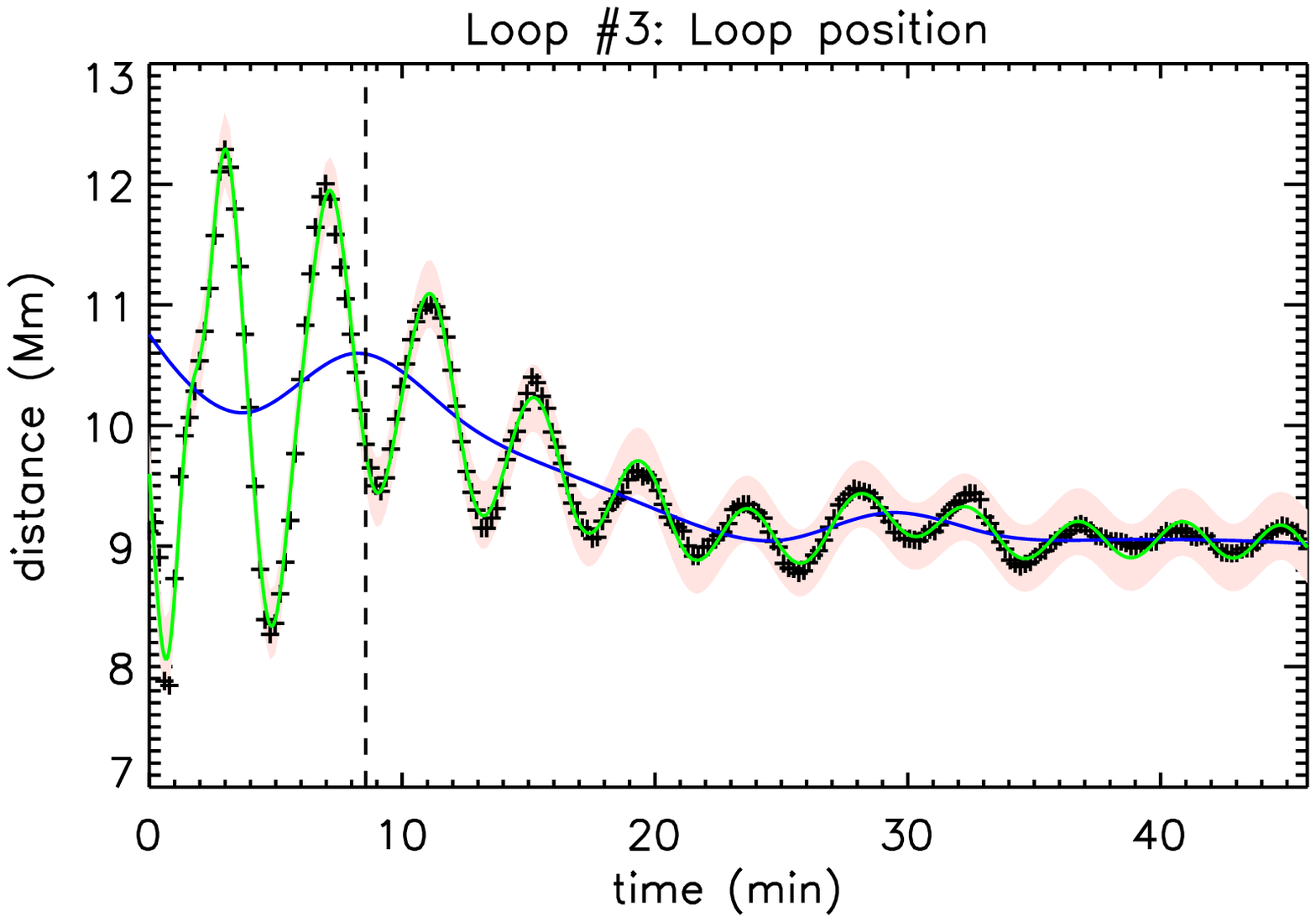}
\includegraphics[width=0.49\textwidth,angle=0]{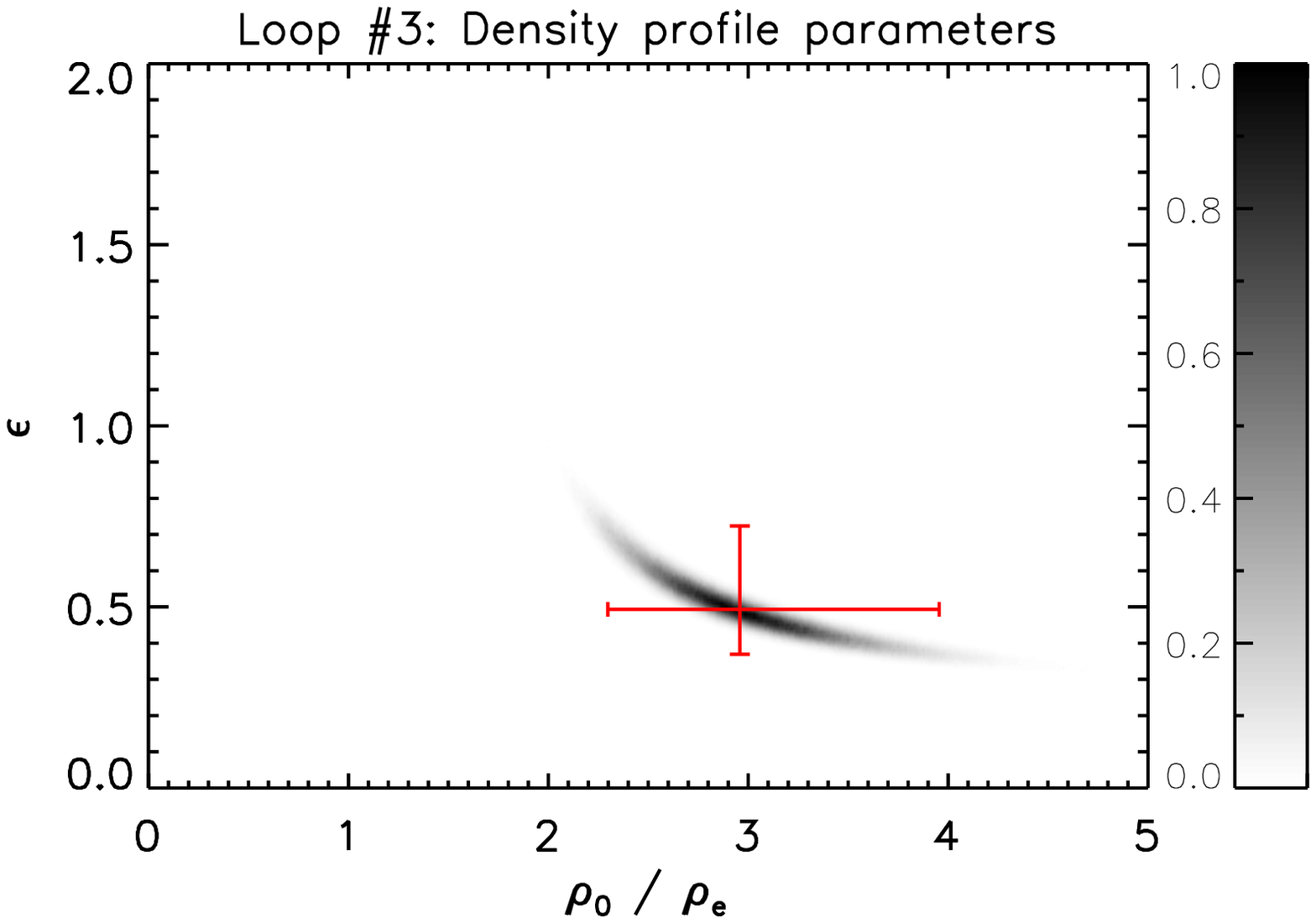}
\caption{Analysis for Loop 3 in \cite{pascoe17a} using a uniform and infinitely thin loop model including the decayless regime. Left: loop position (symbols) as a function of time described by the model (green line) which includes a background trend (blue line). The red shaded region represents the 99\% credible intervals for the loop position predicted by the model, including an estimated noise. The dotted and dashed lines denote the start time of the oscillation and the time when the switch between Gaussian and exponential damping profiles occurs, respectively. Right: density profile parameters determined by the oscillation damping envelope. The red bars are based on the median values and the 95\% credible intervals. Adapted from \cite{pascoe17a}.}
\label{seismologypascoe}
\end{figure*}

The analysed loop oscillations are selected from a catalogue compiled by \cite{zimovets15}. An example is shown in Figure~\ref{seismologypascoe}. The measurement of the loop position in time shows that the aharmonic shape of the signal for the first couple of cycles is reproduced by small amplitude second and third harmonics. The oscillation also contains a decayless component. Evidence is gathered by an additional analysis involving Bayesian model comparison and the computation of Bayes factors. There is strong evidence  for the stratified model over the uniform loop model with Bayes factor $\sim$ 12.6, and strong evidence for the uniform loop model over the expanding model with Bayes factor $\sim$ 6.1. The right panel in Figure~\ref{seismologypascoe} shows the seismologically inferred density contrast and transverse inhomogeneity length scale.  We note that the thin boundary approximation was used in the theoretical modelling, while the inference result leads to not so thin layer estimates, with $l/R\sim 0.6$. This example shows how the combination of Bayesian inference, spline interpolation of the background trend, and Markov Chain Monte Carlo sampling offers an accurate estimate of a range of model parameters and an efficient handling of models with a large number of parameters. In the example here shown, there are 26 varied parameters describing the four components of the oscillation, the damping, the background trend, and the estimated noise. The use of Bayes factors is also useful to test for the possible existence of additional harmonics.

\begin{figure*}
\centering
\includegraphics[width=0.49\textwidth,angle=0]{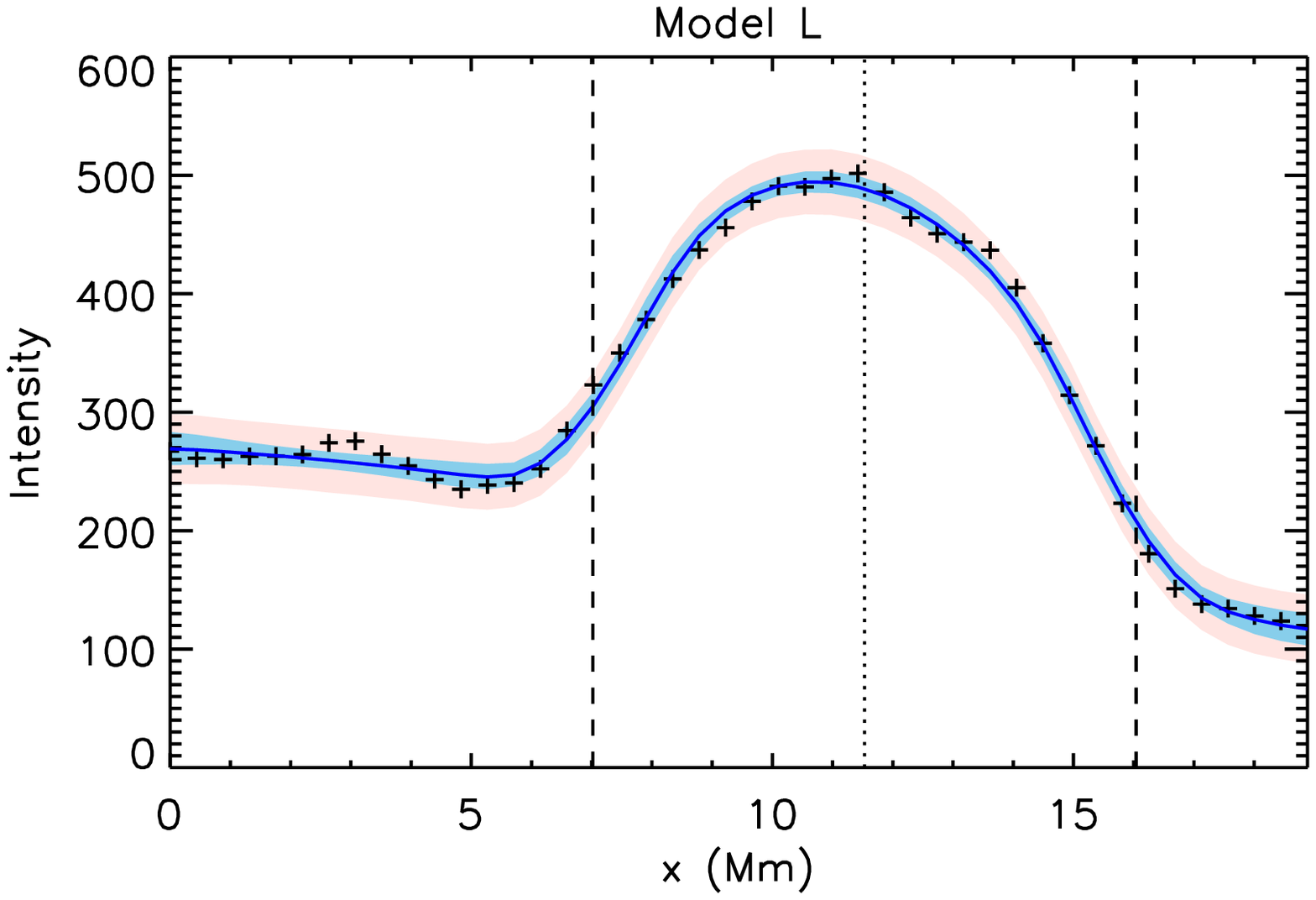}
\includegraphics[width=0.49\textwidth,angle=0]{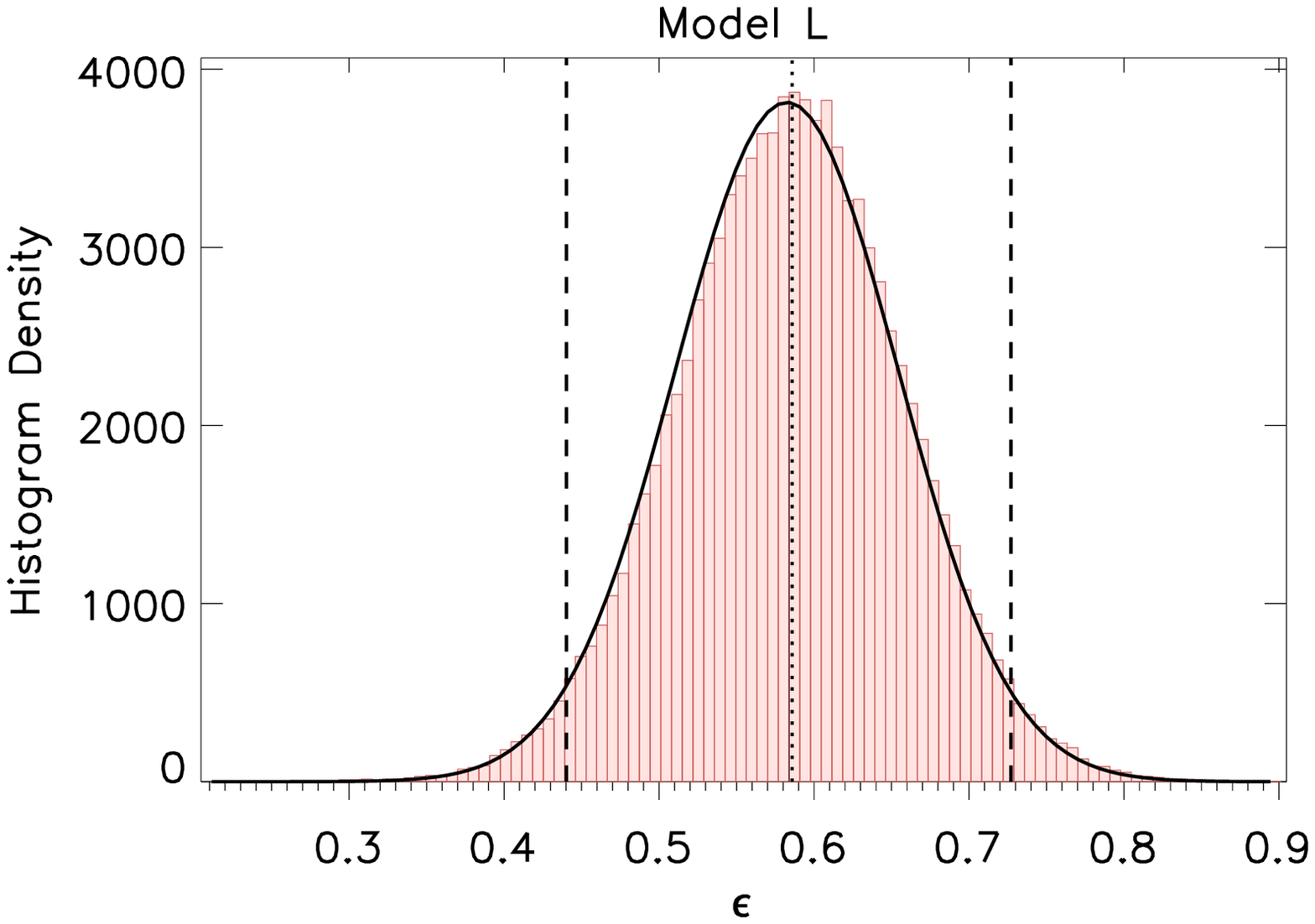}
\caption{Left: SDO/AIA 171 Å EUV intensity (left) across the loop is described by Model L (blue line) which includes a background trend described by a second order polynomial. The shaded regions represent the 99\% confidence intervals for the intensity predicted by the model, with (red) and without (blue) modelled noise. The vertical dotted and dashed lines denote x$_0$ and x0 ± R, respectively. Right: Histogram of normalised layer width $\epsilon=l/R$ based on 105 samples. The vertical dotted and dashed lines denote the median value and the 95\% credible interval, respectively. Adapted from \cite{pascoe17b}.}
\label{forwardpascoe}
\end{figure*}

As a follow-up study,  \cite{pascoe17b} have analysed to what extent the derived density profile accounts for the observed intensity profile of the loop and how the transverse intensity profile may be used to complement the seismological technique. This was done by forward modelling of the EUV intensity using the isothermal and optically transparent approximations for which the intensity of EUV emission is proportional to the square of the plasma density integrated along the line of sight. 

Four alternative models for the transverse loop structure were considered: the step density profile (S); the generalised Epstein density profile (E); the linear transition layer profile (L); and a Gaussian density profile (G). For each of these models, the integrated loop intensity is calculated by numerically constructing a 2D Cartesian density profile for the radial profiles. The emission from the background plasma is incorporated as well. Then, the intensity profile is smoothed to simulate the effect of the point spread function. The modelled intensity profile is then compared with the observational data and values for the Bayes factors are computed.

The method is applied to the same Loop 3 in the discussion above. Figure~\ref{forwardpascoe} shows the obtained results. The left panel shows the observational intensity profile (symbols) and the forward modelled linear profile (blue line). The right panel shows the histogram for the transverse inhomogeneity layer. By comparing the obtained result for $l/R$ using this method with the one obtained by \cite{pascoe17a}, the authors find that these two different approaches give consistent estimates for this density structure parameter. The size of the layer is consistent with the linear model, but also with the Epstein profile. This application shows how the density profile parameter $l/R$ may be estimated by forward modelling of the transverse intensity profile. Note that it this case the loop oscillation is not a requirement for the method to be applied. This study also  shows how Bayes factors can be useful to compare different density profile models.

%%%%%%%%%%%%%%%%%%%%%%%%%%%%%%%%%%%%%%%%%%%%%%%%%%%%%%%%%%%%%%%%%%%%%
\section{Summary}\label{summary}
%%%%%%%%%%%%%%%%%%%%%%%%%%%%%%%%%%%%%%%%%%%%%%%%%%%%%%%%%%%%%%%%%%%%%

We have presented the fundamental concepts of Bayesian methods and their first applications to seismology of the solar atmosphere.  We argued that because seismology diagnostics of the physical conditions and processes in the solar atmosphere has to be performed under conditions of limited and uncertain information, our conclusions need to be given in terms of probability statements. These probability statements are obtained in the form of marginal posterior distributions for the grade of belief on unknown parameters taking on given values, marginal likelihoods for the plausibility of the measured data under the assumption of a given theoretical model, and Bayes factors to measure the relative performance of alternative models in explaining the same data.

The application of Bayesian seismology techniques has been successful in the task of inferring parameters such as the internal Alfv\'en speed and the transverse inhomogeneity length-scale in coronal loops, the coronal density scale height, or the magnetic field expansion rate, all of them conditional on the measured wave properties and taking into account their uncertainty.  The method incorporates consistently calculated credible intervals and propagation of uncertainty from measured wave properties to inferred physical parameters.
Bayesian analysis also enables to perform model comparison. Initial applications have provided the assessment of alternative plasma density structuring models along and across the magnetic field in coronal loops and prominence fine structures.  When the evidence does not support a particular model strong enough, model averaging offers the most general inference. This is the most general inference one can perform. It uses all the available information in a consistent way: prior information, data with uncertainty, model predictions, and model evidence in view of data.

The future of Bayesian inversion techniques in solar atmospheric seismology looks encouraging. As we improve our analytical and theoretical models, considering more realistic representations of MHD waves, and gather higher quality data from space observatories the need to develop self-consistent methods for a proper comparison between theory and observations increases. The future prospects in the application of Bayesian methods to another unsolved problems related to solar atmospheric wave dynamics are promising. Possible future applications of the Bayesian methodology should consider problems such as the assessment of the wave vs. flow character of the observed disturbances or the discrimination between alternative heating mechanisms, both wave-based or otherwise. 

\vspace{0.2cm}
Acknowledgements: I am grateful to the organisers of the first Dynamic Sun Meeting in Varanasi, where this work was presented.  I acknowledge financial support from the Spanish Ministry of Economy and Competitiveness (MINECO) through projects AYA2014-55456-P (Bayesian Analysis of the Solar Corona), AYA2014-60476-P (Solar Magnetometry in the Era of Large Telescopes), from FEDER funds, and through a Ram\'on y Cajal fellowship. I am grateful to the referees and to David J. Pascoe for comments that helped to improve the manuscript. I want to express my gratitude to Andr\'es Asensio 
Ramos for helping me initiate the field of Bayesian coronal seismology.

\section{References}
%\bibliographystyle{elsarticle-harv}
%\bibliography{/Users/arregui/Dropbox/work/arregui}

%%%%%%%%%%%%%%%%%%%%%%%%%%%%%%%%%%%%%%%%%%%%%%%%%%%%%%%%%%%%%%%%%%%%%%%%%%%%%
%% Appendices
% \appendix

\end{document}